\documentclass[11pt]{scrartcl}
\DeclareOldFontCommand{\rm}{\normalfont\rmfamily}{\mathrm}
\DeclareOldFontCommand{\sf}{\normalfont\sffamily}{\mathsf}
\DeclareOldFontCommand{\tt}{\normalfont\ttfamily}{\mathtt}
\DeclareOldFontCommand{\bf}{\normalfont\bfseries}{\mathbf}
\DeclareOldFontCommand{\it}{\normalfont\itshape}{\mathit}
\DeclareOldFontCommand{\sl}{\normalfont\slshape}{\@nomath\sl}
\DeclareOldFontCommand{\sc}{\normalfont\scshape}{\@nomath\sc}
\DeclareRobustCommand*\cal{\@fontswitch\relax\mathcal}
\DeclareRobustCommand*\mit{\@fontswitch\relax\mathnormal}

\usepackage{titling}
\predate{}
\postdate{}

\usepackage[utf8]{inputenc}
\usepackage{color}
\usepackage{type1cm}
\usepackage{graphicx}
\usepackage[usenames,svgnames]{xcolor}
\usepackage[hyphens]{url}
\usepackage[colorlinks=true,linkcolor=blue,citecolor=blue,breaklinks=true]{hyperref}
\usepackage[numbers,sort&compress]{natbib}

\usepackage{amsmath,amssymb,bm,amsthm,cases}
\usepackage{physics}
\usepackage{mathtools}
\usepackage{cleveref}

\usepackage{authblk}

\newcommand{\e}{e}
\newcommand{\im}{i}

\newcommand{\HF}{\hat{H}_\mathrm{F}}
\newcommand{\HS}{\hat{H}_\mathrm{S}}
\newcommand{\HB}{\hat{H}_\mathrm{B}}
\newcommand{\HI}{\hat{H}_\mathrm{I}}
\newcommand{\HT}{\hat{H}_\mathrm{T}}

\newcommand{\rhoS}{\rho_\mathrm{S}}
\newcommand{\rhoB}{\rho_\mathrm{B}}
\newcommand{\rhoT}{\rho_\mathrm{T}}
\newcommand{\tilderhoS}{\tilde{\rho}_\mathrm{S}}
\newcommand{\TrB}{\Tr_\mathrm{B}}
\newcommand{\LS}{\mathcal{L}_\mathrm{S}}
\newcommand{\LB}{\mathcal{L}_\mathrm{B}}
\newcommand{\LI}{\mathcal{L}_\mathrm{I}}

\newcommand{\tauS}{\tau_\mathrm{S}}
\newcommand{\tauB}{\tau_\mathrm{B}}
\newcommand{\tauR}{\tau_\mathrm{R}}
\newcommand{\tauH}{\tau_\mathrm{H}}

\begin{document}
\nocite{apsrev41control}
\title{\LARGE Floquet States in Open Quantum Systems}
\author[1]{\normalsize Takashi Mori}
\affil[1]{\it RIKEN Center for Emergent Matter Science (CEMS), Wako 351-0198, Japan}
\date{}
\maketitle

\begin{abstract}
In Floquet engineering, periodic driving is used to realize novel phases of matter which are inaccessible in thermal equilibrium.
For this purpose, the Floquet theory provides us a recipe of obtaining a static effective Hamiltonian.
Although many existing works have treated closed systems, it is important to consider the effect of dissipation, which is ubiquitous in nature.
Understanding interplay of periodic driving and dissipation is not only a fundamental problem of nonequilibrium statistical physics, but also receiving growing interest because of the fact that experimental advances have allowed us to engineer dissipation in a controllable manner.
In this review, we give a detailed exposition on the formalism of quantum master equations for open Floquet systems and highlight recent works investigating whether equilibrium statistical mechanics applies to Floquet states.
\end{abstract}

\tableofcontents

\section{INTRODUCTION}
Nonequilibrium states are not well understood as compared with equilibrium states, except for weakly driven systems described by the linear response theory~\citep{Kubo1957}.
One of the recent progresses of nonequilibrium statistical physics is a discovery of fluctuation theorems~\citep{Evans1993, Gallavotti1995, Jarzynski1997}, which yield nontrivial equalities for arbitrarily far-from-equilibrium systems.
Although fluctuation theorems reveal universal properties of atypical fluctuations in nonequilibrium states, they do not give much insights into typical behavior of the system.

There is another important class of nonequilibrium problems for which we have gained a deeper understanding of generic properties through recent studies, that is, periodically driven (Floquet) systems in the high-frequency regime.
Nonequilibrium dynamics of closed Floquet systems have intensively been studied~\citep{Bukov2015_review, Eckardt2017_review}.
In a closed system, the second law of thermodynamics implies that the system absorbs the energy from periodic driving and eventually heats up to an infinitely high temperature.
However, it is revealed that the heating process is exponentially slow in the high-frequency regime and we can have nontrivial transient states which last in a relatively long timescale~\citep{Mori2016b, Kuwahara2016, Abanin2017, Abanin2017a}.
Theoretical understanding of this phenomenon gives a solid basis for the ``Floquet engineering'', which is an attempt to realize desired states by utilizing engineered periodic driving~\citep{Oka2019_review}.
Periodic driving has been used to realize nonequilibrium phase transitions~\citep{Eckardt2005, Zenesini2009, Bastidas2012}, topologically nontrivial systems~\citep{Oka2009, Kitagawa2010, Lindner2011, Jotzu2014, Aidelsburger2015}, artificial gauge fields~\citep{Aidelsburger2011, Struck2012, Bermudez2011}, and discrete time crystals~\citep{Else2016, Else2017, Yao2017}.

In this review, we discuss open Floquet systems.
In the context of the Floquet engineering, the motivation of considering open systems is two-fold.
First, although the closed-system description would be adequate in ultracold atomic systems, where high isolation from the environment is achieved, dissipation is inevitable in condensed matter physics.
Dissipation can largely affect many-body states, and hence it is clearly important to formulate the problem in an open-system setup to develop the theory of Floquet engineering in condensed matter systems~\citep{Tsuji2008, Tsuji2009, Dehghani2014, Dehghani2015, Seetharam2015, Iadecola2015, Murakami2017}.
For instance, the importance of dissipation has been pointed out in recent experimental~\citep{McIver2020} and theoretical~\citep{Sato2019} studies on the light-induced anomalous Hall effect in graphene.
Second, it has been recognized that dissipation can be used as a resource for creating novel nonequilibrium states. Indeed, the balance of periodic driving and dissipation can yield a variety of nonequilibrium steady states and phase transitions in various systems including cavity-QED systems~\citep{Drummond1980, Baumann2010, DallaTorre2013, Shirai2014, Foss-Feig2017}, cold atoms~\citep{Diehl2008, Diehl2011}, ideal Bose gases~\citep{Vorberg2013, Vorberg2015, Schnell2018}, and so on.
This direction of research has attracted growing attention in the field of ultracold atomic physics, reflecting recent experimental development that enables us to engineer dissipation in a highly controllable manner~\citep{Barreiro2011, Barontini2013, Tomita2017}.

The interplay of periodic driving and dissipation is also intriguing from a purely theoretical perspective.
The Floquet theory tells us that a Floquet system is described by an effective static Hamiltonian, and hence it is a fundamental question to what extent the notion of equilibrium statistical thermodynamics applies to the effective Hamiltonian description of Floquet systems~\citep{Breuer1991, Breuer2000, Kohn2001, Hone2009, Ketzmerick2010, Liu2015, Shirai2015, Shirai2016}, especially beyond the linear response regime.

The purpose of this review is to give a detailed description of theoretical formulation of open Floquet systems and an overview of recent studies on statistical mechanics for Floquet systems.
The organization of this review is as follows.
In \cref{sec:Floquet}, we present an introduction to the Floquet theory for closed systems.
We explain the method of the high-frequency expansion and discuss its generic properties.
A brief exposition on recent theoretical studies is also given.
In \cref{sec:master}, we give a detailed description of the master equation approach to open Floquet systems.
We introduce several kinds of quantum master equations and discuss their limitations.
In \cref{sec:stat}, we focus on nonequilibrium periodic steady states under fast driving.
Here we deal with a fundamental question to what extent the method of statistical mechanics applies.
In \cref{sec:outlook}, we discuss some related recent topics and provide an outlook for future studies.

\section{FLOQUET THEORY FOR CLOSED SYSTEMS}
\label{sec:Floquet}

\subsection{Floquet Theorem}
First, let us start with discussion on how to describe Floquet systems without dissipation.
We denote by $\hat{H}_0$ and $\hat{V}(t)$ the reference Hamiltonian and the driving field of a quantum system, respectively.
We assume that $\hat{V}(t)$ is periodic, $\hat{V}(t)=\hat{V}(t+T)$, with the period $T$.
Without loss of generality, we can put $\int_0^T\dd t\,\hat{V}(t)=0$.
We define $\omega=2\pi/T$, which is simply called the frequency of the driving field, although the driving field is not necessarily monochromatic and generally written as $\hat{V}(t)=\sum_{m\neq 0}V_m\e^{-\im m\omega t}$.
The state $\ket{\psi(t)}$ of the system evolves under the Schr\"odinger equation
\begin{equation}
\im\dv{t}\ket{\psi(t)}=\hat{H}(t)\ket{\psi(t)},
\label{eq:sch}
\end{equation}
where $\hat{H}(t)=\hat{H}_0+\hat{V}(t)$ and we set $\hbar=1$.

Throughout this article, we consider a local Hamiltonian, i.e., a short-range interacting system, on a regular lattice.
The characteristic local energy scale is denoted by $g$.
More explicitly, in a lattice system, we define $g$ as the maximum single-site energy~\citep{Kuwahara2016, Mori2016b}.

A convenient theoretical tool to deal with a set of time-periodic linear differential equations like \cref{eq:sch} is the Floquet theorem.
The Floquet theorem states that the time evolution operator from time $t_0$ to time $t$ is written in the following form~\citep{Bukov2015_review}:
\begin{align}\label{eq:decomposition}
\hat{U}_{t,t_0}=\e^{-\im\hat{K}(t)}\e^{-\im\hat{H}_\mathrm{F}(t-t_0)}\e^{\im\hat{K}(t_0)},
\end{align}
where $\hat{K}(t)$ is called the micromotion operator or the kick operator, which is periodic in time $\hat{K}(t)=\hat{K}(t+T)$, and $\hat{H}_\mathrm{F}$ is the time-independent Floquet Hamiltonian.
Physically, $\hat{K}(t)$ describes fast periodic motion of the system, whereas $\hat{H}_\mathrm{F}$ determines the long-time behavior of the system.
In closed systems, we can always choose $\hat{K}(t)$ and $\hat{H}_\mathrm{F}$ to be Hermitian.

Eigenstates $\qty{\ket{u_a}}$ and eigenvalues $\qty{\varepsilon_a}$ of $\HF$ are called Floquet eigenstates and quasi-energies, respectively.
We also define $\ket{u_a(t)}=\e^{-\im\hat{K}(t)}\ket{u_a}$, which satisfies $\ket{u_a(t)}=\ket{u_a(t+T)}$.
A solution of the Schr\"odinger equation is then written in the form $\ket{\psi(t)}=\sum_ac_a\e^{-\im\varepsilon_at}\ket{u_a(t)}$, which can be regarded as an analogue of the Bloch theorem in condensed matter physics.
It should be noted that quasi-energies are defined modulo $\omega$: \cref{eq:decomposition} does not change if we replace $\varepsilon_a\to\varepsilon_a+m\omega$ with $m$ an arbitrary integer.
In literature, $\varepsilon_a$ is often chosen within the ``first Brillouin zone'',  $\varepsilon_a\in [-\omega/2,\omega/2)$ or $\varepsilon_a\in[0,\omega)$.
It is also convenient to choose $\varepsilon_a$ so that the value of $\varepsilon_a$ is closest to the mean energy $\bar{E}_a=(1/T)\int_0^T\dd t\,\expval{\hat{H}(t)}{u_a(t)}$, which will be used in \cref{sec:stat}.

By substituting \cref{eq:decomposition} into the Schr\"odinger equation $\im\dd \hat{U}_{t,t_0}/\dd t=\hat{H}(t)\hat{U}_{t,t_0}$, we obtain
\begin{align}
\hat{H}_\mathrm{F}=\e^{\im\hat{K}(t)}\qty[\hat{H}(t)-\im\dv{t}]\e^{-\im\hat{K}(t)}.
\label{eq:HF_K}
\end{align}
This expression is equivalent to \cref{eq:decomposition}.

We remark that the decomposition of \cref{eq:decomposition} is not unique: there are infinitely many variants of the decomposition, which are equivalent with each other but offer different approximation schemes.
Indeed, one can confirm that \cref{eq:decomposition} or \cref{eq:HF_K} is invariant under the transformation $\HF'=\hat{U}\HF\hat{U}^\dagger$ and $\e^{\im\hat{K}'(t)}=\hat{U}\e^{\im\hat{K}(t)}$ for any unitary operator $\hat{U}$.
One familiar choice is putting $\hat{K}(t_0)=0$ for a fixed $t_0\in[0,T)$.
In this choice, the time evolution operator over a cycle $U_{t_0+T,t_0}=\mathcal{T}\e^{-\im\int_{t_0}^{t_0+T}\dd t\,\hat{H}(t)}$, where $\mathcal{T}$ denotes the time-ordering operation, is expressed as $\hat{U}_{t_0+T,t_0}=\e^{-\im\HF T}$.
Thus, $\HF$ exactly describes stroboscopic evolution, which is defined at $t=t_0+nT$ with $n$ being an integer.

\subsection{High-Frequency Expansion}
Although the decomposition of \cref{eq:decomposition} is exact, it is hard to calculate $\hat{H}_\mathrm{F}$ and $\hat{K}(t)$.
It is also difficult to gain some insights into the nonequilibrium dynamics from $\hat{K}(t)$ and $\hat{H}_\mathrm{F}$ since no analytic expression is available.

In the high-frequency regime, we can obtain approximate analytic expressions via the method of the high-frequency expansion, which is written in the form
\begin{align}
\hat{K}(t)=\sum_{k=1}^\infty\frac{\hat{\Lambda}_k(t)}{\omega^k} \qq{and} \hat{H}_\mathrm{F}=\hat{H}_0+\sum_{k=1}^\infty\frac{\hat{\Omega}_k}{\omega^k}.
\label{eq:high-freq}
\end{align}
By substituting \cref{eq:high-freq} into \cref{eq:HF_K}, we can determine $\hat{\Lambda}_k(t)$ and $\hat{\Omega}_k$ order by order.

Reflecting the non-uniqueness of $\hat{K}(t)$ and $\hat{H}_\mathrm{F}$, there are several high-frequency expansions~\citep{Mikami2016}.
A familiar one is the high-frequency expansion with the boundary condition $\hat{K}(t_0)=0$, which is referred to as the Floquet-Magnus expansion~\citep{Blanes2009}.
In the order of $\omega^{-1}$, we have
\begin{align}
\hat{\Lambda}_1^\mathrm{FM}(t)=\im\sum_{m\neq 0}\frac{\hat{V}_m}{m}\qty(\e^{-\im m\omega t}-\e^{-\im m\omega t_0}), \quad
\hat{\Omega}_1^\mathrm{FM}=\sum_{m\neq 0}\frac{[\hat{V}_{-m},\hat{V}_m]}{2m}+\sum_{m\neq 0}\frac{[\hat{V}_m,\hat{H}_0]}{m}\e^{-\im m\omega t_0}.
\end{align}
It should be noted that the high-frequency expansion of $\HF$ depends on $t_0$, which is sometimes not desirable because we are interested in the long-time physics independent of the particular choice of $t_0$.
We can also construct the high-frequency expansion so that this undesirable $t_0$ dependence is completely removed.
It is known as the van Vleck expansion~\citep{Rahav2003, Goldman2014, Eckardt2015} and characterized by the boundary condition $\int_0^T\dd t\,\hat{K}(t)=0$.
In the order of $\omega^{-1}$, the van Vleck expansion is given by
\begin{align}
\hat{\Lambda}_1^\mathrm{vV}(t)=\im\sum_{m\neq 0}\frac{\hat{V}_m}{m}\e^{-\im m\omega t}, \quad
\hat{\Omega}_1^\mathrm{vV}=\sum_{m\neq 0}\frac{\qty[\hat{V}_{-m},\hat{V}_m]}{2m}.
\end{align}

Since analytic expressions of $\hat{\Omega}_k$ and $\hat{\Lambda}_k(t)$ are available, the high-frequency expansion provides us much insights into long-time dynamics of Floquet systems.
A generic strategy of the Floquet engineering is to engineer periodic driving so that a truncation of the Floquet Hamiltonian 
\begin{align}
\hat{H}_\mathrm{F}^{(n)}=\hat{H}_0+\sum_{k=1}^n\frac{\hat{\Omega}_k}{\omega^k}
\end{align}
has a desired property~\citep{Oka2019_review}.

It should be noted that if we simply increase the frequency, the leading-order term in \cref{eq:high-freq} is dominant and thus we approximately have $\hat{H}_\mathrm{F}^{(n)}\approx\hat{H}_0$: the periodic driving does not play any role.
In order to realize nontrivial states by using fast periodic driving, we should either (i) strongly or (ii) resonantly drive the system, which we explain in the following.

\subsubsection{Strong Driving}
\label{sec:high_strong}
Let us consider a class of periodic fields of the form $\hat{V}(t)=\omega f(\omega t)\hat{V}$, where $\hat{V}$ is a time-independent operator and $f(\omega t)$ is an arbitrary periodic function satisfying $f(\theta)=f(\theta+2\pi)$ and $\int_0^{2\pi}\dd\theta\, f(\theta)=0$. 
A monochromatic driving $f(\omega t)=\cos(\omega t)$ is often used.
Since the driving amplitude scales with $\omega$, a large $\omega$ compared with a local energy scale $g$ of $\hat{H}_0$ corresponds to fast and strong driving.
The Hamiltonian is given by
\begin{align}
\hat{H}(t)=\hat{H}_0+\omega f(\omega t)\hat{V}.
\label{eq:H_strong}
\end{align}
Under such a strong driving field, the convergence of the high-frequency expansion is very slow.
This difficulty is addressed by moving to a rotating frame via the unitary transformation $\e^{-\im\int_0^t\dd s\,\hat{V}(s)}=\e^{-\im F(\omega t)\hat{V}}$ with $F(\theta)=\int_0^\theta\dd s\, f(s)$.
The Hamiltonian in the rotating frame is given by
\begin{align}
\hat{H}^\mathrm{R}(t)=\e^{-\im F(\omega t)\hat{V}}\hat{H}_0\e^{\im F(\omega t)\hat{V}}=\hat{H}_0^\mathrm{R}+\hat{V}^\mathrm{R}(t),
\end{align}
where $\hat{H}_0^\mathrm{R}=(1/T)\int_0^T\dd t\,\hat{H}^\mathrm{R}(t)$ and $\hat{V}^\mathrm{R}(t)$ are the static Hamiltonian and the driving field in the rotating frame, respectively. 
It is emphasized that $\hat{H}_0^\mathrm{R}$ contains nonperturbative effects of periodic driving.
Since $\hat{H}^\mathrm{R}(t)=\hat{H}^\mathrm{R}(t+T)$ is satisfied, the Floquet theory is still applicable in the rotating frame.
Remarkably, the periodic driving in the rotating frame is no longer strong.
It is therefore expected that the high-frequency expansion quickly converges.
The static part $\hat{H}_0^\mathrm{R}$ gives the leading order term of the high-frequency expansion of the Floquet Hamiltonian.

This technique has been used for Floquet engineering.
As an example, let us consider a one-dimensional Bose-Hubbard system in a shaken optical lattice~\citep{Eckardt2005}:
\begin{align}
\hat{H}_0=-J\sum_{j=1}^{L-1}\qty(\hat{b}_{j+1}^\dagger\hat{b}_j+\hat{b}_j^\dagger\hat{b}_{j+1})+\frac{U}{2}\sum_{j=1}^L\hat{n}_j\qty(\hat{n}_j-1),
\quad
\hat{V}(t)=\omega\xi\cos(\omega t)\sum_{j=1}^Lj\hat{n}_j,
\label{eq:BH}
\end{align}
where $\hat{b}_j^\dagger$ and $\hat{b}_j$ are creation and annihilation operators of bosons at site $j$, and $\hat{n}_j=\hat{b}_j^\dagger\hat{b}_j$.
By putting $f(\theta)=\xi\cos\theta$ and $\hat{V}=\sum_{j=1}^Lj\hat{n}_j$, this model is written in the form of \cref{eq:H_strong}.

In the rotating frame, we truncate the high-frequency expansion in the lowest order:
\begin{align}
\hat{K}(t)\approx 0 \qq{and} \HF\approx\hat{H}_0^\mathrm{R}=\frac{1}{T}\int_0^T\dd t\, \e^{-\im\xi\sin(\omega t)\hat{V}}\hat{H}_0\e^{\im\xi\sin(\omega t)\hat{V}}.
\end{align}
Explicit calculations of $\hat{H}_0^\mathrm{R}$ show that it is identical to the Hamiltonian of the Bose-Hubbard model with an effective tunneling amplitude $J_\mathrm{eff}=J\mathcal{J}_0(\xi)$, where $\mathcal{J}_0(\cdot)$ denotes the Bessel function of order zero.
When $\xi$ is tuned to one of the zeros of the Bessel function, the tunneling is suppressed and bosons are localized.
This is regarded as a many-body variant of the dynamical localization~\citep{Dunlap1986} or the coherent destruction of tunneling~\citep{Grossmann1991}.
From this observation, Eckardt et al.~\cite{Eckardt2005} proposed that we can induce a quantum phase transition between the superfluid phase and the Mott insulator phase by controlling the driving amplitude $\xi$.
This theoretical prediction was confirmed in experiment~\citep{Lignier2007, Zenesini2009}

\subsubsection{Resonant Driving}
\label{sec:high_resonant}
We can also realize a nontrivial truncated Floquet Hamiltonian by using resonant driving~\citep{Eckardt2007, Goldman2015}. 
Suppose that the Hamiltonian $\hat{H}(t)$ is written in the following form: 
\begin{align}
\hat{H}(t)=\hat{H}_0+\omega\hat{N}+\hat{V}(t),
\label{eq:H_resonant}
\end{align}
where $\hat{N}$ is an operator with eigenvalues $\{n_a\}$ such that $n_a-n_b$ is integer for any pair of eigenvalues $n_a$ and $n_b$.
Because of the presence of $\omega\hat{N}$, the high-frequency expansion converges very slowly even for large $\omega$.
We again move to a rotating frame via the unitary transformation $\e^{-\im \omega\hat{N}t}$.
The Hamiltonian $\hat{H}^\mathrm{R}(t)$ in the rotating frame reads
\begin{align}
\hat{H}^\mathrm{R}(t)=\e^{\im\omega\hat{N}t}\qty(\hat{H}_0+\hat{V}(t))\e^{-\im\omega\hat{N}t}
=\hat{H}_0^\mathrm{R}+\hat{V}^\mathrm{R}(t),
\label{eq:H_resonant}
\end{align}
where $\hat{H}^\mathrm{R}(t)=\hat{H}^\mathrm{R}(t+T)$ is satisfied, and $\hat{H}_0^\mathrm{R}=(1/T)\int_0^T\dd t\,\hat{H}^\mathrm{R}(t)$ is the static part of the Hamiltonian in the rotating frame.
Again, $\hat{H}_0^\mathrm{R}$ contains nonperturbative effects of periodic driving.
The high-frequency expansion will quickly converge in the rotating frame.

Later, we will see in \cref{sec:resonant} that an interplay of resonant driving and dissipation enables us to implement a simple quantum master equation with a nontrivial steady state.

\subsection{Floquet Prethermalization}
\label{sec:pre}

\begin{figure}[t]
\centering
\includegraphics[width=0.8\linewidth]{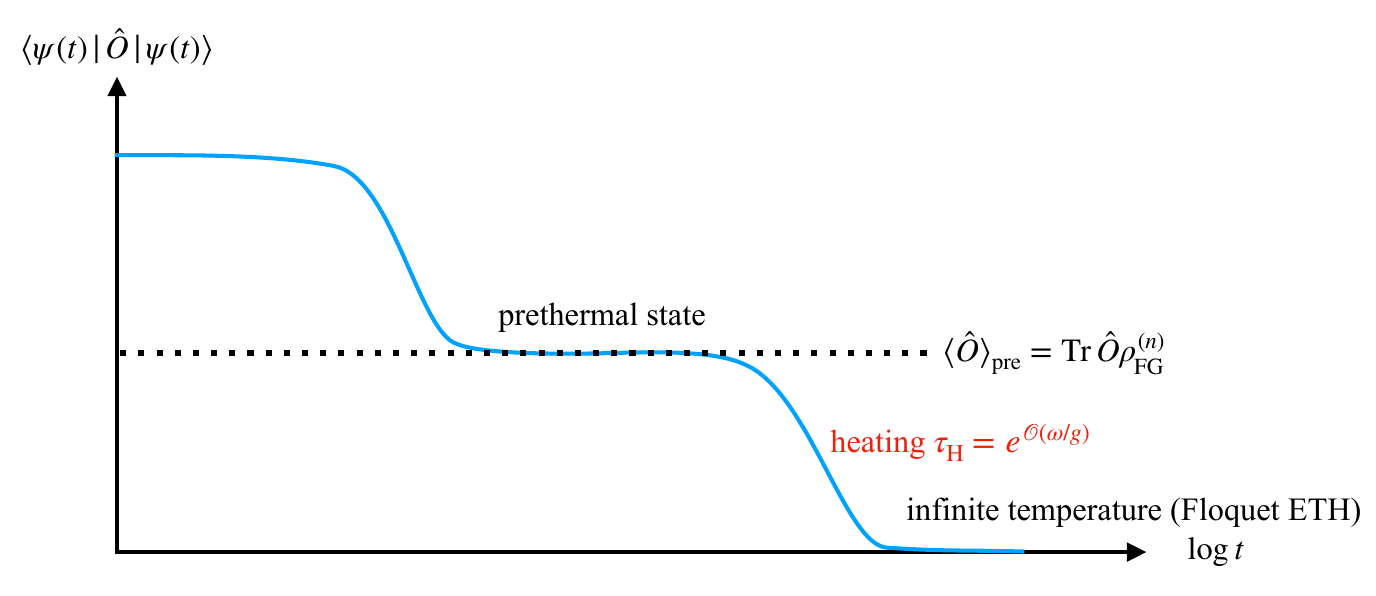}
\caption{Schematic picture of Floquet prethermalization.
The time evolution of $\expval{\hat{O}}{\psi(t)}$ for a local observable $\hat{O}$ exhibits a two-step relaxation.
After the first relaxation, the system reaches a prethermal state described by $\rho_\mathrm{FG}^{(n)}$.
The second relaxation is due to heating, which occurs in a timescale $\tauH$, which is large exponentially in $\omega/g$.}
\label{fig:pre}
\end{figure}

The high-frequency expansion is a powerful tool, but we should be careful about its applicability.
D'Alessio and Rigol~\cite{DAlessio2014} and Lazarides et al.~\cite{Lazarides2014} conjectured through exact diagonalization of $\HF$ that any eigenstate of $\HF$ in a nonintegrable system is locally equivalent to the infinite temperature ensemble for any finite $\omega$ in the thermodynamic limit.
This conjecture has been evidenced by numerical calculations for various nonintegrable models~\citep{DAlessio2014, Lazarides2014, Kim2014}, and is now called the Floquet eigenstate thermalization hypothesis (ETH)~\citep{Mori2018_review} (but also see References~\cite{Das2010, Haldar2018, Haldar2021} for dynamical freezing phenomena).
The Floquet ETH implies that $\hat{H}_\mathrm{F}$ is a highly nonlocal operator although $\HF^{(n)}$ is local for not too large $n$.
It means that the exact Floquet Hamiltonian $\hat{H}_\mathrm{F}$ and a truncation $\hat{H}_\mathrm{F}^{(n)}$ of its high-frequency expansion with a finite $n$ have qualitatively distinct properties.
Indeed, the non-locality of $\HF$ is understood as the divergence of the high-frequency expansion for any finite $\omega$ in the thermodynamic limit~\citep{DAlessio2014, Lazarides2014}.
A rigorous analysis~\citep{Mori2016b} indicates that the divergence begins at an order $n_0\propto\omega/g$, which is large in the high-frequency regime.
When $n<n_0$, the high-frequency expansion looks convergent.

Physically, the Floquet ETH is related to the fact that the system eventually heats up to infinite temperature.
The locality of $\hat{H}_\mathrm{F}^{(n)}$ implies that a truncation of the high-frequency expansion does not capture the heating process.
Heating is a nonperturbative phenomenon in $1/\omega$, and $\HF^{(n)}$ may describe the system before heating takes place.
Indeed, rigorous analysis of the high-frequency expansion has revealed that heating in quantum~\citep{Kuwahara2016, Mori2016b, Abanin2017, Abanin2017a} and classical~\citep{Mori2018} spin systems is exponentially slow in frequency, and a system driven by quickly oscillating external fields generically shows two-step relaxation referred to as the Floquet prethermalization~\citep{Mori2018_review}.
In a timescale much shorter than the heating time $\tauH=\e^{\order{\omega/g}}$, the dynamics is described by $\HF^{(n)}$.
Therefore, the system will first thermalize under $\HF^{(n)}$, and a prethermal state is described by a truncated Floquet-Gibbs state $\rho_\mathrm{FG}^{(n)}=\e^{-\beta\HF^{(n)}}/\Tr\e^{-\beta\HF^{(n)}}$, where $n$ is arbitrary as long as it is smaller than $n_0$.
Here, $\beta$ is the inverse temperature of the system at the initial state $\ket{\psi_\mathrm{ini}}$, which is determined by $\Tr(\HF^{(n)}\rho_\mathrm{FG}^{(n)})=\expval{\HF^{(n)}}{\psi_\mathrm{ini}}$.
In a timescale longer than $\tauH$, the system will heat up to infinite temperature as predicted by the Floquet ETH for $\HF$.
See \cref{fig:pre} for a schematic picture of the Floquet prethermalization.

The Floquet prethermalization has also been shown to occur in classical systems~\citep{Rajak2018, Rajak2019, Hodson2021} and boson systems~\citep{Bukov2015, DallaTorre2021}, which are not covered by rigorous results.
Floquet prethermalization has also been observed in recent experiments~\citep{Rubio-Abadal2020, Peng2021}.

\subsection{Floquet Reference Frame}
\label{sec:reference}

\begin{figure}[t]
\centering
\includegraphics[width=0.9\linewidth]{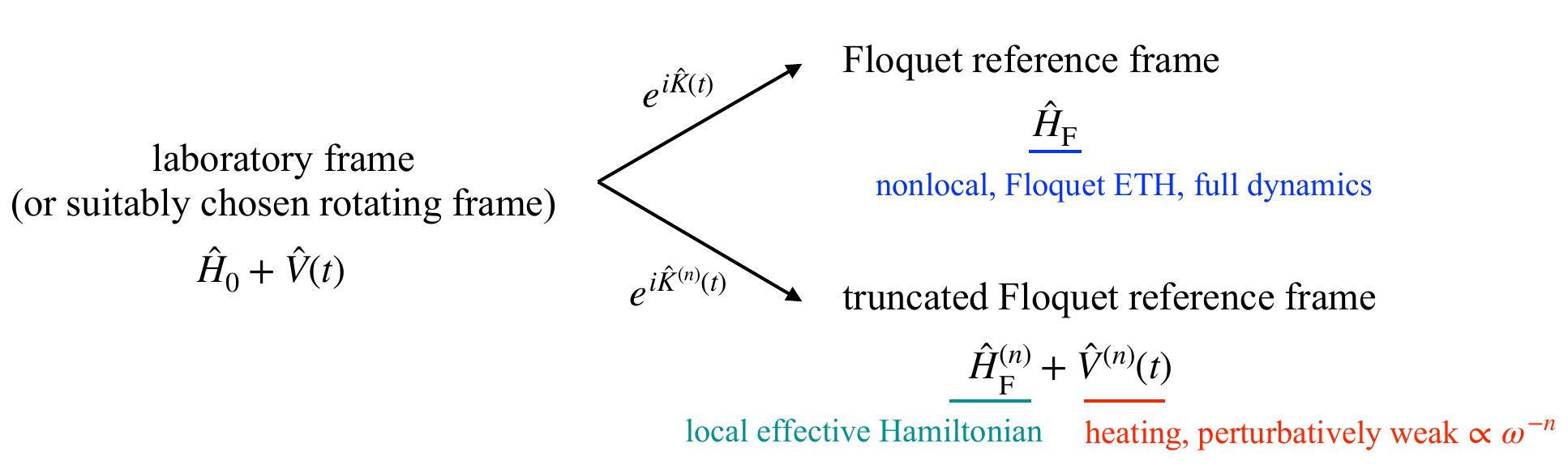}
\caption{Brief summary of the transformation to the (truncated) Floquet reference frame.
In the Floquet reference frame, the Hamiltonian $\HF$ becomes completely time-independent, but is highly nonlocal and obeys the Floquet ETH.
In the truncated Floquet reference frame, the Hamiltonian consists of an effective static Hamiltonian $\HF^{(n)}$ and a perturbatively weak driving field $\hat{V}^{(n)}(t)$.}
\label{fig:frame}
\end{figure}

The decomposition of \cref{eq:decomposition} means that if we move to a special rotating frame via the periodic unitary transformation as $\ket{\psi'(t)}=\e^{\im\hat{K}(t)}\ket{\psi(t)}$, $\HF$ plays the role of the Hamiltonian in that rotating frame: $\im\dd\ket{\psi'(t)}/\dd t=\HF\ket{\psi'(t)}$.
We call this special rotating frame the ``Floquet reference frame'' to distinguish it from other rotating frames introduced in \cref{sec:high_strong,sec:high_resonant}.
The decomposition of \cref{eq:decomposition} is therefore interpreted as follows: we can always find the Floquet reference frame in which the time dependence of the Hamiltonian is completely washed out.

When we consider a many-body system in which the Floquet ETH holds, $\HF$ is too complicated and  the Floquet reference frame is not so convenient.
Mori~\cite{Mori2022} introduced a ``truncated Floquet reference frame'' associated with a periodic unitary transformation $\e^{\im\hat{K}^{(n)}(t)}$, where $\hat{K}^{(n)}(t)$ is defined by truncating the high-frequency expansion for the micromotion operator:
\begin{align}
\hat{K}^{(n)}(t)=\sum_{k=1}^n\frac{\Lambda_k(t)}{\omega^k}.
\end{align}
The Hamiltonian in the truncated Floquet reference frame, which is called the ``dressed Hamiltonian'' reads
\begin{align}
\tilde{H}^{(n)}(t)=\e^{\im\hat{K}^{(n)}(t)}\qty[\hat{H}(t)-\im\dv{t}]\e^{-\im\hat{K}^{(n)}(t)}
\approx\HF^{(n)}+\hat{V}^{(n)}(t),
\end{align}
where we ignore terms of $\order{(g/\omega)^{n+1}}$ or higher, and
\begin{align}
\hat{V}^{(n)}(t)=\frac{1}{\omega^{n+1}}\dv{\hat{\Lambda}_{n+1}(t)}{t}
\end{align}
is called the dressed driving field satisfying $\hat{V}^{(n)}(t)=\hat{V}^{(n)}(t+T)$ and $\int_0^T\dd t\,\hat{V}^{(n)}(t)=0$.
Remarkably, the truncated Floquet Hamiltonian $\HF^{(n)}$ appears as a static part of the dressed Hamiltonian.
Unlike the Floquet reference frame, the time-dependence of the Hamiltonian is not completely removed.
However, the dressed driving field $\hat{V}^{(n)}(t)$ is effectively weakened ($\hat{V}^{(n)}(t)\propto\omega^{-n}$), and we can treat it perturbatively even if the original driving field $\hat{V}(t)$ is strong~\citep{Mori2022}.
As we argued in \cref{sec:pre}, $\HF^{(n)}$ does not induce heating.
The dressed driving field $\hat{V}^{(n)}(t)$ is therefore responsible for heating in the truncated Floquet reference frame.

We summarize the above discussion in \cref{fig:frame}.

\section{FLOQUET MASTER EQUATION FORMALISM}
\label{sec:master}

\begin{figure}[t]
\centering
\includegraphics[width=0.8\linewidth]{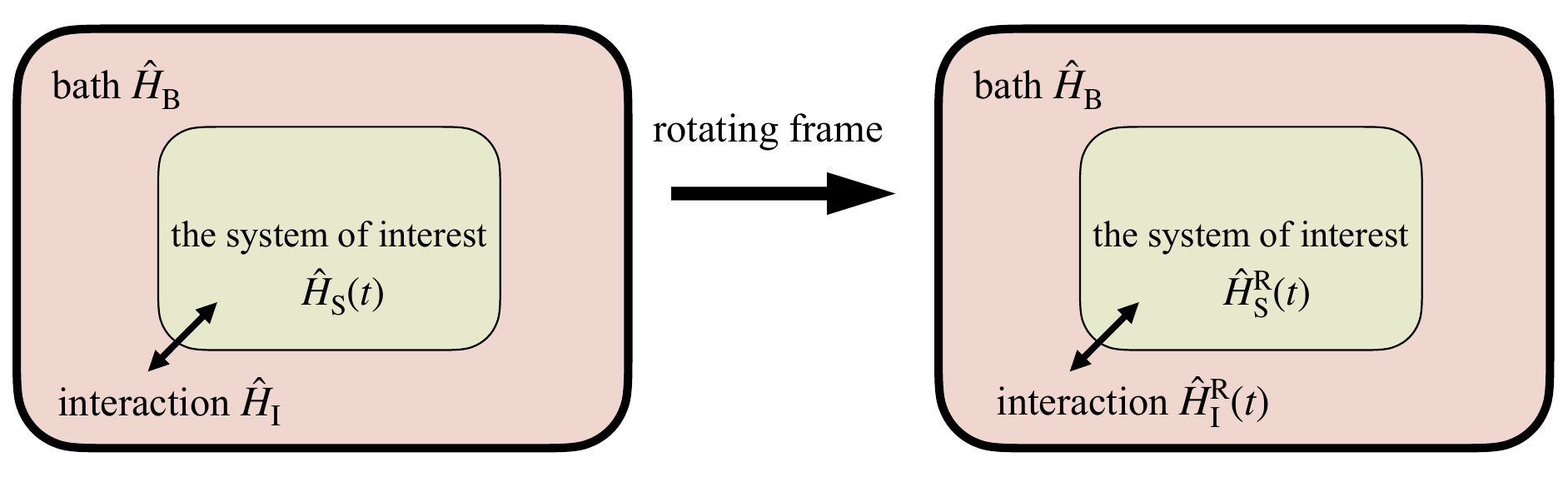}
\caption{Setup of open Floquet systems.
Consider an isolated system consisting of the system of interest (S) and the bath (B) interacting with each other via $\HI$.
Periodic driving is applied only to the system, and hence only $\HS(t)$ depends on time.
When periodic driving is strong or resonant, it is often convenient to consider the problem in an appropriate rotating frame, where the interaction Hamiltonian can also periodically depend on time.
}
\label{fig:setup}
\end{figure}

Now we explain how the effect of dissipation is taken into account.
A standard setup in considering dissipation is to suppose that the system of interest and the environment (or the bath) constitute a closed system.
See \cref{fig:setup} for an illustrative picture of the setup.
The dynamics of the system of interest is obtained by tracing out the bath degrees of freedom.

Theoretical formulation of open Floquet systems is almost parallel with static open systems~\citep{Breuer_text}.
We first perform the Born-Markov approximation for exact equations of motion.
Consequently, we obtain a quantum master equation of the Redfield form~\citep{Redfield1957}.
By considering an ideal limit (the weak-coupling limit or the singular-coupling limit), we obtain a quantum master equation of the Lindblad form~\citep{Lindblad1976, Gorini1976}.\footnote{The quantum master equation of the Lindblad form is referred to as the Gorini-Kossakowski-Sudarshan-Lindblad (GKSL) equation, or simply the Lindblad equation.}

\subsection{Born-Markov Approximation}

Let us denote by $\HS(t)$, $\HB$, $\HI(t)$ the Hamiltonian of the system of interest, that of the bath, and the interaction Hamiltonian, respectively.
It is assumed that $\HS(t)=\HS(t+T)$ and $\HI(t)=\HI(t+T)$.
The Hamiltonian of the total system is given by $\HT(t)=\HS(t)+\HB+\HI(t)$.
Without loss of generality, we can assume $\TrB[\HI(t)]=0$.

It is assumed that the periodic driving is applied only to the system of interest, but we allow that the interaction Hamiltonian $\HI(t)$ depends on time to consider applications to the Floquet engineering.
Indeed, as is explained in \cref{sec:high_strong,sec:high_resonant}, it is sometimes convenient to move to a suitable rotating frame.
The interaction Hamiltonian in a rotating frame generally depends on time, even though it is time-independent in the laboratory frame.
Thus, allowing the time dependence of $\HI(t)$ enables us to consider the problem in a rotating frame.
We also assume that the bath is in thermal equilibrium at the inverse temperature $\beta$, which is expressed by the density matrix $\rhoB=\e^{-\beta\HB}/Z_\mathrm{B}$, where $Z_\mathrm{B}$ is the partition function.

The density matrix of the total system $\rho_\mathrm{T}$ obeys the Liouville-von Neumann equation:
\begin{align}
\dv{t}\rho_\mathrm{T}(t)=-\im [\hat{H}_\mathrm{T}(t),\rho_\mathrm{T}(t)]=\qty(\mathcal{L}_\mathrm{S}(t)+\mathcal{L}_\mathrm{B}+\mathcal{L}_\mathrm{I}(t))\rhoT(t),
\label{eq:Liouville}
\end{align}
where $\mathcal{L}_\mathrm{X}(t)=-\im[\hat{H}_\mathrm{X}(t),\cdot]$ for $\mathrm{X}\in\{\mathrm{S},\mathrm{B},\mathrm{I}\}$.
The reduced density matrix is defined as $\rhoS(t)=\Tr_\mathrm{B}\rho_\mathrm{T}(t)$.
We now assume a timescale separation $\tau_\mathrm{B}\ll\tau_\mathrm{R}$, where $\tau_\mathrm{B}$ and $\tau_\mathrm{R}$ denote the bath correlation time and the relaxation time due to dissipation.
Then, the Born approximation is justified, which yields the following equation of motion for $\rhoS(t)$~\citep{Breuer_text} starting with an initial time $t_0$:
\begin{align}
\dv{t}\rhoS(t)=\LS(t)\rhoS(t)+\int_0^{t-t_0}\dd s\, \TrB\qty[\LI(t)\LI(t,t-s)\rhoS(t)\otimes\rhoB]+\mathcal{I}_{t,t_0}[\rhoT(t_0)],
\label{eq:g_master2}
\end{align}
where $\LI(t,t-s)=-i[\HI(t,t-s),\cdot]$ with $\HI(t,t-s)=\mathcal{T}\qty(\e^{\int_{t-s}^t\dd t'\,[\mathcal{L}_S(t')+\LB]})\HI(t-s)$ and $\mathcal{I}_{t,t_0}[\rhoT(t_0)]$ stands for the effect of the initial correlation between the system and the bath.
Explicitly, $\mathcal{I}_{t,t_0}[\rhoT(t_0)]$ is expressed as, in the leading order of $\tauB/\tauR$,
\begin{align}
\mathcal{I}_{t,t_0}[\rhoT(t_0)]=\TrB\LI(t)\mathcal{T}\qty[\e^{\int_{t_0}^t\dd t'\, [\LS(t')+\LB]}]\delta\rhoT(t_0),
\end{align}
where $\delta\rhoT(t_0)$ describes the system-bath correlations:
\begin{align}
\delta\rhoT(t_0)=\rhoT(t_0)-\rhoS(t_0)\otimes\rhoB.
\end{align}

Next, we perform the Markov approximation, which states that we can take the limit of $t-t_0\to\infty$ in \cref{eq:g_master2}:
\begin{align}
\int_0^{t-t_0}\dd s\, \TrB\qty[\LI(t)\LI(t,t-s)\rhoS(t)\otimes\rhoB]+\mathcal{I}_{t,t_0}[\rhoT(t_0)] \nonumber \\
\approx\int_0^\infty\dd s\, \TrB\qty[\LI(t)\LI(t,t-s)\rhoS(t)\otimes\rhoB],
\label{eq:Markov}
\end{align}
which is also justified when $\tauB/\tauR\ll 1$ as well as $t-t_0\gg\tauB$~\citep{vanKampen_text}.

Now we put $\HI(t)=\sum_i\hat{X}_i(t)\otimes\hat{Y}_i$, where $\hat{X}_i(t)$ is an operator acting to the system of interest and $\hat{Y}_i$ is an operator acting to the bath.
Here, the periodicity of $\HI(t)$ implies $\hat{X}_i(t)=\hat{X}_i(t+T)$, and we can always choose $\hat{X}_i(t)$ and $\hat{Y}_i$ as Hermitian: $\hat{X}_i(t)=\hat{X}_i(t)^\dagger$ and $\hat{Y}_i=\hat{Y}_i^\dagger$
We also define the bath correlation function as $\Phi_{ij}(t)=\TrB\qty[\hat{Y}_i(t)\hat{Y}_j\rhoB]$ with $\hat{Y}_i(t)=\e^{\im\HB t}\hat{Y}_i\e^{-\im\HB t}$ ($\tauB$ corresponds to a characteristic decay time of $\Phi_{ij}(t)$).
The Born-Markov quantum master equation is then obtained:
\begin{align}
\dv{t}\rhoS(t)=-\im\qty[\HS(t),\rhoS(t)]-\sum_i\qty[\hat{X}_i(t),\hat{R}_i(t)\rhoS(t)-\rhoS(t)\hat{R}_i^\dagger(t)],
\label{eq:Redfield}
\end{align}
where
\begin{align}
\hat{R}_i(t)=\sum_j\int_0^\infty\dd s\,\hat{X}_j(t,t-s)\Phi_{ij}(s),\quad \hat{X}_j(t,t-s)=\mathcal{T}\e^{\int_{t-s}^t\dd t'\,\LS(t')}\hat{X}_j(t-s).
\label{eq:R}
\end{align}
\Cref{eq:Redfield} is called the Floquet-Redfield equation since its right-hand side is known as the Redfield form~\citep{Redfield1957}.

It should be emphasized that $\hat{R}_i(t)$ depends on $\HS(t)$.
Consequently, when periodic driving is applied, we cannot use the dissipator derived for an undriven system.
The dissipator should be modified by periodic driving.
Kohler et al.~\cite{Kohler1997} investigated this effect and concluded that it becomes increasingly important when we consider stronger driving and lower temperatures.

Finally, we make a remark on the Markov approximation.
Suppose that \cref{eq:Markov} is satisfied at a certain time $t$: the time evolution of the system of interest is Markovian at time $t$.
Now this $t$ is chosen as a new initial time: we put $t_0=t$.
Since the time evolution should be independent of our choice of the initial time, \cref{eq:Markov} must be satisfied at $t_0=t$, which yields the following equality:
\begin{align}
\TrB\LI(t)\delta\rhoT(t)\approx\int_0^\infty\dd s\, \TrB\qty[\LI(t)\LI(t,t-s)\rhoS(t)\otimes\rhoB].
\label{eq:natural}
\end{align}
The system-bath correlation $\delta\rhoT(t)$ must satisfy the above nontrivial relation during the Markovian time evolution.
Such correlations are called ``natural correlations''~\citep{Mori2014}.
For example, the reduced equilibrium state $\rhoS^\mathrm{eq}=\TrB\e^{-\beta\HT}/\Tr\e^{-\beta\HT}$ is shown to satisfy \cref{eq:natural}~\citep{Mori2008}.
However, natural correlations do not exist for some $\rhoS$.
For example, when the system of interest is in a pure state, the state of the total system must be of the product form, and thus \cref{eq:natural} cannot hold.
For such an initial state, the time evolution is inevitably non-Markovian at short times $t\lesssim\tauB$~\citep{Suarez1992, Gaspard1999} due to the violation of \cref{eq:natural}.
Then, the initial state of the Redfield equation should be restricted to a subset of density matrices every member of which allows natural correlations with the bath.

\Cref{eq:Redfield} is not of the Lindblad form.
It is sometimes argued that a ``physical'' Markovian generator must be of the Lindblad form because otherwise the complete positivity is violated~\citep{Lindblad1976, Gorini1976}.
It should be noted that this argument implicitly assumes that the complete positivity should be imposed to an arbitrary initial state.
As we saw above, possible initial states of the Redfield equation are restricted to a certain subset of density matrices.
Then, it is not necessary to require the complete positivity for an arbitrary initial state, and hence physical Markovian generator may not be of the Lindblad form.
Indeed, it has numerically been demonstrated that the positivity in a Redfield equation holds for suitably restricted initial states~\citep{Mori2014}, although there is no rigorous proof.

\subsection{Floquet-Redfield Equation in the Floquet Reference Frame}
\label{sec:Redfield}

We rewrite \cref{eq:Redfield} in a useful form.
Let us move to the Floquet reference frame:
\begin{align}
\left\{
\begin{aligned}
&\tilderhoS(t)=\e^{\im\hat{K}(t)}\rhoS(t)\e^{-\im\hat{K}(t)},\\
&\tilde{X}_i(t)=\e^{\im\hat{K}(t)}\hat{X}_i(t)\e^{-\im\hat{K}(t)}=\sum_{m=-\infty}^\infty\tilde{X}_{i,m}\e^{\im m\omega t},\\
&\tilde{R}_i(t)=\e^{\im\hat{K}(t)}\hat{R}_i(t)\e^{-\im\hat{K}(t)}=\sum_{m=-\infty}^\infty\tilde{R}_{i,m}\e^{\im m\omega t}.
\end{aligned}
\right.
\label{eq:Floquet_reference}
\end{align}
By using \cref{eq:decomposition,eq:R}, it is shown that $\tilde{R}_{i,m}$ is expressed as follows:
\begin{align}
\tilde{R}_{i,m}=\sum_j\int_0^\infty\dd s\,\e^{-\im\HF s}\tilde{X}_{j,m}\e^{\im\HF s}\e^{-\im m\omega s}\Phi_{ij}(s).
\label{eq:R_im}
\end{align}
\Cref{eq:Redfield} is then expressed as
\begin{align}
\dv{t}\tilderhoS(t)=-\im\qty[\HF,\tilderhoS(t)]-\sum_i\sum_{m,l=-\infty}^\infty\e^{\im(m+l)\omega t}\qty[\tilde{X}_{i,m},\tilde{R}_{i,l}\tilderhoS(t)-\tilderhoS(t)\tilde{R}_{i,-l}^\dagger].
\label{eq:Floquet_Redfield}
\end{align}

When $\omega$ is large enough, we may further expect that contributions from $l\neq-m$ in \cref{eq:Floquet_Redfield} are negligible because of a quickly oscillating factor $\e^{\im(m+l)\omega t}$.
We then have
\begin{align}
\dv{t}\tilderhoS(t)=-\im\qty[\HF,\tilderhoS(t)]-\sum_i\sum_{m=-\infty}^\infty\qty[\tilde{X}_{i,m},\tilde{R}_{i,-m}\tilderhoS(t)-\tilderhoS(t)\tilde{R}_{i,m}^\dagger],
\label{eq:Floquet_QME}
\end{align}
which can be used as an approximation of the Floquet-Redfield equation~\citep{Kohler1998, Hone2009}.

If we denote by $\tilde{\rho}_\mathrm{ss}$ the stationary solution of \cref{eq:Floquet_QME}, the density matrix $\rhoS(t)$ in the original frame will relax to a periodic steady state $\rho_\mathrm{ss}(t)=\e^{-\im\hat{K}(t)}\tilde{\rho}_\mathrm{ss}\e^{\im\hat{K}(t)}$ satisfying $\rho_\mathrm{ss}(t)=\rho_\mathrm{ss}(t+T)$.

We can also write down the Floquet-Redfield equation in a truncated Floquet reference frame.
It is obtained by replacing $\HF$ by $\HF^{(n)}+\hat{V}^{(n)}(t)$ and $\hat{K}(t)$ by $\hat{K}^{(n)}(t)$ in \cref{eq:Floquet_reference}.

\subsection{Quantum Master Equations of the Lindblad Form}

We have derived the Floquet-Redfield equation under the assumption of $\tauB\ll\tauR$.
In the argument so far, the timescale $\tauS$ of the intrinsic evolution of the system of interest did not matter.
In the following, we derive quantum master equations of the Lindblad form in some limiting cases depending on $\tauS$.
In \cref{sec:weak_coupling}, we consider the weak-coupling limit $\tauR\to\infty$, implying $\tauS\ll\tauR$ in addition to $\tauB\ll\tauR$.
In \cref{sec:singular_coupling}, we consider another limit called the singular coupling limit $\tauB\to 0$, implying $\tauB\ll\tauS$ in addition to $\tauB\ll\tauR$.

As we will see in \cref{sec:weak_coupling}, the weak-coupling limit leads to a Lindblad equation with nonlocal dissipator.
It is argued that the weak-coupling approximation cannot describe nontrivial balance of heating and dissipation.
On the other hand, the singular-coupling limit leads to a Lindblad equation with local dissipator.
However, its steady state is the trivial infinite-temperature ensemble, which is not of great interest in practice.
In \cref{sec:resonant} we show that a local Lindblad equation with a non-trivial steady state is obtained under certain conditions by using resonant driving.

\subsubsection{Weak Coupling Limit}
\label{sec:weak_coupling}
If we assume $\tauS\ll\tauR$ in addition to $\tauB\ll\tauR$, \cref{eq:Floquet_QME} is further simplified.
This situation is realized in the weak-coupling limit: we consider the scaling $\HT(t)=\HS(t)+\HB+\lambda\HI(t)$ and take the limit of $\lambda\to 0$ with $\lambda^2 t$ held fixed, which is known as the van Hove limit~\citep{vanHove1957}.
The scaling parameter $\lambda$ characterizes the strength of dissipation, and $\tauR\sim\lambda^{-2}$.

In order to investigate the weak-coupling limit, let us consider the problem in the interaction picture: $\tilderhoS^\mathrm{(I)}(t)=\e^{-\im\HF t}\tilderhoS(t)\e^{\im\HF t}$.
The operators $\tilde{X}_{i,m}$ and $\tilde{R}_{i,m}$ in \cref{eq:Floquet_QME} is transformed as
\begin{align}
\left\{
\begin{aligned}
&\tilde{X}_{i,m}^\mathrm{(I)}(t)=\e^{\im\HF t}\tilde{X}_{i,m}\e^{-\im\HF t}=\sum_\Omega\tilde{X}_{i,m}[\Omega]\e^{\im\Omega t},& &\tilde{X}_{i,m}[\Omega]=\sum_{a,b:\varepsilon_a-\varepsilon_b=\Omega}\mel{u_a}{\tilde{X}_{i,m}}{u_b}, \\
&\tilde{R}_{i,m}^\mathrm{(I)}(t)=\e^{\im\HF t}\tilde{R}_{i,m}\e^{-\im\HF t}=\sum_\Omega\tilde{R}_{i,m}[\Omega]\e^{\im\Omega t},& &\tilde{R}_{i,m}[\Omega]=\sum_{a,b:\varepsilon_a-\varepsilon_b=\Omega}\mel{u_a}{\tilde{R}_{i,m}}{u_b},
\end{aligned}
\right.
\label{eq:X_R_int}
\end{align}
where recall that $\{\varepsilon_a\}$ are quasi-energies and $\{\ket{u_a}\}$ are Floquet eigenstates.
By using a rescaled time $\tau=\lambda^2 t$, \cref{eq:Floquet_QME} is written in the interaction picture as
\begin{align}
\dv{\tau}\tilderhoS^\mathrm{(I)}=-\sum_i\sum_{m=-\infty}^\infty\sum_{\Omega,\Omega'}\exp\qty(\im\frac{\Omega+\Omega'}{\lambda^2}\tau)\qty[\tilde{X}_{i,m}[\Omega],\tilde{R}_{i,-m}[\Omega']\tilderhoS^\mathrm{(I)}-\tilderhoS^\mathrm{(I)}\tilde{R}_{i,m}[-\Omega']^\dagger].
\label{eq:Floquet_QME_I}
\end{align}
For small $\lambda$, the factor $\exp\qty(\im\frac{\Omega+\Omega'}{\lambda^2}\tau)$ will rapidly oscillate unless $\Omega'=-\Omega$.
It is therefore a reasonable approximation that this factor is averaged out in the limit of $\lambda\to 0$.
This approximation is called the secular approximation or the rotating-wave approximation in literature~\citep{Breuer_text}.
Actually, the secular approximation becomes exact in the van Hove limit under certain conditions~\citep{Spohn1980_review}.

By using \cref{eq:R_im},  we can express $\tilde{R}_{i,m}[\Omega]$ as
\begin{align}
\tilde{R}_{i,m}[\Omega]&=\sum_j\tilde{X}_{j,m}[\Omega]\int_0^\infty\dd s\,\Phi_{ij}(s)\e^{-\im(\Omega+m\omega)s} \nonumber \\
&=\sum_j\tilde{X}_{j,m}[\Omega]\qty[\frac{1}{2}\gamma_{ij}(\Omega+m\omega)+\im\eta_{ij}(\Omega+m\omega)],
\label{eq:R_Omega}
\end{align}
where $\gamma_{ij}(\varepsilon)$ and $\eta_{ij}(\varepsilon)$ are defined as
$\gamma_{ij}(\varepsilon)=\int_{-\infty}^\infty\dd t\,\Phi_{ij}(t)\e^{-\im\omega t}$ and $\eta_{ij}(\varepsilon)=\mathrm{P}\int_{-\infty}^\infty\frac{\dd\varepsilon'}{2\pi}\frac{\gamma_{ij}(\varepsilon')}{\varepsilon'-\varepsilon}$.
Here, $\mathrm{P}\int$ denotes Cauchy's principale value integral.
By using the Hermiticity of $\{\hat{Y}_i\}$, it is shown that $\gamma_{ij}(\varepsilon)$ and $\eta_{ij}(\varepsilon)$ are Hermitian matrices: $\gamma_{ij}(\varepsilon)=\gamma_{ji}(\varepsilon)^*$ and $\eta_{ij}(\varepsilon)=\eta_{ji}(\varepsilon)^*$.
It is also shown that the matrix $\gamma_{ij}(\omega)$ is positive semidefinite.
When the bath is in thermal equilibrium at the inverse temperature $\beta$, the Kubo-Martin-Schwinger (KMS) relation holds: \begin{align}
\gamma_{ij}(\varepsilon)=\gamma_{ji}(-\varepsilon)\e^{-\beta\varepsilon}.
\label{eq:KMS}
\end{align}

By performing the secular approximation and substituting \cref{eq:R_Omega} into \cref{eq:Floquet_QME_I}, we obtain a quantum master equation of the Lindblad form, which is expressed in the Sch\"odinger picture as
\begin{align}
&\dv{t}\tilderhoS(t)=-\im\qty[\HF+\hat{H}_\mathrm{LS},\tilderhoS]\nonumber \\
&+\lambda^2\sum_{ij}\sum_{m=-\infty}^\infty\sum_{\Omega}\gamma_{ij}(\Omega+m\omega)\qty(\tilde{X}_{j,m}[\Omega]\tilderhoS\tilde{X}_{i,m}[\Omega]^\dagger-\frac{1}{2}\qty{\tilde{X}_{i,m}[\Omega]^\dagger\tilde{X}_{j,m}[\Omega],\tilderhoS}),
\label{eq:Floquet_Lindblad}
\end{align}
where $\hat{H}_\mathrm{LS}$ is the Lamb-shift Hamiltonian given by
\begin{align}
\hat{H}_\mathrm{LS}=\lambda^2\sum_{ij}\sum_{m=-\infty}^\infty\sum_\Omega\eta_{ij}(\Omega+m\omega)\tilde{X}_{i,m}[\Omega]^\dagger\tilde{X}_{j,n}[\Omega].
\end{align}
\Cref{eq:Floquet_Lindblad} is called the Floquet-Lindblad equation.

Generally, the dissipator of a Lindblad equation obtained by taking the weak-coupling limit is nonlocal even in the absence of driving field~\citep{Spohn1980_review}.
This nonlocality can be physically interpreted in the following way: a local excitation of the system will spread over the entire system until it dissipates into the bath.

It should be emphasized that the validity of the Floquet-Lindblad equation is severely limited due to the nature of the secular approximation.
We have assumed $\tauR\gg\tauS$ for any relevant timescale $\tauS$ of the system of interest.
In macroscopic open Floquet systems, however, we are typically interested in the situation in which this assumption does not hold.
When the system is driven by periodic fields, heating is obviously an important process, and hence its timescale $\tauH$ enters into $\tauS$.
We expect that some nontrivial (periodic) stationary state is realized by a balance between dissipation and heating.
Such a balance requires $\tauR\lesssim\tauH$, but it contradicts the weak-coupling limit $\tauR\to\infty$. 
In other words, the weak-coupling limit implicitly assumes that heating is always faster than dissipation, and hence any nontrivial balance of heating and dissipation cannot be described by the weak-coupling Floquet-Lindblad equation.

\subsubsection{Singular Coupling Limit}
\label{sec:singular_coupling}
We consider another limiting procedure: $\tauB\ll\tauS$ and $\tauB\ll\tauR$.
This situation is dealt with the scaling $\HT(t)=\HS(t)+\lambda^{-2}\HB+\lambda^{-1}\HI(t)$ and $\beta=\lambda^2\tilde{\beta}$ ($\tilde{\beta}$ is independent of $\lambda$).
We take the limit of $\lambda\to 0$ with this scaling, which is called the singular coupling limit.
The bath correlation function is written as
\begin{align}
\Phi_{ij}(t)=\TrB\qty[\e^{\im\HB t/\lambda^2}\hat{Y}_i\e^{-\im\HB t/\lambda^2}\hat{Y}_j\frac{\e^{-\tilde{\beta}\HB}}{\TrB\e^{-\tilde{\beta}\HB}}]\eqqcolon\tilde{\Phi}_{ij}\qty(\frac{t}{\lambda^2}).
\end{align}
In the limit of $\lambda\to 0$, $\Phi_{ij}(t)$ becomes a delta function.
The operator $\hat{R}_i(t)$ in \cref{eq:R} is therefore simplified as
\begin{align}
\hat{R}_i(t)\approx\lambda^2\sum_j\hat{X}_j\int_0^\infty\dd\tau\,\tilde{\Phi}_{ij}(\tau)
\eqqcolon\lambda^2\sum_j\hat{X}_j\qty(\frac{1}{2}\gamma_{ij}+\im\eta_{ij}),
\end{align}
where $\gamma_{ij}=\gamma_{ji}\geq 0$ and $\eta_{ij}=\eta_{ji}^*$ are satisfied.
\Cref{eq:Redfield} then becomes
\begin{align}
\dv{t}\rhoS(t)=-\im\qty[\HS(t)+\sum_{ij}\eta_{ij}\hat{X}_i\hat{X}_j,\rhoS(t)]+\sum_{ij}\gamma_{ij}\qty(\hat{X}_j\rhoS(t)\hat{X}_i-\frac{1}{2}\qty{\hat{X}_i\hat{X}_j,\rhoS(t)}).
\label{eq:Lindblad_singular}
\end{align}
This is a quantum master equation of the Lindblad form.
The dissipator is local in contrast to the Lindblad equation in the weak-coupling limit: any local excitation will immediately dissipate into the bath before it spreads over the system.

\Cref{eq:Lindblad_singular} has the trivial infinite-temperature ensemble as a steady state.
Therefore, the singular-coupling limit cannot describe any nontrivial steady state, unless there are other nontrivial steady states due to some conserved quantities~\citep{Tindall2019}. 

\subsubsection{Phenomenological Lindblad Equation Using Resonant Driving}
\label{sec:resonant}
In the weak-coupling limit, the system is described by the Floquet-Lindblad equation with quite complicated highly nonlocal dissipator.
In the study of open quantum systems~\citep{Prosen2011, Znidaric2015, Sponselee2018}, we sometimes consider a more intuitive ``phenomenological'' Lindblad equation with dissipator
\begin{align}
\mathcal{D}(\rho)=\sum_i\qty(\hat{L}_i\rhoS\hat{L}_i^\dagger-\frac{1}{2}\qty{\hat{L}_i^\dagger\hat{L}_i,\rhoS}),
\end{align}
where Lindblad jump operators $\qty{\hat{L}_i}$ are phenomenologically introduced local operators, which may be non-Hermitian in contrast to those in the singular-coupling limit.
For example, in a two-level atom described by the Pauli matrices $\qty{\hat{\sigma}^x,\hat{\sigma}^y,\hat{\sigma}^z}$ in contact with a photon bath, an excitation of the atom by absorbing a photon is described by a Lindblad jump operator $\hat{L}=\hat{\sigma}^+=\qty(\hat{\sigma}^x+\im\hat{\sigma}^y)/2$, whereas a de-excitation by emitting a photon is described by $\hat{L}=\hat{\sigma}^-=\qty(\hat{\sigma}^x-\im\hat{\sigma}^y)/2$.
In a Bose or Fermi particle described by the creation and annihilation operators $\hat{a}^\dagger$ and $\hat{a}$, respectively, the particle loss is described by a Lindblad jump operator $\hat{L}=\hat{a}$ and the dephasing is described by $\hat{L}=\hat{a}^\dagger\hat{a}$.

In this section, we present another route to the Lindblad equation from \cref{eq:Floquet_QME}, and demonstrate that a certain kind of ``phenomenological'' Lindblad equations is microscopically derived.

By using \cref{eq:X_R_int}, we have $\e^{-\im\HF s}\tilde{X}_{j,m}\e^{\im\HF s}=\sum_\Omega\tilde{X}_{j,m}[\Omega]\e^{-\im\Omega s}$.
Let us substitute this expression into \cref{eq:R_im}.
We then have
\begin{align}
\tilde{R}_{i,m}&=\sum_j\sum_\Omega\tilde{X}_{j,m}[\Omega]\int_0^\infty\dd s\,\e^{-\im(\Omega+m\omega)s}\Phi_{ij}(s) \nonumber \\
&=\sum_j\sum_\Omega\tilde{X}_{j,m}[\Omega]\qty(\frac{1}{2}\gamma_{ij}(\Omega+m\omega)+\im\eta_{ij}(\Omega+m\omega)).
\label{eq:R_gamma_eta}
\end{align}
Now let us assume that contributions from $\abs{\Omega}\ll\omega$ are dominant.
Since $\Omega$ typically satisfies $\abs{\Omega}\lesssim g$, where recall that $g$ denotes a characteristic local energy of $\HS(t)$, this assumption will be justified as long as $g\ll\omega$.
Under this assumption, we further perform the following approximations in \cref{eq:R_gamma_eta}:
\begin{align}
\gamma_{ij}(\Omega+m\omega)\approx\gamma_{ij}(m\omega) \qq{and} \eta_{ij}(\Omega+m\omega)\approx\eta_{ij}(m\omega).
\label{eq:bath_approx}
\end{align}
It should be noted that \cref{eq:bath_approx} is justified for $m=0$ only when the temperature of the bath is high enough or $\tilde{X}_{j,0}\approx 0$ in \cref{eq:R_gamma_eta}.
Indeed, the KMS relation $\gamma_{ij}(\varepsilon)=\gamma_{ji}(-\varepsilon)\e^{-\beta\varepsilon}$ implies that $\gamma_{ij}(\Omega)$ can be approximated by a constant value only when $\beta\abs{\Omega}\ll 1$.
Therefore, behind \cref{eq:bath_approx}, it is implicitly assumed that
\begin{align}
\beta g\ll 1 \qq{or} \tilde{X}_{j,0}\approx 0.
\end{align}
By using \cref{eq:bath_approx}, we have
\begin{align}
\tilde{R}_{i,m}\approx\sum_j\sum_\Omega\tilde{X}_{j,m}[\Omega]\qty(\frac{1}{2}\gamma_{ij}(m\omega)+\im\eta_{ij}(m\omega))
=\sum_j\tilde{X}_{j,m}\qty(\frac{1}{2}\gamma_{ij}(m\omega)+\im\eta_{ij}(m\omega)).
\end{align}
Substituting it into \cref{eq:Floquet_QME}, we obtain
\begin{align}
\dv{t}\tilderhoS(t)=&-\im\qty[\HF+\hat{H}_\mathrm{LS},\tilderhoS(t)] \nonumber \\
&+\sum_{ij}\sum_{m=-\infty}^\infty\gamma_{ij}(m\omega)\qty(\tilde{X}_{j,m}\tilderhoS(t)\tilde{X}_{i,m}^\dagger-\frac{1}{2}\qty{\tilde{X}_{i,m}^\dagger\tilde{X}_{j,m},\tilderhoS(t)}),
\label{eq:Lindblad_resonant}
\end{align}
where the Lamb-shift Hamiltonian is given by
$\hat{H}_\mathrm{LS}=\sum_{ij}\sum_{m=-\infty}^\infty\eta_{ij}(m\omega)\tilde{X}_{i,m}^\dagger\tilde{X}_{j,m}$.
\Cref{eq:Lindblad_resonant} is of the Lindblad form, which has a nontrivial steady state when $\beta\omega=\order{1}$.

We now present a simple example, which leads to a phenomenological Lindblad equation~\citep{DallaTorre2013}.
Let us consider a spin chain driven by circularly polarized fields
\begin{align}
\HS(t)=\sum_{i=1}^N\qty[\frac{\omega+\Delta}{2}\hat{\sigma}_i^z-J\hat{\sigma}_i^z\hat{\sigma}_{i+1}^z-h\qty(\e^{-\im\omega t}\hat{\sigma}_i^++\e^{\im\omega t}\hat{\sigma}_i^-)],
\end{align}
which is in contact with a free-boson bath independently at each site:
\begin{align}
\HB=\sum_{i=1}^N\sum_k\omega_k\hat{b}_{i,k}^\dagger\hat{b}_{i,k} \qq{and}
\HI=\sum_{i=1}^N\hat{\sigma}_i^x\sum_k\lambda_k\qty(\hat{b}_{i,k}+\hat{b}_{i,k}^\dagger),
\end{align}
where $\omega_k>0$ and $\lambda_k\in\mathbb{R}$.
The creation and annihilation operators of bosons of the mode $k$ in the $i$th bath are denoted by $b_{i,k}^\dagger$ and $b_{i,k}$, respectively.
This model is regarded as an example of the systems treated in \ref{sec:high_resonant}: $\sum_{i=1}^N\hat{\sigma}_i^z/2$ corresponds to $\hat{N}$ in \cref{eq:H_resonant}.

We shall move to a rotating frame via the unitary transformation $\e^{-\im(\omega/2)\sum_{i=1}^N\hat{\sigma}_i^zt}$.
The Hamiltonian in the rotating frame is given by
\begin{align}
\HS^\mathrm{R}=\sum_{i=1}^N\qty(\frac{\Delta}{2}\hat{\sigma}_i^z-J\hat{\sigma}_i^z\hat{\sigma}_{i+1}^z-h\hat{\sigma}_i^x),
\quad
\HB^\mathrm{R}=\HB,
\quad
\HI^\mathrm{R}(t)=\sum_{i=1}^N\hat{X}_i(t)\otimes\hat{Y}_i,
\end{align}
where $\hat{X}_i(t)=\hat{\sigma}_i^+\e^{\im\omega t}+\hat{\sigma}_i^-\e^{-\im\omega t}$ and $\hat{Y}_i=\sum_k\lambda_k\qty(\hat{b}_{i,k}+\hat{b}_{i,k}^\dagger)$.
In this model, $\HS^\mathrm{R}$ does not depend on time, and hence there is no micromotion: $\hat{K}(t)=0$ and we can simply put $\HF=\HS^\mathrm{R}$, $\tilde{X}_{i,1}=\hat{\sigma}_i^+$, and $\tilde{X}_{i,-1}=\hat{\sigma}_i^-$.
For $m\neq \pm 1$, $\tilde{X}_{i,m}=0$.

\Cref{eq:Lindblad_resonant} is then written as
\begin{align}
\dv{t}\rhoS^\mathrm{R}(t)=-\im\qty[\HS^\mathrm{R}+\hat{H}_\mathrm{LS},\rhoS^\mathrm{R}(t)]
&+\sum_{i=1}^N\gamma(\omega)\qty(\hat{\sigma}_i^+\rhoS^\mathrm{R}(t)\hat{\sigma}_i^--\frac{1}{2}\qty{\hat{\sigma}_i^-\hat{\sigma}_i^+,\rhoS^\mathrm{R}(t)}) \nonumber \\
&+\sum_{i=1}^N\gamma(-\omega)\qty(\hat{\sigma}_i^-\rhoS^\mathrm{R}(t)\hat{\sigma}_i^+-\frac{1}{2}\qty{\hat{\sigma}_i^+\hat{\sigma}_i^-,\rhoS^\mathrm{R}(t)}).
\label{eq:Lindblad_resonant_example}
\end{align}
Here, $\gamma(\varepsilon)$ is calculated as $\gamma(\varepsilon)=\pi\mathrm{sgn}(\varepsilon)J(\abs{\varepsilon})/(\e^{\beta\varepsilon}-1)$, where the bath spectral density $J(\varepsilon)$ is defined as $J(\varepsilon)=\sum_k\delta(\omega_k-\varepsilon)\lambda_k^2$.

In \cref{eq:Lindblad_resonant_example}, the Lindblad jump operators $\qty{\hat{\sigma}_i^\pm}$ are very simple and intuitive: they represent an excitation and a de-excitation of a two-level atom via the coupling to the bath.
Moreover, the KMS relation $\gamma(\omega)=\gamma(-\omega)\e^{-\beta\omega}$ implies that \cref{eq:Lindblad_resonant_example} has a nontrivial steady state when $\beta\omega=\order{1}$.

\section{STATISTICAL MECHANICS OF FLOQUET SYSTEMS}
\label{sec:stat}

When an undriven system is in a weak contact with a thermal bath at the inverse temperature $\beta$, equilibrium statistical mechanics predicts that the steady state is described by a Gibbs state $\e^{-\beta\HS}/\Tr_\mathrm{S}\e^{-\beta\HS}$.
A natural question is to what extent the method of equilibrium statistical mechanics is extended to Floquet systems.
Since the Floquet theorem offers a static description of the long-time behavior of a closed Floquet system via the Floquet Hamiltonian $\HF$ or its approximation $\HF^{(n)}$, it is natural to consider a Floquet-Gibbs state $\rho_\mathrm{FG}=\e^{-\beta\HF}/\Tr_\mathrm{S}\e^{-\beta\HF}$ or a truncated Floquet-Gibbs state $\rho_\mathrm{FG}^{(n)}=\e^{-\beta\HF^{(n)}}/\Tr_\mathrm{S}\e^{-\beta\HF^{(n)}}$ as a candidate of the steady state in an open system.

We remark that there is an arbitrariness in the definition of $\rho_\mathrm{FG}$ due to the indefiniteness of quasi-energies $\qty{\varepsilon_a}$.
In the following discussion, we choose $\varepsilon_a$ so that it is closest to the mean energy $\bar{E}_a=(1/T)\int_0^T\dd t\expval{\hat{H}(t)}{u_a(t)}$.
On the other hand, there is no arbitrariness in the definition of a truncated Floquet-Gibbs state since the high-frequency expansion naturally fixes the quasi-energies.

The possibility of extending equilibrium statistical mechanics to Floquet systems has been investigated in various previous works.
Breuer and Holthaus~\cite{Breuer1991} showed that a periodic steady state of the Floquet-Lindblad equation for a driven harmonic oscillator is given by the Floquet-Gibbs state.
On the other hand, Breuer et al.~\cite{Breuer2000} showed that, for a periodically driven particle in a box, the population over Floquet states is significantly distinct from the Boltzmann-Gibbs distribution.
Ketzmerick and Wustmann~\cite{Ketzmerick2010} made a similar observation in a driven unharmonic oscillator.
In this way, the validity of equilibrium statistical mechanics depends on the specific model Hamiltonian, and hence general considerations have been desirable.
Breuer et al.~\cite{Breuer2000} and Kohn~\cite{Kohn2001} pointed out that a transition between two Floquet states with the quasi-energy difference $\Delta\varepsilon$ will be accompanied by bath transitions with many different energy changes $\Delta E=-\Delta\varepsilon+m\omega$ with $m$ being an arbitrary integer, which results in the violation of the detailed-balance condition.
Thus, in general, the steady state will be distinct from the Floquet-Gibbs state.

In this section, based on the formalism developed in \cref{sec:master}, we review recent studies~\citep{Shirai2015, Liu2015, Shirai2016} investigating general conditions for the validity of equilibrium statistical mechanics to open Floquet systems.
In \cref{sec:FG_Lindblad}, we investigate the steady state of the Floquet-Lindblad equation in the weak-coupling limit.
In \cref{sec:FG_Redfield}, we investigate the steady state of the Floquet-Redfield equation at a finite system-bath coupling.

\subsection{Steady States in the Weak-Coupling Limit}
\label{sec:FG_Lindblad}

We investigate the steady state of the Floquet-Lindblad equation in the weak-coupling limit given by \cref{eq:Floquet_Lindblad}.
For simplicity, we assume that $\HF$ has no degeneracy.
We then find that the dynamics of the diagonal matrix elements $P_a(t)=\expval{\tilderhoS(t)}{u_a}$ in the basis diagonalizing $\HF$ is decoupled from that of the off-diagonal matrix elements.
The time evolution of $P_a(t)$ obeys the following Pauli master equation:
\begin{align}
\dv{t}P_a(t)=\sum_b\qty(W_{ab}P_b(t)-W_{ba}P_a(t)),
\label{eq:Pauli}
\end{align}
where $W_{ab}$ stands for the transition rate from the state $\ket{u_b}$ to the state $\ket{u_a}$, and is explicitly given by
\begin{align}
W_{ab}=\sum_{ij}\sum_{m=-\infty}^\infty\gamma_{ij}(\varepsilon_a-\varepsilon_b+m\omega)\mel{u_a}{\tilde{X}_{j,m}}{u_b}\mel{u_a}{\tilde{X}_{i,m}}{u_b}^*
\label{eq:transition}
\end{align}
for $a\neq b$, and $W_{aa}=0$ for any $a$.

Without any special reason, every off-diagonal matrix element exponentially decays to zero.
Therefore, the steady state $\tilde{\rho}_\mathrm{ss}$ is written in the diagonal form $\tilde{\rho}_\mathrm{ss}=\sum_aP_a^\mathrm{ss}\dyad{u_a}$, where $\qty{P_a^\mathrm{ss}}$ is the steady solution of \cref{eq:Pauli} satisfying $\sum_b\qty(W_{ab}P_b^\mathrm{ss}-W_{ba}P_a^\mathrm{ss})=0$.

When no driving field is applied, there is no summation over $m$ in \cref{eq:transition}.
In this case, the KMS relation yields the detailed balance condition:
\begin{align}
W_{ab}\e^{-\beta E_b}=W_{ba}\e^{-\beta E_a},
\end{align}
where $\qty{E_a}$ are eigenvalues of $\HS$ without driving field.
The detailed balance condition ensures that the steady state is given by the Gibbs distribution $P_a^\mathrm{ss}=\e^{-\beta E_a}/\sum_b\e^{-\beta E_b}$.

On the other hand, under periodic driving, the system can undergo a transition between Floquet eigenstates by absorbing or emitting $m$ energy quanta.
This is the physical meaning of the additional sum over $m$ in \cref{eq:transition}.
Importantly, the sum over $m$ in \cref{eq:transition} in general breaks the detailed balance~\citep{Breuer2000, Kohn2001}, and the steady state is not necessarily given by the Floquet-Gibbs state.

Here, we find the following important observation.
Even when the system is subject to periodic driving, the detailed balance condition is approximately satisfied if only $m=0$ is dominant in \cref{eq:transition} for any $a$ and $b$.
Shirai et al.~\cite{Shirai2015} and Liu~\cite{Liu2015} investigated the condition for it to happen, which we now explain below.

We first assume that 
\begin{align}
\frac{1}{T}\int_0^T\dd t\,\norm{\HS(t)}\ll\omega.
\label{eq:condition_norm}
\end{align}
In this case, the convergence of the high-frequency expansion is guaranteed.
It should be noted that if we consider strong driving (\cref{sec:high_strong}) or resonant driving (\cref{sec:high_resonant}), we must move to the rotating frame so that \cref{eq:condition_norm} is satisfied.
Under \cref{eq:condition_norm}, the mean energy $\bar{E}_a$ for any Floquet eigenstate satisfies $\abs{\bar{E}_a}\ll\omega$, which implies $\abs{\varepsilon_a}\ll\omega$ in our choice of the quasi-energies.
Moreover, the micromotion operator $\hat{K}(t)$ is almost zero: $\hat{K}(t)=\order{\omega^{-1}}$.
Thus, $\tilde{X}_i(t)\approx\hat{X}_i(t)$, and hence $\tilde{X}_{i,m}$ is approximately given by
\begin{align}
\tilde{X}_{i,m}\approx\frac{1}{T}\int_0^T\dd t\,\hat{X}_i(t)\e^{-\im m\omega t}.
\label{eq:X_approx}
\end{align}

When \cref{eq:condition_norm} is satisfied in the laboratory frame, we do not have to move to the rotating frame.
In this case, $\hat{X}_i$ does not depend on $t$, and hence $\tilde{X}_{i,m}\approx\delta_{m,0}\hat{X}_i$.
Thus, $m=0$ is dominant in \cref{eq:transition} for any $a$ and $b$, which ensures that the Floquet-Gibbs state is realized as a steady state.
On the other hand, for strong driving or resonant driving, we must move to the rotating frame to guarantee \cref{eq:condition_norm}, and the interaction Hamiltonian $\HI^\mathrm{R}(t)$ can depend on $t$.
In this case, there are two situations where $m=0$ is dominant for any $a$ and $b$ in \cref{eq:transition}: either $\tilde{X}_{i,m}\approx 0$ for all $m\neq 0$ or $\gamma_{ij}(\varepsilon_a-\varepsilon_b+m\omega)\approx 0$ for all $m\neq 0$.
In addition, we have to assume that $(1/T)\int_0^T\dd t\,\HI(t)\approx\sum_i\tilde{X}_{i,0}\otimes\hat{Y}_i$ does not vanish.
Otherwise, we cannot say that the term of $m=0$ is dominant.

First, we consider the possibility of $\tilde{X}_{i,m}\approx 0$ for all $m\neq 0$.
This condition is satisfied when the interaction Hamiltonian does not depend on $t$ even in the rotating frame: $\tilde{X}_i(t)=\hat{X}_i$, and hence \cref{eq:X_approx} leads to $\tilde{X}_{i,m}\approx\delta_{m,0}\hat{X}_i$.
Under strong driving $\HS(t)=\hat{H}_0+\omega f(\omega t)\hat{V}$, the interaction Hamiltonian in the rotating frame is given by $\HI^\mathrm{R}(t)=\e^{\im F(\omega t)\hat{V}}\HI\e^{-\im F(\omega t)\hat{V}}$, where $F(\theta)=\int_0^\theta \dd s\,f(\theta)$.
We see that it is independent of time when $\HI$ commutes with $\hat{V}$.
This situation is realized when the periodic field is only applied to a part of the system which is not directly coupled to the bath~\citep{Shirai2015}.
Similarly, under resonant driving $\HS(t)=\hat{H}_0+\omega\hat{N}+\hat{V}(t)$, the interaction Hamiltonian does not depend on $t$ when $\HI$ commutes with $\hat{N}$.

Next, we consider the possibility of $\gamma_{ij}(\varepsilon_a-\varepsilon_b+m\omega)\approx 0$ for all $m\neq 0$.
Because of \cref{eq:condition_norm}, $\abs{\varepsilon_a-\varepsilon_b}\ll \omega$, and hence $\varepsilon_a-\varepsilon_b+m\omega\approx m\omega$.
Generally, $\gamma_{ij}(\varepsilon)$ decays for $\varepsilon$ greater than a cutoff frequency $\omega_c$, which corresponds to a characteristic local energy scale of the bath.
If $\omega\gg\omega_c$, which physically means that periodic driving is much faster than the motion of the bath, $\gamma_{ij}(\varepsilon_a-\varepsilon_b+m\omega)\approx 0$ for all $m\neq 0$ is satisfied because $\abs{\varepsilon_a-\varepsilon_b+m\omega}\gg\omega_c$.

Let us summarize the condition for the realization of the Floquet-Gibbs state:
\begin{description}
\item[(i)] $(1/T)\int_0^T\dd t\,\norm{\HS(t)}\ll\omega$.
\item[(ii)] Either $\qty[\HI,\hat{A}]=0$ or $\omega_c\ll\omega$.
Here, $\hat{A}=\hat{V}$ in the case of strong driving (\cref{eq:H_strong}) and $\hat{A}=\hat{N}$ in the case of resonant driving (\cref{eq:H_resonant}).
\item[(iii)] $(1/T)\int_0^T\dd t\, \HI(t)$ is not vanishingly small.
\end{description}
The condition (i) ensures that the system does not suffer from heating.
However, even when (i) is satisfied, periodic driving may indirectly induce excitations in the bath.
The conditions (ii) and (iii) ensure that this effect is negligible~\citep{Shirai2018}.

\subsection{Steady States at Finite System-Bath Coupling}
\label{sec:FG_Redfield}

In macroscopic systems, the condition (i) cannot be satisfied, and the system heats up due to periodic driving.
In a closed system, a truncated Floquet-Gibbs state $\rho_\mathrm{FG}^{(n)}$ is realized only in a prethermal regime.
In an open system, on the other hand, it is hopefully expected that dissipation suppresses heating and stabilizes the truncated Floquet-Gibbs state.

Shirai et al.~\cite{Shirai2016} investigated the conditions under which the above scenario is realized.
To stabilize the truncated Floquet-Gibbs state, we must consider a finite system-bath coupling, where the Floquet-Lindblad equation is not appropriate.
We assume that the Born-Markov approximation is still valid, i.e. $\tauB\ll\tauR$.
In this case, the dynamics is described by the Floquet-Redfield equation.

Since we want to consider the stability of the truncated Floquet-Gibbs state, it is convenient to employ a truncated Floquet reference frame introduced in \cref{sec:reference}.
The Floquet-Redfield equation in the truncated Floquet reference frame is given by
\begin{align}
\dv{t}\tilderhoS^{(n)}=-\im\qty[\HF^{(n)}+\hat{V}^{(n)}(t),\tilderhoS^{(n)}]-\sum_i\sum_{m=-\infty}^\infty\qty[\tilde{X}_{i,m}^{(n)},\tilde{R}_{i,-m}^{(n)}\tilderhoS^{(n)}-\tilderhoS^{(n)}\tilde{R}_{i,m}^{(n)\dagger}],
\label{eq:truncated_Redfield}
\end{align}
where $\qty{\tilderhoS^{(n)}(t), \tilde{X}_{i,m}^{(n)}, \tilde{R}_{i,m}^{(n)}}$ are given by replacing $\hat{K}(t)$ by $\hat{K}^{(n)}(t)$ in \cref{eq:Floquet_reference}.
Let us denote by $\tauH$ the heating time when the system is isolated from the bath.
If $\tauR\ll\tauH$ is satisfied, the energy absorbed from the periodic driving field immediately dissipates into the bath and heating is suppressed.
In this case, we can drop the term $-\im\qty[\hat{V}^{(n)}(t),\tilderhoS^{(n)}(t)]$ in \cref{eq:truncated_Redfield}, and in the same approximation,
\begin{align}
\tilde{R}_{i,m}^{(n)}\approx\sum_j\int_0^\infty\dd s\,\e^{-\im\HF^{(n)}s}\tilde{X}_{j,m}^{(n)}\e^{\im\HF^{(n)}s}\e^{-\im m\omega s}\Phi_{ij}(s).
\end{align}

After dropping $\hat{V}^{(n)}(t)$ in \cref{eq:truncated_Redfield}, the steady state $\tilde{\rho}_\mathrm{ss}$ satisfies
\begin{align}
-\im\qty[\HF^{(n)},\tilde{\rho}_\mathrm{ss}]+\mathcal{D}^{(n)}\qty[\tilde{\rho}_\mathrm{ss}]=0,
\end{align}
where dissipator $\mathcal{D}^{(n)}[\cdot]$ is defined as the last term of \cref{eq:truncated_Redfield}.
We now assume $g^{-1}\ll\tauR$: dissipation is much weaker than a characteristic local energy scale.
We then expect that we can perform the perturbative expansion of $\tilde{\rho}_\mathrm{ss}$ in terms of dissipation strength~\citep{Shirai2020}: $\tilde{\rho}_\mathrm{ss}=\sum_{l=0}^\infty\rho_l$, where
\begin{align}
&-\im\qty[\HF^{(n)},\rho_0]=0, \label{eq:rho_0}\\
&-\im\qty[\HF^{(n)},\rho_{l+1}]+\mathcal{D}^{(n)}\qty[\rho_l]=0 \qq{for $l=1,2,\dots$.}
\label{eq:rho_l}
\end{align}
In the following, we put $\tilde{\rho}_\mathrm{ss}\approx\rho_0$.
From \cref{eq:rho_0}, we find that $\rho_0$ is diagonal in the basis diagonalizing $\HF^{(n)}$: $\rho_0=\sum_aP_a\dyad*{u_a^{(n)}}$.
From \cref{eq:rho_l} with $l=0$, we find $\expval*{\mathcal{D}^{(n)}[\rho_0]}{u_a^{(n)}}=0$, which yields
\begin{align}
\sum_b\qty(W_{ab}^{(n)}P_b-W_{ba}^{(n)}P_a)=0,
\label{eq:Pauli_n}
\end{align}
where the transition rate $W_{ab}^{(n)}$ is given by
\begin{align}
W_{ab}^{(n)}=\sum_{ij}\sum_{m=-\infty}^\infty\gamma_{ij}(E_a^{(n)}-E_b^{(n)}+m\omega)\mel{u_a^{(n)}}{\tilde{X}_{j,m}^{(n)}}{u_b^{(n)}}\mel{u_a^{(n)}}{\tilde{X}_{i,m}^{(n)}}{u_b^{(n)}}^*.
\label{eq:transition_n}
\end{align}
\Cref{eq:Pauli_n} shows that $\qty{P_a}$ is nothing but the steady solution of the Pauli master equation under the transition rates $\qty{W_{ab}^{(n)}}$.
Thus, the truncated Floquet-Gibbs state is realized if the transition rates satisfy the detailed balance condition.

We can repeat the same argument as in \cref{sec:FG_Lindblad}: The detailed balance condition is satisfied if the sum over $m$ in \cref{eq:transition_n} is dominated by $m=0$ for all $a$ and $b$.
Since the high-frequency expansion is divergent, $\HF$ and $\hat{K}(t)$ are highly nonlocal operators.
On the other hand, the truncated ones $\HF^{(n)}$ and $\hat{K}^{(n)}(t)$ are local when $n<n_c\sim\omega/g$, which plays an essential role in the following analysis.
First, we assume that $\qty{\hat{X}_i(t)}$ are local.
Since $\hat{K}^{(n)}(t)=\order{\omega^{-1}}\approx 0$ for large $\omega$, $\tilde{X}_{i,m}^{(n)}\approx(1/T)\int_0^T\dd t\,\hat{X}_i(t)\e^{-\im m\omega t}$, which is also a local operator.
In general, if both $\tilde{X}_{i,m}^{(n)}$ and $\HF^{(n)}$ are local, matrix elements $\mel*{u_a^{(n)}}{\tilde{X}_{i,m}^{(n)}}{u_b^{(n)}}$ decay quickly for $\abs{\Omega}=\abs*{E_a^{(n)}-E_b^{(n)}}>g$~\citep{Mori2018_review}.
It means that $\tilde{X}_{i,m}^{(n)}$ for any $\Omega\gg g$ is negligible.
Since $\omega\gg g\sim\Omega$, we can assume that $\Omega+m\omega\approx m\omega$ for any $m\neq 0$ in \cref{eq:transition_n}.
If $(1/T)\int_0^T\dd t\,\HI(t)\approx\sum_i\tilde{X}_{i,0}^{(n)}\otimes\hat{Y}_i$ is not vanishingly small, only $m=0$ is dominant in \cref{eq:transition_n} under the same condition (ii) in \cref{sec:FG_Lindblad}.

Let us summarize the conditions under which the truncated Floquet-Gibbs state is realized in the steady state:
\begin{description}
\item[(i')] $g^{-1}\ll\tauR\ll\tauH$.
\item[(ii)] Either $\qty[\HI,\hat{A}]=0$ or $\omega_c\ll\omega$.
Here, $\hat{A}=\hat{V}$ in the case of strong driving (\cref{eq:H_strong}) and $\hat{A}=\hat{N}$ in the case of resonant driving (\cref{eq:H_resonant}).
\item[(iii)] $(1/T)\int_0^T\dd t\, \HI(t)$ is not vanishingly small.
\end{description}

The condition (i') can be realized when $g\ll\omega$ and dissipation is weak compared with $g$ but strong enough to suppress heating.
Because of the locality of $\HF^{(n)}$ and $\hat{K}^{(n)}(t)$, the condition (i) $(1/T)\int_0^T\dd t\,\norm*{\HS(t)}\ll\omega$ is replaced by $g\ll\omega$.
It should be emphasized that increasing the strength of dissipation alone is not enough to stabilize a truncated Floquet-Gibbs state: the conditions (ii) and (iii) are also necessary~\citep{Shirai2016}.

\section{CONCLUSION AND OUTLOOK}
\label{sec:outlook}

In this review, we have presented in detail a master-equation formalism for open Floquet systems and discussed statistical mechanics for Floquet states.
It is found that some conditions are necessary to apply the method of equilibrium statistical mechanics.

When one or more of these conditions are violated, the system will relax to a periodic nonthermal steady state~\citep{Iadecola2015_floquet, Iwahori2016}, which is not described by equilibrium statistical mechanics for $\HF$ or $\HF^{(n)}$.
Such steady states will be of great importance since they can exhibit novel phases that are inaccessible in thermal equilibrium.
Thus, it should be an important future problem to systematically study nonthermal steady states in open Floquet systems that do not satisfy one of the conditions given in \cref{sec:stat}.

In this review, we have applied the Floquet theory to a closed system: we have investigated the effect of dissipation by using the Floquet reference frame that is constructed without dissipation.
Instead, the Floquet theory may be directly applied to the quantum master equation with periodic time dependence~\citep{Haddadfarshi2015, Dai2016, Hartmann2017}: $\dd\rhoS/\dd t=\mathcal{L}(t)\rhoS$, where the generator $\mathcal{L}(t)=\mathcal{L}(t+T)$ is simply called the Liouvillian.
Correspondingly to \cref{eq:decomposition} for closed systems, we obtain
$\mathcal{T}\e^{\int_{t_0}^t\dd s\,\mathcal{L}(s)}=\e^{-\im\mathcal{G}(t)}\e^{\mathcal{L}_\mathrm{F}(t-t_0)}\e^{\im\mathcal{G}(t_0)}$,
where $\mathcal{G}(t)$ is the micromotion superoperator with periodicity $\mathcal{G}(t)=\mathcal{G}(t+T)$ and $\mathcal{L}_\mathrm{F}$ is the Floquet Liouvillian.

Properties of $\mathcal{G}(t)$ and $\mathcal{L}_\mathrm{F}$ or their high-frequency expansions have recently been studied when $\mathcal{L}(t)$ is of the Lindblad form~\citep{Schnell2020, Mizuta2021, Schnell2021, Ikeda2021_nonequilibrium}.
It is known that $\mathcal{L}_\mathrm{F}$ may not be of the Lindblad form even if $\mathcal{L}(t)$ is of the Lindblad form for all times $t$~\citep{Wolf2008}.
Schnell et al.~\cite{Schnell2020} demonstrated that a two-level system already has a phase in which the ``Lindbladianity'' of the Floquet Liouvillian breaks down.
It is pointed out that the non-Lindbladianity is a consequence of the non-unitarity of the micromotion $\mathcal{G}(t)$.
Mizuta et al.~\cite{Mizuta2021} showed that the Floquet-Magnus high-frequency expansion of $\mathcal{L}_\mathrm{F}$ may not be of the Lindblad form.
Schnell et al.~\cite{Schnell2021} showed that the Lindbladianity of the high-frequency expansion of $\mathcal{L}_\mathrm{F}$ depends on the expansion technique (Floquet-Magnus or van Vleck) and the reference frame (the laboratory frame or a rotating frame).
Ikeda et al.~\cite{Ikeda2021_nonequilibrium} showed that although an effective Liouvillian obtained by truncating the high-frequency expansion of $\mathcal{L}_\mathrm{F}$ may not be of the Lindblad form, a periodic steady state is guaranteed to exist at each order of the high-frequency expansion.
However, it is still an open problem to fully understand general properties of $\mathcal{L}_\mathrm{F}$ and $\mathcal{G}(t)$ and their high-frequency expansions.

Another important direction of research, which could not be discussed in this review, is on nonequilibrium dynamics at a single trajectory level.
Under continuous monitoring of a quantum system, its state $\ket{\psi(t)}$ undergoes stochastic time evolution reflecting randomness of measurement outcomes.
The dynamics of $\ket{\psi(t)}$ is described by the stochastic Schr\"odinger equation~\citep{Breuer_text}, whereas its ensemble average $\rho(t)=\mathrm{Ave}[\dyad{\psi(t)}]$ follows a Lindblad equation, where $\mathrm{Ave}[\cdot]$ denotes the average over infinitely many realizations of $\ket{\psi(t)}$.
Remarkably, even if the ensemble average does not exhibit any singularity, a phase transition regarding the entanglement entropy can occur at a single trajectory level~\citep{Li2018_quantum, Chan2019, Skinner2019, Li2019_measurement, Cao2019_entanglement, Bao2020, Gullans2020, Fuji2020, Ippoliti2021, Alberton2021}.
This new type of phase transitions is called the measurement-induced phase transition or the entanglement phase transition.
Elucidating interplay of periodic driving and continuous monitoring at a single trajectory level would also be an intriguing future problem.

Finally, although we focus on Markovian dynamics in this review, it is definitely an important future problem to explore rich physics in the non-Markovian regime.

\section*{ACKNOWLEDGMENTS}
This work was supported by JSPS KAKENHI Grant Numbers JP19K14622, JP21H05185.
The author is grateful to Yoshihiro Michishita, Masaya Nakagawa, and Tatsuhiko Shirai for their valuable comments on the manuscript.
%

\bibliography{apsrevcontrol,physics}

\begin{thebibliography}{120}%
\makeatletter
\providecommand \@ifxundefined [1]{%
 \@ifx{#1\undefined}
}%
\providecommand \@ifnum [1]{%
 \ifnum #1\expandafter \@firstoftwo
 \else \expandafter \@secondoftwo
 \fi
}%
\providecommand \@ifx [1]{%
 \ifx #1\expandafter \@firstoftwo
 \else \expandafter \@secondoftwo
 \fi
}%
\providecommand \natexlab [1]{#1}%
\providecommand \enquote  [1]{``#1''}%
\providecommand \bibnamefont  [1]{#1}%
\providecommand \bibfnamefont [1]{#1}%
\providecommand \citenamefont [1]{#1}%
\providecommand \href@noop [0]{\@secondoftwo}%
\providecommand \href [0]{\begingroup \@sanitize@url \@href}%
\providecommand \@href[1]{\@@startlink{#1}\@@href}%
\providecommand \@@href[1]{\endgroup#1\@@endlink}%
\providecommand \@sanitize@url [0]{\catcode `\\12\catcode `\$12\catcode
  `\&12\catcode `\#12\catcode `\^12\catcode `\_12\catcode `\%12\relax}%
\providecommand \@@startlink[1]{}%
\providecommand \@@endlink[0]{}%
\providecommand \url  [0]{\begingroup\@sanitize@url \@url }%
\providecommand \@url [1]{\endgroup\@href {#1}{\urlprefix }}%
\providecommand \urlprefix  [0]{URL }%
\providecommand \Eprint [0]{\href }%
\providecommand \doibase [0]{https://doi.org/}%
\providecommand \selectlanguage [0]{\@gobble}%
\providecommand \bibinfo  [0]{\@secondoftwo}%
\providecommand \bibfield  [0]{\@secondoftwo}%
\providecommand \translation [1]{[#1]}%
\providecommand \BibitemOpen [0]{}%
\providecommand \bibitemStop [0]{}%
\providecommand \bibitemNoStop [0]{.\EOS\space}%
\providecommand \EOS [0]{\spacefactor3000\relax}%
\providecommand \BibitemShut  [1]{\csname bibitem#1\endcsname}%
\let\auto@bib@innerbib\@empty
\bibitem [{\citenamefont {Kubo}(1957)}]{Kubo1957}%
  \BibitemOpen
  \bibfield  {author} {\bibinfo {author} {\bibfnamefont {R.}~\bibnamefont
  {Kubo}},\ }\bibfield  {title} {\bibinfo {title} {{Statistical Mechanical
  Theory of Irreversible Processes. I. General Theory and Simple Applications
  to Magnetic and Conduction Problems}},\ }\href
  {https://doi.org/10.1143/JPSJ.12.570} {\bibfield  {journal} {\bibinfo
  {journal} {J. Phys. Soc. Japan}\ }\textbf {\bibinfo {volume} {12}},\ \bibinfo
  {pages} {570--586} (\bibinfo {year} {1957})}\BibitemShut {NoStop}%
\bibitem [{\citenamefont {Evans}\ \emph {et~al.}(1993)\citenamefont {Evans},
  \citenamefont {Cohen},\ and\ \citenamefont {Morriss}}]{Evans1993}%
  \BibitemOpen
  \bibfield  {author} {\bibinfo {author} {\bibfnamefont {D.~J.}\ \bibnamefont
  {Evans}}, \bibinfo {author} {\bibfnamefont {E.~G.}\ \bibnamefont {Cohen}},\
  and\ \bibinfo {author} {\bibfnamefont {G.~P.}\ \bibnamefont {Morriss}},\
  }\bibfield  {title} {\bibinfo {title} {{Probability of second law violations
  in shearing steady states}},\ }\href
  {https://doi.org/10.1103/PhysRevLett.71.2401} {\bibfield  {journal} {\bibinfo
   {journal} {Phys. Rev. Lett.}\ }\textbf {\bibinfo {volume} {71}},\ \bibinfo
  {pages} {2401} (\bibinfo {year} {1993})}\BibitemShut {NoStop}%
\bibitem [{\citenamefont {Gallavotti}\ and\ \citenamefont
  {Cohen}(1995)}]{Gallavotti1995}%
  \BibitemOpen
  \bibfield  {author} {\bibinfo {author} {\bibfnamefont {G.}~\bibnamefont
  {Gallavotti}}\ and\ \bibinfo {author} {\bibfnamefont {E.~G.}\ \bibnamefont
  {Cohen}},\ }\bibfield  {title} {\bibinfo {title} {{Dynamical ensembles in
  nonequilibrium statistical mechanics}},\ }\href
  {https://doi.org/10.1103/PhysRevLett.74.2694} {\bibfield  {journal} {\bibinfo
   {journal} {Phys. Rev. Lett.}\ }\textbf {\bibinfo {volume} {74}},\ \bibinfo
  {pages} {2694} (\bibinfo {year} {1995})}\BibitemShut {NoStop}%
\bibitem [{\citenamefont {Jarzynski}(1997)}]{Jarzynski1997}%
  \BibitemOpen
  \bibfield  {author} {\bibinfo {author} {\bibfnamefont {C.}~\bibnamefont
  {Jarzynski}},\ }\bibfield  {title} {\bibinfo {title} {{Nonequilibrium
  Equality for Free Energy Differences}},\ }\href
  {https://doi.org/10.1103/PhysRevLett.78.2690} {\bibfield  {journal} {\bibinfo
   {journal} {Phys. Rev. Lett.}\ }\textbf {\bibinfo {volume} {78}},\ \bibinfo
  {pages} {2690} (\bibinfo {year} {1997})}\BibitemShut {NoStop}%
\bibitem [{\citenamefont {Bukov}\ \emph
  {et~al.}(2015{\natexlab{a}})\citenamefont {Bukov}, \citenamefont
  {D'Alessio},\ and\ \citenamefont {Polkovnikov}}]{Bukov2015_review}%
  \BibitemOpen
  \bibfield  {author} {\bibinfo {author} {\bibfnamefont {M.}~\bibnamefont
  {Bukov}}, \bibinfo {author} {\bibfnamefont {L.}~\bibnamefont {D'Alessio}},\
  and\ \bibinfo {author} {\bibfnamefont {A.}~\bibnamefont {Polkovnikov}},\
  }\bibfield  {title} {\bibinfo {title} {{Universal high-frequency behavior of
  periodically driven systems: From dynamical stabilization to Floquet
  engineering}},\ }\href {https://doi.org/10.1080/00018732.2015.1055918}
  {\bibfield  {journal} {\bibinfo  {journal} {Adv. Phys.}\ }\textbf {\bibinfo
  {volume} {64}},\ \bibinfo {pages} {139--226} (\bibinfo {year}
  {2015}{\natexlab{a}})}\BibitemShut {NoStop}%
\bibitem [{\citenamefont {Eckardt}(2017)}]{Eckardt2017_review}%
  \BibitemOpen
  \bibfield  {author} {\bibinfo {author} {\bibfnamefont {A.}~\bibnamefont
  {Eckardt}},\ }\bibfield  {title} {\bibinfo {title} {{Colloquium: Atomic
  quantum gases in periodically driven optical lattices}},\ }\href
  {https://doi.org/10.1103/RevModPhys.89.011004} {\bibfield  {journal}
  {\bibinfo  {journal} {Rev. Mod. Phys.}\ }\textbf {\bibinfo {volume} {89}},\
  \bibinfo {pages} {011004} (\bibinfo {year} {2017})}\BibitemShut {NoStop}%
\bibitem [{\citenamefont {Mori}\ \emph {et~al.}(2016)\citenamefont {Mori},
  \citenamefont {Kuwahara},\ and\ \citenamefont {Saito}}]{Mori2016b}%
  \BibitemOpen
  \bibfield  {author} {\bibinfo {author} {\bibfnamefont {T.}~\bibnamefont
  {Mori}}, \bibinfo {author} {\bibfnamefont {T.}~\bibnamefont {Kuwahara}},\
  and\ \bibinfo {author} {\bibfnamefont {K.}~\bibnamefont {Saito}},\ }\bibfield
   {title} {\bibinfo {title} {{Rigorous Bound on Energy Absorption and Generic
  Relaxation in Periodically Driven Quantum Systems}},\ }\href
  {https://doi.org/10.1103/PhysRevLett.116.120401} {\bibfield  {journal}
  {\bibinfo  {journal} {Phys. Rev. Lett.}\ }\textbf {\bibinfo {volume} {116}},\
  \bibinfo {pages} {120401} (\bibinfo {year} {2016})}\BibitemShut {NoStop}%
\bibitem [{\citenamefont {Kuwahara}\ \emph {et~al.}(2016)\citenamefont
  {Kuwahara}, \citenamefont {Mori},\ and\ \citenamefont
  {Saito}}]{Kuwahara2016}%
  \BibitemOpen
  \bibfield  {author} {\bibinfo {author} {\bibfnamefont {T.}~\bibnamefont
  {Kuwahara}}, \bibinfo {author} {\bibfnamefont {T.}~\bibnamefont {Mori}},\
  and\ \bibinfo {author} {\bibfnamefont {K.}~\bibnamefont {Saito}},\ }\bibfield
   {title} {\bibinfo {title} {{Floquet-Magnus theory and generic transient
  dynamics in periodically driven many-body quantum systems}},\ }\href
  {https://doi.org/10.1016/j.aop.2016.01.012} {\bibfield  {journal} {\bibinfo
  {journal} {Ann. Phys. (N. Y).}\ }\textbf {\bibinfo {volume} {367}},\ \bibinfo
  {pages} {96--124} (\bibinfo {year} {2016})}\BibitemShut {NoStop}%
\bibitem [{\citenamefont {Abanin}\ \emph
  {et~al.}(2017{\natexlab{a}})\citenamefont {Abanin}, \citenamefont {{De
  Roeck}}, \citenamefont {Ho},\ and\ \citenamefont {Huveneers}}]{Abanin2017}%
  \BibitemOpen
  \bibfield  {author} {\bibinfo {author} {\bibfnamefont {D.}~\bibnamefont
  {Abanin}}, \bibinfo {author} {\bibfnamefont {W.}~\bibnamefont {{De Roeck}}},
  \bibinfo {author} {\bibfnamefont {W.~W.}\ \bibnamefont {Ho}},\ and\ \bibinfo
  {author} {\bibfnamefont {F.}~\bibnamefont {Huveneers}},\ }\bibfield  {title}
  {\bibinfo {title} {{A Rigorous Theory of Many-Body Prethermalization for
  Periodically Driven and Closed Quantum Systems}},\ }\href
  {https://doi.org/10.1007/s00220-017-2930-x} {\bibfield  {journal} {\bibinfo
  {journal} {Commun. Math. Phys.}\ }\textbf {\bibinfo {volume} {354}},\
  \bibinfo {pages} {809--827} (\bibinfo {year}
  {2017}{\natexlab{a}})}\BibitemShut {NoStop}%
\bibitem [{\citenamefont {Abanin}\ \emph
  {et~al.}(2017{\natexlab{b}})\citenamefont {Abanin}, \citenamefont {{De
  Roeck}}, \citenamefont {Ho},\ and\ \citenamefont {Huveneers}}]{Abanin2017a}%
  \BibitemOpen
  \bibfield  {author} {\bibinfo {author} {\bibfnamefont {D.~A.}\ \bibnamefont
  {Abanin}}, \bibinfo {author} {\bibfnamefont {W.}~\bibnamefont {{De Roeck}}},
  \bibinfo {author} {\bibfnamefont {W.~W.}\ \bibnamefont {Ho}},\ and\ \bibinfo
  {author} {\bibfnamefont {F.}~\bibnamefont {Huveneers}},\ }\bibfield  {title}
  {\bibinfo {title} {{Effective Hamiltonians, prethermalization, and slow
  energy absorption in periodically driven many-body systems}},\ }\href
  {https://doi.org/10.1103/PhysRevB.95.014112} {\bibfield  {journal} {\bibinfo
  {journal} {Phys. Rev. B}\ }\textbf {\bibinfo {volume} {95}},\ \bibinfo
  {pages} {014112} (\bibinfo {year} {2017}{\natexlab{b}})}\BibitemShut
  {NoStop}%
\bibitem [{\citenamefont {Oka}\ and\ \citenamefont
  {Kitamura}(2019)}]{Oka2019_review}%
  \BibitemOpen
  \bibfield  {author} {\bibinfo {author} {\bibfnamefont {T.}~\bibnamefont
  {Oka}}\ and\ \bibinfo {author} {\bibfnamefont {S.}~\bibnamefont {Kitamura}},\
  }\bibfield  {title} {\bibinfo {title} {{Floquet engineering of quantum
  materials}},\ }\href
  {https://doi.org/10.1146/annurev-conmatphys-031218-013423} {\bibfield
  {journal} {\bibinfo  {journal} {Annu. Rev. Condens. Matter Phys.}\ }\textbf
  {\bibinfo {volume} {10}},\ \bibinfo {pages} {387--408} (\bibinfo {year}
  {2019})}\BibitemShut {NoStop}%
\bibitem [{\citenamefont {Eckardt}\ \emph {et~al.}(2005)\citenamefont
  {Eckardt}, \citenamefont {Weiss},\ and\ \citenamefont
  {Holthaus}}]{Eckardt2005}%
  \BibitemOpen
  \bibfield  {author} {\bibinfo {author} {\bibfnamefont {A.}~\bibnamefont
  {Eckardt}}, \bibinfo {author} {\bibfnamefont {C.}~\bibnamefont {Weiss}},\
  and\ \bibinfo {author} {\bibfnamefont {M.}~\bibnamefont {Holthaus}},\
  }\bibfield  {title} {\bibinfo {title} {{Superfluid-insulator transition in a
  periodically driven optical lattice}},\ }\href
  {https://doi.org/10.1103/PhysRevLett.95.260404} {\bibfield  {journal}
  {\bibinfo  {journal} {Phys. Rev. Lett.}\ }\textbf {\bibinfo {volume} {95}},\
  \bibinfo {pages} {260404} (\bibinfo {year} {2005})}\BibitemShut {NoStop}%
\bibitem [{\citenamefont {Zenesini}\ \emph {et~al.}(2009)\citenamefont
  {Zenesini}, \citenamefont {Lignier}, \citenamefont {Ciampini}, \citenamefont
  {Morsch},\ and\ \citenamefont {Arimondo}}]{Zenesini2009}%
  \BibitemOpen
  \bibfield  {author} {\bibinfo {author} {\bibfnamefont {A.}~\bibnamefont
  {Zenesini}}, \bibinfo {author} {\bibfnamefont {H.}~\bibnamefont {Lignier}},
  \bibinfo {author} {\bibfnamefont {D.}~\bibnamefont {Ciampini}}, \bibinfo
  {author} {\bibfnamefont {O.}~\bibnamefont {Morsch}},\ and\ \bibinfo {author}
  {\bibfnamefont {E.}~\bibnamefont {Arimondo}},\ }\bibfield  {title} {\bibinfo
  {title} {{Coherent control of dressed matter waves}},\ }\href
  {https://doi.org/10.1103/PhysRevLett.102.100403} {\bibfield  {journal}
  {\bibinfo  {journal} {Phys. Rev. Lett.}\ }\textbf {\bibinfo {volume} {102}},\
  \bibinfo {pages} {100403} (\bibinfo {year} {2009})}\BibitemShut {NoStop}%
\bibitem [{\citenamefont {Bastidas}\ \emph {et~al.}(2012)\citenamefont
  {Bastidas}, \citenamefont {Emary}, \citenamefont {Regler},\ and\
  \citenamefont {Brandes}}]{Bastidas2012}%
  \BibitemOpen
  \bibfield  {author} {\bibinfo {author} {\bibfnamefont {V.~M.}\ \bibnamefont
  {Bastidas}}, \bibinfo {author} {\bibfnamefont {C.}~\bibnamefont {Emary}},
  \bibinfo {author} {\bibfnamefont {B.}~\bibnamefont {Regler}},\ and\ \bibinfo
  {author} {\bibfnamefont {T.}~\bibnamefont {Brandes}},\ }\bibfield  {title}
  {\bibinfo {title} {{Nonequilibrium quantum phase transitions in the Dicke
  model}},\ }\href {https://doi.org/10.1103/PhysRevLett.108.043003} {\bibfield
  {journal} {\bibinfo  {journal} {Phys. Rev. Lett.}\ }\textbf {\bibinfo
  {volume} {108}},\ \bibinfo {pages} {043003} (\bibinfo {year}
  {2012})}\BibitemShut {NoStop}%
\bibitem [{\citenamefont {Oka}\ and\ \citenamefont {Aoki}(2009)}]{Oka2009}%
  \BibitemOpen
  \bibfield  {author} {\bibinfo {author} {\bibfnamefont {T.}~\bibnamefont
  {Oka}}\ and\ \bibinfo {author} {\bibfnamefont {H.}~\bibnamefont {Aoki}},\
  }\bibfield  {title} {\bibinfo {title} {{Photovoltaic Hall effect in
  graphene}},\ }\href {https://doi.org/10.1103/PhysRevB.79.081406} {\bibfield
  {journal} {\bibinfo  {journal} {Phys. Rev. B}\ }\textbf {\bibinfo {volume}
  {79}},\ \bibinfo {pages} {081406} (\bibinfo {year} {2009})}\BibitemShut
  {NoStop}%
\bibitem [{\citenamefont {Kitagawa}\ \emph {et~al.}(2010)\citenamefont
  {Kitagawa}, \citenamefont {Berg}, \citenamefont {Rudner},\ and\ \citenamefont
  {Demler}}]{Kitagawa2010}%
  \BibitemOpen
  \bibfield  {author} {\bibinfo {author} {\bibfnamefont {T.}~\bibnamefont
  {Kitagawa}}, \bibinfo {author} {\bibfnamefont {E.}~\bibnamefont {Berg}},
  \bibinfo {author} {\bibfnamefont {M.}~\bibnamefont {Rudner}},\ and\ \bibinfo
  {author} {\bibfnamefont {E.}~\bibnamefont {Demler}},\ }\bibfield  {title}
  {\bibinfo {title} {{Topological characterization of periodically driven
  quantum systems}},\ }\href {https://doi.org/10.1103/PhysRevB.82.235114}
  {\bibfield  {journal} {\bibinfo  {journal} {Phys. Rev. B}\ }\textbf {\bibinfo
  {volume} {82}},\ \bibinfo {pages} {235114} (\bibinfo {year}
  {2010})}\BibitemShut {NoStop}%
\bibitem [{\citenamefont {Lindner}\ \emph {et~al.}(2011)\citenamefont
  {Lindner}, \citenamefont {Refael},\ and\ \citenamefont
  {Galitski}}]{Lindner2011}%
  \BibitemOpen
  \bibfield  {author} {\bibinfo {author} {\bibfnamefont {N.~H.}\ \bibnamefont
  {Lindner}}, \bibinfo {author} {\bibfnamefont {G.}~\bibnamefont {Refael}},\
  and\ \bibinfo {author} {\bibfnamefont {V.}~\bibnamefont {Galitski}},\
  }\bibfield  {title} {\bibinfo {title} {{Floquet topological insulator in
  semiconductor quantum wells}},\ }\href {https://doi.org/10.1038/nphys1926}
  {\bibfield  {journal} {\bibinfo  {journal} {Nat. Phys.}\ }\textbf {\bibinfo
  {volume} {7}},\ \bibinfo {pages} {490--495} (\bibinfo {year}
  {2011})}\BibitemShut {NoStop}%
\bibitem [{\citenamefont {Jotzu}\ \emph {et~al.}(2014)\citenamefont {Jotzu},
  \citenamefont {Messer}, \citenamefont {Desbuquois}, \citenamefont {Lebrat},
  \citenamefont {Uehlinger}, \citenamefont {Greif},\ and\ \citenamefont
  {Esslinger}}]{Jotzu2014}%
  \BibitemOpen
  \bibfield  {author} {\bibinfo {author} {\bibfnamefont {G.}~\bibnamefont
  {Jotzu}}, \bibinfo {author} {\bibfnamefont {M.}~\bibnamefont {Messer}},
  \bibinfo {author} {\bibfnamefont {R.}~\bibnamefont {Desbuquois}}, \bibinfo
  {author} {\bibfnamefont {M.}~\bibnamefont {Lebrat}}, \bibinfo {author}
  {\bibfnamefont {T.}~\bibnamefont {Uehlinger}}, \bibinfo {author}
  {\bibfnamefont {D.}~\bibnamefont {Greif}},\ and\ \bibinfo {author}
  {\bibfnamefont {T.}~\bibnamefont {Esslinger}},\ }\bibfield  {title} {\bibinfo
  {title} {{Experimental realization of the topological Haldane model with
  ultracold fermions}},\ }\href {https://doi.org/10.1038/nature13915}
  {\bibfield  {journal} {\bibinfo  {journal} {Nature}\ }\textbf {\bibinfo
  {volume} {515}},\ \bibinfo {pages} {237--240} (\bibinfo {year}
  {2014})}\BibitemShut {NoStop}%
\bibitem [{\citenamefont {Aidelsburger}\ \emph {et~al.}(2015)\citenamefont
  {Aidelsburger}, \citenamefont {Lohse}, \citenamefont {Schweizer},
  \citenamefont {Atala}, \citenamefont {Barreiro}, \citenamefont
  {Nascimb{\`{e}}ne}, \citenamefont {Cooper}, \citenamefont {Bloch},\ and\
  \citenamefont {Goldman}}]{Aidelsburger2015}%
  \BibitemOpen
  \bibfield  {author} {\bibinfo {author} {\bibfnamefont {M.}~\bibnamefont
  {Aidelsburger}}, \bibinfo {author} {\bibfnamefont {M.}~\bibnamefont {Lohse}},
  \bibinfo {author} {\bibfnamefont {C.}~\bibnamefont {Schweizer}}, \bibinfo
  {author} {\bibfnamefont {M.}~\bibnamefont {Atala}}, \bibinfo {author}
  {\bibfnamefont {J.~T.}\ \bibnamefont {Barreiro}}, \bibinfo {author}
  {\bibfnamefont {S.}~\bibnamefont {Nascimb{\`{e}}ne}}, \bibinfo {author}
  {\bibfnamefont {N.~R.}\ \bibnamefont {Cooper}}, \bibinfo {author}
  {\bibfnamefont {I.}~\bibnamefont {Bloch}},\ and\ \bibinfo {author}
  {\bibfnamefont {N.}~\bibnamefont {Goldman}},\ }\bibfield  {title} {\bibinfo
  {title} {{Measuring the Chern number of Hofstadter bands with ultracold
  bosonic atoms}},\ }\href {https://doi.org/10.1038/nphys3171} {\bibfield
  {journal} {\bibinfo  {journal} {Nat. Phys.}\ }\textbf {\bibinfo {volume}
  {11}},\ \bibinfo {pages} {162--166} (\bibinfo {year} {2015})}\BibitemShut
  {NoStop}%
\bibitem [{\citenamefont {Aidelsburger}\ \emph {et~al.}(2011)\citenamefont
  {Aidelsburger}, \citenamefont {Atala}, \citenamefont {Nascimb{\`{e}}ne},
  \citenamefont {Trotzky}, \citenamefont {Chen},\ and\ \citenamefont
  {Bloch}}]{Aidelsburger2011}%
  \BibitemOpen
  \bibfield  {author} {\bibinfo {author} {\bibfnamefont {M.}~\bibnamefont
  {Aidelsburger}}, \bibinfo {author} {\bibfnamefont {M.}~\bibnamefont {Atala}},
  \bibinfo {author} {\bibfnamefont {S.}~\bibnamefont {Nascimb{\`{e}}ne}},
  \bibinfo {author} {\bibfnamefont {S.}~\bibnamefont {Trotzky}}, \bibinfo
  {author} {\bibfnamefont {Y.~A.}\ \bibnamefont {Chen}},\ and\ \bibinfo
  {author} {\bibfnamefont {I.}~\bibnamefont {Bloch}},\ }\bibfield  {title}
  {\bibinfo {title} {{Experimental realization of strong effective magnetic
  fields in an optical lattice}},\ }\href
  {https://doi.org/10.1103/PhysRevLett.107.255301} {\bibfield  {journal}
  {\bibinfo  {journal} {Phys. Rev. Lett.}\ }\textbf {\bibinfo {volume} {107}},\
  \bibinfo {pages} {255301} (\bibinfo {year} {2011})}\BibitemShut {NoStop}%
\bibitem [{\citenamefont {Struck}\ \emph {et~al.}(2012)\citenamefont {Struck},
  \citenamefont {Olschl{\"{a}}ger}, \citenamefont {Weinberg}, \citenamefont
  {Hauke}, \citenamefont {Simonet}, \citenamefont {Eckardt}, \citenamefont
  {Lewenstein}, \citenamefont {Sengstock},\ and\ \citenamefont
  {Windpassinger}}]{Struck2012}%
  \BibitemOpen
  \bibfield  {author} {\bibinfo {author} {\bibfnamefont {J.}~\bibnamefont
  {Struck}}, \bibinfo {author} {\bibfnamefont {C.}~\bibnamefont
  {Olschl{\"{a}}ger}}, \bibinfo {author} {\bibfnamefont {M.}~\bibnamefont
  {Weinberg}}, \bibinfo {author} {\bibfnamefont {P.}~\bibnamefont {Hauke}},
  \bibinfo {author} {\bibfnamefont {J.}~\bibnamefont {Simonet}}, \bibinfo
  {author} {\bibfnamefont {A.}~\bibnamefont {Eckardt}}, \bibinfo {author}
  {\bibfnamefont {M.}~\bibnamefont {Lewenstein}}, \bibinfo {author}
  {\bibfnamefont {K.}~\bibnamefont {Sengstock}},\ and\ \bibinfo {author}
  {\bibfnamefont {P.}~\bibnamefont {Windpassinger}},\ }\bibfield  {title}
  {\bibinfo {title} {{Tunable gauge potential for neutral and spinless
  particles in driven optical lattices}},\ }\href
  {https://doi.org/10.1103/PhysRevLett.108.225304} {\bibfield  {journal}
  {\bibinfo  {journal} {Phys. Rev. Lett.}\ }\textbf {\bibinfo {volume} {108}},\
  \bibinfo {pages} {225304} (\bibinfo {year} {2012})}\BibitemShut {NoStop}%
\bibitem [{\citenamefont {Bermudez}\ \emph {et~al.}(2011)\citenamefont
  {Bermudez}, \citenamefont {Schaetz},\ and\ \citenamefont
  {Porras}}]{Bermudez2011}%
  \BibitemOpen
  \bibfield  {author} {\bibinfo {author} {\bibfnamefont {A.}~\bibnamefont
  {Bermudez}}, \bibinfo {author} {\bibfnamefont {T.}~\bibnamefont {Schaetz}},\
  and\ \bibinfo {author} {\bibfnamefont {D.}~\bibnamefont {Porras}},\
  }\bibfield  {title} {\bibinfo {title} {{Synthetic gauge fields for
  vibrational excitations of trapped ions}},\ }\href
  {https://doi.org/10.1103/PhysRevLett.107.150501} {\bibfield  {journal}
  {\bibinfo  {journal} {Phys. Rev. Lett.}\ }\textbf {\bibinfo {volume} {107}},\
  \bibinfo {pages} {150501} (\bibinfo {year} {2011})}\BibitemShut {NoStop}%
\bibitem [{\citenamefont {Else}\ \emph {et~al.}(2016)\citenamefont {Else},
  \citenamefont {Bauer},\ and\ \citenamefont {Nayak}}]{Else2016}%
  \BibitemOpen
  \bibfield  {author} {\bibinfo {author} {\bibfnamefont {D.~V.}\ \bibnamefont
  {Else}}, \bibinfo {author} {\bibfnamefont {B.}~\bibnamefont {Bauer}},\ and\
  \bibinfo {author} {\bibfnamefont {C.}~\bibnamefont {Nayak}},\ }\bibfield
  {title} {\bibinfo {title} {{Floquet Time Crystals}},\ }\href
  {https://doi.org/10.1103/PhysRevLett.117.090402} {\bibfield  {journal}
  {\bibinfo  {journal} {Phys. Rev. Lett.}\ }\textbf {\bibinfo {volume} {117}},\
  \bibinfo {pages} {090402} (\bibinfo {year} {2016})}\BibitemShut {NoStop}%
\bibitem [{\citenamefont {Else}\ \emph {et~al.}(2017)\citenamefont {Else},
  \citenamefont {Bauer},\ and\ \citenamefont {Nayak}}]{Else2017}%
  \BibitemOpen
  \bibfield  {author} {\bibinfo {author} {\bibfnamefont {D.~V.}\ \bibnamefont
  {Else}}, \bibinfo {author} {\bibfnamefont {B.}~\bibnamefont {Bauer}},\ and\
  \bibinfo {author} {\bibfnamefont {C.}~\bibnamefont {Nayak}},\ }\bibfield
  {title} {\bibinfo {title} {{Prethermal phases of matter protected by
  time-translation symmetry}},\ }\href
  {https://doi.org/10.1103/PhysRevX.7.011026} {\bibfield  {journal} {\bibinfo
  {journal} {Phys. Rev. X}\ }\textbf {\bibinfo {volume} {7}},\ \bibinfo {pages}
  {011026} (\bibinfo {year} {2017})}\BibitemShut {NoStop}%
\bibitem [{\citenamefont {Yao}\ \emph {et~al.}(2017)\citenamefont {Yao},
  \citenamefont {Potter}, \citenamefont {Potirniche},\ and\ \citenamefont
  {Vishwanath}}]{Yao2017}%
  \BibitemOpen
  \bibfield  {author} {\bibinfo {author} {\bibfnamefont {N.~Y.}\ \bibnamefont
  {Yao}}, \bibinfo {author} {\bibfnamefont {A.~C.}\ \bibnamefont {Potter}},
  \bibinfo {author} {\bibfnamefont {I.~D.}\ \bibnamefont {Potirniche}},\ and\
  \bibinfo {author} {\bibfnamefont {A.}~\bibnamefont {Vishwanath}},\ }\bibfield
   {title} {\bibinfo {title} {{Discrete Time Crystals: Rigidity, Criticality,
  and Realizations}},\ }\href {https://doi.org/10.1103/PhysRevLett.118.030401}
  {\bibfield  {journal} {\bibinfo  {journal} {Phys. Rev. Lett.}\ }\textbf
  {\bibinfo {volume} {118}},\ \bibinfo {pages} {030401} (\bibinfo {year}
  {2017})}\BibitemShut {NoStop}%
\bibitem [{\citenamefont {Tsuji}\ \emph {et~al.}(2008)\citenamefont {Tsuji},
  \citenamefont {Oka},\ and\ \citenamefont {Aoki}}]{Tsuji2008}%
  \BibitemOpen
  \bibfield  {author} {\bibinfo {author} {\bibfnamefont {N.}~\bibnamefont
  {Tsuji}}, \bibinfo {author} {\bibfnamefont {T.}~\bibnamefont {Oka}},\ and\
  \bibinfo {author} {\bibfnamefont {H.}~\bibnamefont {Aoki}},\ }\bibfield
  {title} {\bibinfo {title} {{Correlated electron systems periodically driven
  out of equilibrium: Floquet+DMFT formalism}},\ }\href
  {https://doi.org/10.1103/PhysRevB.78.235124} {\bibfield  {journal} {\bibinfo
  {journal} {Phys. Rev. B}\ }\textbf {\bibinfo {volume} {78}},\ \bibinfo
  {pages} {235124} (\bibinfo {year} {2008})}\BibitemShut {NoStop}%
\bibitem [{\citenamefont {Tsuji}\ \emph {et~al.}(2009)\citenamefont {Tsuji},
  \citenamefont {Oka},\ and\ \citenamefont {Aoki}}]{Tsuji2009}%
  \BibitemOpen
  \bibfield  {author} {\bibinfo {author} {\bibfnamefont {N.}~\bibnamefont
  {Tsuji}}, \bibinfo {author} {\bibfnamefont {T.}~\bibnamefont {Oka}},\ and\
  \bibinfo {author} {\bibfnamefont {H.}~\bibnamefont {Aoki}},\ }\bibfield
  {title} {\bibinfo {title} {{Nonequilibrium steady state of photoexcited
  correlated electrons in the presence of dissipation}},\ }\href
  {https://doi.org/10.1103/PhysRevLett.103.047403} {\bibfield  {journal}
  {\bibinfo  {journal} {Phys. Rev. Lett.}\ }\textbf {\bibinfo {volume} {103}},\
  \bibinfo {pages} {047403} (\bibinfo {year} {2009})}\BibitemShut {NoStop}%
\bibitem [{\citenamefont {Dehghani}\ \emph {et~al.}(2014)\citenamefont
  {Dehghani}, \citenamefont {Oka},\ and\ \citenamefont {Mitra}}]{Dehghani2014}%
  \BibitemOpen
  \bibfield  {author} {\bibinfo {author} {\bibfnamefont {H.}~\bibnamefont
  {Dehghani}}, \bibinfo {author} {\bibfnamefont {T.}~\bibnamefont {Oka}},\ and\
  \bibinfo {author} {\bibfnamefont {A.}~\bibnamefont {Mitra}},\ }\bibfield
  {title} {\bibinfo {title} {{Dissipative Floquet topological systems}},\
  }\href {https://doi.org/10.1103/PhysRevB.90.195429} {\bibfield  {journal}
  {\bibinfo  {journal} {Phys. Rev. B}\ }\textbf {\bibinfo {volume} {90}},\
  \bibinfo {pages} {195429} (\bibinfo {year} {2014})}\BibitemShut {NoStop}%
\bibitem [{\citenamefont {Dehghani}\ \emph {et~al.}(2015)\citenamefont
  {Dehghani}, \citenamefont {Oka},\ and\ \citenamefont {Mitra}}]{Dehghani2015}%
  \BibitemOpen
  \bibfield  {author} {\bibinfo {author} {\bibfnamefont {H.}~\bibnamefont
  {Dehghani}}, \bibinfo {author} {\bibfnamefont {T.}~\bibnamefont {Oka}},\ and\
  \bibinfo {author} {\bibfnamefont {A.}~\bibnamefont {Mitra}},\ }\bibfield
  {title} {\bibinfo {title} {{Out-of-equilibrium electrons and the Hall
  conductance of a Floquet topological insulator}},\ }\href
  {https://doi.org/10.1103/PhysRevB.91.155422} {\bibfield  {journal} {\bibinfo
  {journal} {Phys. Rev. B}\ }\textbf {\bibinfo {volume} {91}},\ \bibinfo
  {pages} {155422} (\bibinfo {year} {2015})}\BibitemShut {NoStop}%
\bibitem [{\citenamefont {Seetharam}\ \emph {et~al.}(2015)\citenamefont
  {Seetharam}, \citenamefont {Bardyn}, \citenamefont {Lindner}, \citenamefont
  {Rudner},\ and\ \citenamefont {Refael}}]{Seetharam2015}%
  \BibitemOpen
  \bibfield  {author} {\bibinfo {author} {\bibfnamefont {K.~I.}\ \bibnamefont
  {Seetharam}}, \bibinfo {author} {\bibfnamefont {C.~E.}\ \bibnamefont
  {Bardyn}}, \bibinfo {author} {\bibfnamefont {N.~H.}\ \bibnamefont {Lindner}},
  \bibinfo {author} {\bibfnamefont {M.~S.}\ \bibnamefont {Rudner}},\ and\
  \bibinfo {author} {\bibfnamefont {G.}~\bibnamefont {Refael}},\ }\bibfield
  {title} {\bibinfo {title} {{Controlled Population of Floquet-Bloch States via
  Coupling to Bose and Fermi Baths}},\ }\href
  {https://doi.org/10.1103/PhysRevX.5.041050} {\bibfield  {journal} {\bibinfo
  {journal} {Phys. Rev. X}\ }\textbf {\bibinfo {volume} {5}},\ \bibinfo {pages}
  {041050} (\bibinfo {year} {2015})}\BibitemShut {NoStop}%
\bibitem [{\citenamefont {Iadecola}\ \emph {et~al.}(2015)\citenamefont
  {Iadecola}, \citenamefont {Neupert},\ and\ \citenamefont
  {Chamon}}]{Iadecola2015}%
  \BibitemOpen
  \bibfield  {author} {\bibinfo {author} {\bibfnamefont {T.}~\bibnamefont
  {Iadecola}}, \bibinfo {author} {\bibfnamefont {T.}~\bibnamefont {Neupert}},\
  and\ \bibinfo {author} {\bibfnamefont {C.}~\bibnamefont {Chamon}},\
  }\bibfield  {title} {\bibinfo {title} {{Occupation of topological Floquet
  bands in open systems}},\ }\href {https://doi.org/10.1103/PhysRevB.91.235133}
  {\bibfield  {journal} {\bibinfo  {journal} {Phys. Rev. B}\ }\textbf {\bibinfo
  {volume} {91}},\ \bibinfo {pages} {235133} (\bibinfo {year}
  {2015})}\BibitemShut {NoStop}%
\bibitem [{\citenamefont {Murakami}\ \emph {et~al.}(2017)\citenamefont
  {Murakami}, \citenamefont {Tsuji}, \citenamefont {Eckstein},\ and\
  \citenamefont {Werner}}]{Murakami2017}%
  \BibitemOpen
  \bibfield  {author} {\bibinfo {author} {\bibfnamefont {Y.}~\bibnamefont
  {Murakami}}, \bibinfo {author} {\bibfnamefont {N.}~\bibnamefont {Tsuji}},
  \bibinfo {author} {\bibfnamefont {M.}~\bibnamefont {Eckstein}},\ and\
  \bibinfo {author} {\bibfnamefont {P.}~\bibnamefont {Werner}},\ }\bibfield
  {title} {\bibinfo {title} {{Nonequilibrium steady states and transient
  dynamics of conventional superconductors under phonon driving}},\ }\href
  {https://doi.org/10.1103/PhysRevB.96.045125} {\bibfield  {journal} {\bibinfo
  {journal} {Phys. Rev. B}\ }\textbf {\bibinfo {volume} {96}},\ \bibinfo
  {pages} {045125} (\bibinfo {year} {2017})}\BibitemShut {NoStop}%
\bibitem [{\citenamefont {McIver}\ \emph {et~al.}(2020)\citenamefont {McIver},
  \citenamefont {Schulte}, \citenamefont {Stein}, \citenamefont {Matsuyama},
  \citenamefont {Jotzu}, \citenamefont {Meier},\ and\ \citenamefont
  {Cavalleri}}]{McIver2020}%
  \BibitemOpen
  \bibfield  {author} {\bibinfo {author} {\bibfnamefont {J.~W.}\ \bibnamefont
  {McIver}}, \bibinfo {author} {\bibfnamefont {B.}~\bibnamefont {Schulte}},
  \bibinfo {author} {\bibfnamefont {F.-U.}\ \bibnamefont {Stein}}, \bibinfo
  {author} {\bibfnamefont {T.}~\bibnamefont {Matsuyama}}, \bibinfo {author}
  {\bibfnamefont {G.}~\bibnamefont {Jotzu}}, \bibinfo {author} {\bibfnamefont
  {G.}~\bibnamefont {Meier}},\ and\ \bibinfo {author} {\bibfnamefont
  {A.}~\bibnamefont {Cavalleri}},\ }\bibfield  {title} {\bibinfo {title}
  {{Light-induced anomalous Hall effect in graphene}},\ }\href
  {https://doi.org/10.1038/s41567-019-0698-y} {\bibfield  {journal} {\bibinfo
  {journal} {Nat. Phys.}\ }\textbf {\bibinfo {volume} {16}},\ \bibinfo {pages}
  {38--41} (\bibinfo {year} {2020})}\BibitemShut {NoStop}%
\bibitem [{\citenamefont {Sato}\ \emph {et~al.}(2019)\citenamefont {Sato},
  \citenamefont {McIver}, \citenamefont {Nuske}, \citenamefont {Tang},
  \citenamefont {Jotzu}, \citenamefont {Schulte}, \citenamefont
  {H{\"{u}}bener}, \citenamefont {{De Giovannini}}, \citenamefont {Mathey},
  \citenamefont {Sentef}, \citenamefont {Cavalleri},\ and\ \citenamefont
  {Rubio}}]{Sato2019}%
  \BibitemOpen
  \bibfield  {author} {\bibinfo {author} {\bibfnamefont {S.~A.}\ \bibnamefont
  {Sato}}, \bibinfo {author} {\bibfnamefont {J.~W.}\ \bibnamefont {McIver}},
  \bibinfo {author} {\bibfnamefont {M.}~\bibnamefont {Nuske}}, \bibinfo
  {author} {\bibfnamefont {P.}~\bibnamefont {Tang}}, \bibinfo {author}
  {\bibfnamefont {G.}~\bibnamefont {Jotzu}}, \bibinfo {author} {\bibfnamefont
  {B.}~\bibnamefont {Schulte}}, \bibinfo {author} {\bibfnamefont
  {H.}~\bibnamefont {H{\"{u}}bener}}, \bibinfo {author} {\bibfnamefont
  {U.}~\bibnamefont {{De Giovannini}}}, \bibinfo {author} {\bibfnamefont
  {L.}~\bibnamefont {Mathey}}, \bibinfo {author} {\bibfnamefont {M.~A.}\
  \bibnamefont {Sentef}}, \bibinfo {author} {\bibfnamefont {A.}~\bibnamefont
  {Cavalleri}},\ and\ \bibinfo {author} {\bibfnamefont {A.}~\bibnamefont
  {Rubio}},\ }\bibfield  {title} {\bibinfo {title} {{Microscopic theory for the
  light-induced anomalous Hall effect in graphene}},\ }\href
  {https://doi.org/10.1103/PhysRevB.99.214302} {\bibfield  {journal} {\bibinfo
  {journal} {Phys. Rev. B}\ }\textbf {\bibinfo {volume} {99}},\ \bibinfo
  {pages} {214302} (\bibinfo {year} {2019})}\BibitemShut {NoStop}%
\bibitem [{\citenamefont {Drummond}\ and\ \citenamefont
  {Walls}(1980)}]{Drummond1980}%
  \BibitemOpen
  \bibfield  {author} {\bibinfo {author} {\bibfnamefont {P.~D.}\ \bibnamefont
  {Drummond}}\ and\ \bibinfo {author} {\bibfnamefont {D.~F.}\ \bibnamefont
  {Walls}},\ }\bibfield  {title} {\bibinfo {title} {{Quantum theory of optical
  bistability. I. Nonlinear polarisability model}},\ }\href
  {https://doi.org/10.1088/0305-4470/13/2/034} {\bibfield  {journal} {\bibinfo
  {journal} {J. Phys. A. Math. Gen.}\ }\textbf {\bibinfo {volume} {13}},\
  \bibinfo {pages} {725} (\bibinfo {year} {1980})}\BibitemShut {NoStop}%
\bibitem [{\citenamefont {Baumann}\ \emph {et~al.}(2010)\citenamefont
  {Baumann}, \citenamefont {Guerlin}, \citenamefont {Brennecke},\ and\
  \citenamefont {Esslinger}}]{Baumann2010}%
  \BibitemOpen
  \bibfield  {author} {\bibinfo {author} {\bibfnamefont {K.}~\bibnamefont
  {Baumann}}, \bibinfo {author} {\bibfnamefont {C.}~\bibnamefont {Guerlin}},
  \bibinfo {author} {\bibfnamefont {F.}~\bibnamefont {Brennecke}},\ and\
  \bibinfo {author} {\bibfnamefont {T.}~\bibnamefont {Esslinger}},\ }\bibfield
  {title} {\bibinfo {title} {{Dicke quantum phase transition with a superfluid
  gas in an optical cavity}},\ }\href {https://doi.org/10.1038/nature09009}
  {\bibfield  {journal} {\bibinfo  {journal} {Nature}\ }\textbf {\bibinfo
  {volume} {464}},\ \bibinfo {pages} {1301--1306} (\bibinfo {year}
  {2010})}\BibitemShut {NoStop}%
\bibitem [{\citenamefont {Torre}\ \emph {et~al.}(2013)\citenamefont {Torre},
  \citenamefont {Diehl}, \citenamefont {Lukin}, \citenamefont {Sachdev},\ and\
  \citenamefont {Strack}}]{DallaTorre2013}%
  \BibitemOpen
  \bibfield  {author} {\bibinfo {author} {\bibfnamefont {E.~G.}\ \bibnamefont
  {Torre}}, \bibinfo {author} {\bibfnamefont {S.}~\bibnamefont {Diehl}},
  \bibinfo {author} {\bibfnamefont {M.~D.}\ \bibnamefont {Lukin}}, \bibinfo
  {author} {\bibfnamefont {S.}~\bibnamefont {Sachdev}},\ and\ \bibinfo {author}
  {\bibfnamefont {P.}~\bibnamefont {Strack}},\ }\bibfield  {title} {\bibinfo
  {title} {{Keldysh approach for nonequilibrium phase transitions in quantum
  optics: Beyond the Dicke model in optical cavities}},\ }\href
  {https://doi.org/10.1103/PhysRevA.87.023831} {\bibfield  {journal} {\bibinfo
  {journal} {Phys. Rev. A}\ }\textbf {\bibinfo {volume} {87}},\ \bibinfo
  {pages} {023831} (\bibinfo {year} {2013})}\BibitemShut {NoStop}%
\bibitem [{\citenamefont {Shirai}\ \emph {et~al.}(2014)\citenamefont {Shirai},
  \citenamefont {Mori},\ and\ \citenamefont {Miyashita}}]{Shirai2014}%
  \BibitemOpen
  \bibfield  {author} {\bibinfo {author} {\bibfnamefont {T.}~\bibnamefont
  {Shirai}}, \bibinfo {author} {\bibfnamefont {T.}~\bibnamefont {Mori}},\ and\
  \bibinfo {author} {\bibfnamefont {S.}~\bibnamefont {Miyashita}},\ }\bibfield
  {title} {\bibinfo {title} {{Novel symmetry-broken phase in a driven cavity
  system in the thermodynamic limit}},\ }\href
  {https://doi.org/10.1088/0953-4075/47/2/025501} {\bibfield  {journal}
  {\bibinfo  {journal} {J. Phys. B At. Mol. Opt. Phys.}\ }\textbf {\bibinfo
  {volume} {47}},\ \bibinfo {pages} {025501} (\bibinfo {year}
  {2014})}\BibitemShut {NoStop}%
\bibitem [{\citenamefont {Foss-Feig}\ \emph {et~al.}(2017)\citenamefont
  {Foss-Feig}, \citenamefont {Niroula}, \citenamefont {Young}, \citenamefont
  {Hafezi}, \citenamefont {Gorshkov}, \citenamefont {Wilson},\ and\
  \citenamefont {Maghrebi}}]{Foss-Feig2017}%
  \BibitemOpen
  \bibfield  {author} {\bibinfo {author} {\bibfnamefont {M.}~\bibnamefont
  {Foss-Feig}}, \bibinfo {author} {\bibfnamefont {P.}~\bibnamefont {Niroula}},
  \bibinfo {author} {\bibfnamefont {J.~T.}\ \bibnamefont {Young}}, \bibinfo
  {author} {\bibfnamefont {M.}~\bibnamefont {Hafezi}}, \bibinfo {author}
  {\bibfnamefont {A.~V.}\ \bibnamefont {Gorshkov}}, \bibinfo {author}
  {\bibfnamefont {R.~M.}\ \bibnamefont {Wilson}},\ and\ \bibinfo {author}
  {\bibfnamefont {M.~F.}\ \bibnamefont {Maghrebi}},\ }\bibfield  {title}
  {\bibinfo {title} {{Emergent equilibrium in many-body optical bistability}},\
  }\href {https://doi.org/10.1103/PhysRevA.95.043826} {\bibfield  {journal}
  {\bibinfo  {journal} {Phys. Rev. A}\ }\textbf {\bibinfo {volume} {95}},\
  \bibinfo {pages} {043826} (\bibinfo {year} {2017})}\BibitemShut {NoStop}%
\bibitem [{\citenamefont {Diehl}\ \emph {et~al.}(2008)\citenamefont {Diehl},
  \citenamefont {Micheli}, \citenamefont {Kantian}, \citenamefont {Kraus},
  \citenamefont {B{\"{u}}chler},\ and\ \citenamefont {Zoller}}]{Diehl2008}%
  \BibitemOpen
  \bibfield  {author} {\bibinfo {author} {\bibfnamefont {S.}~\bibnamefont
  {Diehl}}, \bibinfo {author} {\bibfnamefont {A.}~\bibnamefont {Micheli}},
  \bibinfo {author} {\bibfnamefont {A.}~\bibnamefont {Kantian}}, \bibinfo
  {author} {\bibfnamefont {B.}~\bibnamefont {Kraus}}, \bibinfo {author}
  {\bibfnamefont {H.~P.}\ \bibnamefont {B{\"{u}}chler}},\ and\ \bibinfo
  {author} {\bibfnamefont {P.}~\bibnamefont {Zoller}},\ }\bibfield  {title}
  {\bibinfo {title} {{Quantum states and phases in driven open quantum systems
  with cold atoms}},\ }\href {https://doi.org/10.1038/nphys1073} {\bibfield
  {journal} {\bibinfo  {journal} {Nat. Phys.}\ }\textbf {\bibinfo {volume}
  {4}},\ \bibinfo {pages} {878--883} (\bibinfo {year} {2008})}\BibitemShut
  {NoStop}%
\bibitem [{\citenamefont {Diehl}\ \emph {et~al.}(2011)\citenamefont {Diehl},
  \citenamefont {Rico}, \citenamefont {Baranov},\ and\ \citenamefont
  {Zoller}}]{Diehl2011}%
  \BibitemOpen
  \bibfield  {author} {\bibinfo {author} {\bibfnamefont {S.}~\bibnamefont
  {Diehl}}, \bibinfo {author} {\bibfnamefont {E.}~\bibnamefont {Rico}},
  \bibinfo {author} {\bibfnamefont {M.~A.}\ \bibnamefont {Baranov}},\ and\
  \bibinfo {author} {\bibfnamefont {P.}~\bibnamefont {Zoller}},\ }\bibfield
  {title} {\bibinfo {title} {{Topology by dissipation in atomic quantum
  wires}},\ }\href {https://doi.org/10.1038/NPHYS2106} {\bibfield  {journal}
  {\bibinfo  {journal} {Nat. Phys.}\ }\textbf {\bibinfo {volume} {7}},\
  \bibinfo {pages} {971--977} (\bibinfo {year} {2011})}\BibitemShut {NoStop}%
\bibitem [{\citenamefont {Vorberg}\ \emph {et~al.}(2013)\citenamefont
  {Vorberg}, \citenamefont {Wustmann}, \citenamefont {Ketzmerick},\ and\
  \citenamefont {Eckardt}}]{Vorberg2013}%
  \BibitemOpen
  \bibfield  {author} {\bibinfo {author} {\bibfnamefont {D.}~\bibnamefont
  {Vorberg}}, \bibinfo {author} {\bibfnamefont {W.}~\bibnamefont {Wustmann}},
  \bibinfo {author} {\bibfnamefont {R.}~\bibnamefont {Ketzmerick}},\ and\
  \bibinfo {author} {\bibfnamefont {A.}~\bibnamefont {Eckardt}},\ }\bibfield
  {title} {\bibinfo {title} {{Generalized bose-einstein condensation into
  multiple states in driven-dissipative systems}},\ }\href
  {https://doi.org/10.1103/PhysRevLett.111.240405} {\bibfield  {journal}
  {\bibinfo  {journal} {Phys. Rev. Lett.}\ }\textbf {\bibinfo {volume} {111}},\
  \bibinfo {pages} {240405} (\bibinfo {year} {2013})}\BibitemShut {NoStop}%
\bibitem [{\citenamefont {Vorberg}\ \emph {et~al.}(2015)\citenamefont
  {Vorberg}, \citenamefont {Wustmann}, \citenamefont {Schomerus}, \citenamefont
  {Ketzmerick},\ and\ \citenamefont {Eckardt}}]{Vorberg2015}%
  \BibitemOpen
  \bibfield  {author} {\bibinfo {author} {\bibfnamefont {D.}~\bibnamefont
  {Vorberg}}, \bibinfo {author} {\bibfnamefont {W.}~\bibnamefont {Wustmann}},
  \bibinfo {author} {\bibfnamefont {H.}~\bibnamefont {Schomerus}}, \bibinfo
  {author} {\bibfnamefont {R.}~\bibnamefont {Ketzmerick}},\ and\ \bibinfo
  {author} {\bibfnamefont {A.}~\bibnamefont {Eckardt}},\ }\bibfield  {title}
  {\bibinfo {title} {{Nonequilibrium steady states of ideal bosonic and
  fermionic quantum gases}},\ }\href
  {https://doi.org/10.1103/PhysRevE.92.062119} {\bibfield  {journal} {\bibinfo
  {journal} {Phys. Rev. E}\ }\textbf {\bibinfo {volume} {92}},\ \bibinfo
  {pages} {062119} (\bibinfo {year} {2015})}\BibitemShut {NoStop}%
\bibitem [{\citenamefont {Schnell}\ \emph {et~al.}(2018)\citenamefont
  {Schnell}, \citenamefont {Ketzmerick},\ and\ \citenamefont
  {Eckardt}}]{Schnell2018}%
  \BibitemOpen
  \bibfield  {author} {\bibinfo {author} {\bibfnamefont {A.}~\bibnamefont
  {Schnell}}, \bibinfo {author} {\bibfnamefont {R.}~\bibnamefont
  {Ketzmerick}},\ and\ \bibinfo {author} {\bibfnamefont {A.}~\bibnamefont
  {Eckardt}},\ }\bibfield  {title} {\bibinfo {title} {{On the number of
  Bose-selected modes in driven-dissipative ideal Bose gases}},\ }\href
  {https://doi.org/10.1103/PhysRevE.97.032136} {\bibfield  {journal} {\bibinfo
  {journal} {Phys. Rev. E}\ }\textbf {\bibinfo {volume} {97}},\ \bibinfo
  {pages} {032136} (\bibinfo {year} {2018})}\BibitemShut {NoStop}%
\bibitem [{\citenamefont {Barreiro}\ \emph {et~al.}(2011)\citenamefont
  {Barreiro}, \citenamefont {M{\"{u}}ller}, \citenamefont {Schindler},
  \citenamefont {Nigg}, \citenamefont {Monz}, \citenamefont {Chwalla},
  \citenamefont {Hennrich}, \citenamefont {Roos}, \citenamefont {Zoller},\ and\
  \citenamefont {Blatt}}]{Barreiro2011}%
  \BibitemOpen
  \bibfield  {author} {\bibinfo {author} {\bibfnamefont {J.~T.}\ \bibnamefont
  {Barreiro}}, \bibinfo {author} {\bibfnamefont {M.}~\bibnamefont
  {M{\"{u}}ller}}, \bibinfo {author} {\bibfnamefont {P.}~\bibnamefont
  {Schindler}}, \bibinfo {author} {\bibfnamefont {D.}~\bibnamefont {Nigg}},
  \bibinfo {author} {\bibfnamefont {T.}~\bibnamefont {Monz}}, \bibinfo {author}
  {\bibfnamefont {M.}~\bibnamefont {Chwalla}}, \bibinfo {author} {\bibfnamefont
  {M.}~\bibnamefont {Hennrich}}, \bibinfo {author} {\bibfnamefont {C.~F.}\
  \bibnamefont {Roos}}, \bibinfo {author} {\bibfnamefont {P.}~\bibnamefont
  {Zoller}},\ and\ \bibinfo {author} {\bibfnamefont {R.}~\bibnamefont
  {Blatt}},\ }\bibfield  {title} {\bibinfo {title} {{An open-system quantum
  simulator with trapped ions}},\ }\href {https://doi.org/10.1038/nature09801}
  {\bibfield  {journal} {\bibinfo  {journal} {Nature}\ }\textbf {\bibinfo
  {volume} {470}},\ \bibinfo {pages} {486--491} (\bibinfo {year}
  {2011})}\BibitemShut {NoStop}%
\bibitem [{\citenamefont {Barontini}\ \emph {et~al.}(2013)\citenamefont
  {Barontini}, \citenamefont {Labouvie}, \citenamefont {Stubenrauch},
  \citenamefont {Vogler}, \citenamefont {Guarrera},\ and\ \citenamefont
  {Ott}}]{Barontini2013}%
  \BibitemOpen
  \bibfield  {author} {\bibinfo {author} {\bibfnamefont {G.}~\bibnamefont
  {Barontini}}, \bibinfo {author} {\bibfnamefont {R.}~\bibnamefont {Labouvie}},
  \bibinfo {author} {\bibfnamefont {F.}~\bibnamefont {Stubenrauch}}, \bibinfo
  {author} {\bibfnamefont {A.}~\bibnamefont {Vogler}}, \bibinfo {author}
  {\bibfnamefont {V.}~\bibnamefont {Guarrera}},\ and\ \bibinfo {author}
  {\bibfnamefont {H.}~\bibnamefont {Ott}},\ }\bibfield  {title} {\bibinfo
  {title} {{Controlling the dynamics of an open many-body quantum system with
  localized dissipation}},\ }\href
  {https://doi.org/10.1103/PhysRevLett.110.035302} {\bibfield  {journal}
  {\bibinfo  {journal} {Phys. Rev. Lett.}\ }\textbf {\bibinfo {volume} {110}},\
  \bibinfo {pages} {035302} (\bibinfo {year} {2013})}\BibitemShut {NoStop}%
\bibitem [{\citenamefont {Tomita}\ \emph {et~al.}(2017)\citenamefont {Tomita},
  \citenamefont {Nakajima}, \citenamefont {Danshita}, \citenamefont {Takasu},\
  and\ \citenamefont {Takahashi}}]{Tomita2017}%
  \BibitemOpen
  \bibfield  {author} {\bibinfo {author} {\bibfnamefont {T.}~\bibnamefont
  {Tomita}}, \bibinfo {author} {\bibfnamefont {S.}~\bibnamefont {Nakajima}},
  \bibinfo {author} {\bibfnamefont {I.}~\bibnamefont {Danshita}}, \bibinfo
  {author} {\bibfnamefont {Y.}~\bibnamefont {Takasu}},\ and\ \bibinfo {author}
  {\bibfnamefont {Y.}~\bibnamefont {Takahashi}},\ }\bibfield  {title} {\bibinfo
  {title} {{Observation of the Mott insulator to superfluid crossover of a
  driven-dissipative Bose-Hubbard system}},\ }\href
  {https://doi.org/10.1126/sciadv.1701513} {\bibfield  {journal} {\bibinfo
  {journal} {Sci. Adv.}\ }\textbf {\bibinfo {volume} {3}},\ \bibinfo {pages}
  {e1701513} (\bibinfo {year} {2017})}\BibitemShut {NoStop}%
\bibitem [{\citenamefont {Breuer}\ and\ \citenamefont
  {Holthaus}(1991)}]{Breuer1991}%
  \BibitemOpen
  \bibfield  {author} {\bibinfo {author} {\bibfnamefont {H.~P.}\ \bibnamefont
  {Breuer}}\ and\ \bibinfo {author} {\bibfnamefont {M.}~\bibnamefont
  {Holthaus}},\ }\bibfield  {title} {\bibinfo {title} {{A semiclassical theory
  of quasienergies and Floquet wave functions}},\ }\href
  {https://doi.org/10.1016/0003-4916(91)90206-N} {\bibfield  {journal}
  {\bibinfo  {journal} {Ann. Phys. (N. Y).}\ }\textbf {\bibinfo {volume}
  {211}},\ \bibinfo {pages} {249--291} (\bibinfo {year} {1991})}\BibitemShut
  {NoStop}%
\bibitem [{\citenamefont {Breuer}\ \emph {et~al.}(2000)\citenamefont {Breuer},
  \citenamefont {Huber},\ and\ \citenamefont {Petruccione}}]{Breuer2000}%
  \BibitemOpen
  \bibfield  {author} {\bibinfo {author} {\bibfnamefont {H.~P.}\ \bibnamefont
  {Breuer}}, \bibinfo {author} {\bibfnamefont {W.}~\bibnamefont {Huber}},\ and\
  \bibinfo {author} {\bibfnamefont {F.}~\bibnamefont {Petruccione}},\
  }\bibfield  {title} {\bibinfo {title} {{Quasistationary distributions of
  dissipative nonlinear quantum oscillators in strong periodic driving
  fields}},\ }\href {https://doi.org/10.1103/PhysRevE.61.4883} {\bibfield
  {journal} {\bibinfo  {journal} {Phys. Rev. E - Stat. Physics, Plasmas,
  Fluids, Relat. Interdiscip. Top.}\ }\textbf {\bibinfo {volume} {61}},\
  \bibinfo {pages} {4883} (\bibinfo {year} {2000})}\BibitemShut {NoStop}%
\bibitem [{\citenamefont {Kohn}(2001)}]{Kohn2001}%
  \BibitemOpen
  \bibfield  {author} {\bibinfo {author} {\bibfnamefont {W.}~\bibnamefont
  {Kohn}},\ }\bibfield  {title} {\bibinfo {title} {{Periodic thermodynamics}},\
  }\href {https://doi.org/10.1023/A:1010327828445} {\bibfield  {journal}
  {\bibinfo  {journal} {J. Stat. Phys.}\ }\textbf {\bibinfo {volume} {103}},\
  \bibinfo {pages} {417--423} (\bibinfo {year} {2001})}\BibitemShut {NoStop}%
\bibitem [{\citenamefont {Hone}\ \emph {et~al.}(2009)\citenamefont {Hone},
  \citenamefont {Ketzmerick},\ and\ \citenamefont {Kohn}}]{Hone2009}%
  \BibitemOpen
  \bibfield  {author} {\bibinfo {author} {\bibfnamefont {D.~W.}\ \bibnamefont
  {Hone}}, \bibinfo {author} {\bibfnamefont {R.}~\bibnamefont {Ketzmerick}},\
  and\ \bibinfo {author} {\bibfnamefont {W.}~\bibnamefont {Kohn}},\ }\bibfield
  {title} {\bibinfo {title} {{Statistical mechanics of Floquet systems: The
  pervasive problem of near degeneracies}},\ }\href
  {https://doi.org/10.1103/PhysRevE.79.051129} {\bibfield  {journal} {\bibinfo
  {journal} {Phys. Rev. E}\ }\textbf {\bibinfo {volume} {79}},\ \bibinfo
  {pages} {051129} (\bibinfo {year} {2009})}\BibitemShut {NoStop}%
\bibitem [{\citenamefont {Ketzmerick}\ and\ \citenamefont
  {Wustmann}(2010)}]{Ketzmerick2010}%
  \BibitemOpen
  \bibfield  {author} {\bibinfo {author} {\bibfnamefont {R.}~\bibnamefont
  {Ketzmerick}}\ and\ \bibinfo {author} {\bibfnamefont {W.}~\bibnamefont
  {Wustmann}},\ }\bibfield  {title} {\bibinfo {title} {{Statistical mechanics
  of Floquet systems with regular and chaotic states}},\ }\href
  {https://doi.org/10.1103/PhysRevE.82.021114} {\bibfield  {journal} {\bibinfo
  {journal} {Phys. Rev. E}\ }\textbf {\bibinfo {volume} {82}},\ \bibinfo
  {pages} {021114} (\bibinfo {year} {2010})}\BibitemShut {NoStop}%
\bibitem [{\citenamefont {Liu}(2015)}]{Liu2015}%
  \BibitemOpen
  \bibfield  {author} {\bibinfo {author} {\bibfnamefont {D.~E.}\ \bibnamefont
  {Liu}},\ }\bibfield  {title} {\bibinfo {title} {{Classification of the
  Floquet statistical distribution for time-periodic open systems}},\ }\href
  {https://doi.org/10.1103/PhysRevB.91.144301} {\bibfield  {journal} {\bibinfo
  {journal} {Phys. Rev. B}\ }\textbf {\bibinfo {volume} {91}},\ \bibinfo
  {pages} {144301} (\bibinfo {year} {2015})}\BibitemShut {NoStop}%
\bibitem [{\citenamefont {Shirai}\ \emph {et~al.}(2015)\citenamefont {Shirai},
  \citenamefont {Mori},\ and\ \citenamefont {Miyashita}}]{Shirai2015}%
  \BibitemOpen
  \bibfield  {author} {\bibinfo {author} {\bibfnamefont {T.}~\bibnamefont
  {Shirai}}, \bibinfo {author} {\bibfnamefont {T.}~\bibnamefont {Mori}},\ and\
  \bibinfo {author} {\bibfnamefont {S.}~\bibnamefont {Miyashita}},\ }\bibfield
  {title} {\bibinfo {title} {{Condition for emergence of the Floquet-Gibbs
  state in periodically driven open systems}},\ }\href
  {https://doi.org/10.1103/PhysRevE.91.030101} {\bibfield  {journal} {\bibinfo
  {journal} {Phys. Rev. E}\ }\textbf {\bibinfo {volume} {91}},\ \bibinfo
  {pages} {030101} (\bibinfo {year} {2015})}\BibitemShut {NoStop}%
\bibitem [{\citenamefont {Shirai}\ \emph {et~al.}(2016)\citenamefont {Shirai},
  \citenamefont {Thingna}, \citenamefont {Mori}, \citenamefont {Denisov},
  \citenamefont {H{\"{a}}nggi},\ and\ \citenamefont {Miyashita}}]{Shirai2016}%
  \BibitemOpen
  \bibfield  {author} {\bibinfo {author} {\bibfnamefont {T.}~\bibnamefont
  {Shirai}}, \bibinfo {author} {\bibfnamefont {J.}~\bibnamefont {Thingna}},
  \bibinfo {author} {\bibfnamefont {T.}~\bibnamefont {Mori}}, \bibinfo {author}
  {\bibfnamefont {S.}~\bibnamefont {Denisov}}, \bibinfo {author} {\bibfnamefont
  {P.}~\bibnamefont {H{\"{a}}nggi}},\ and\ \bibinfo {author} {\bibfnamefont
  {S.}~\bibnamefont {Miyashita}},\ }\bibfield  {title} {\bibinfo {title}
  {{Effective Floquet-Gibbs states for dissipative quantum systems}},\ }\href
  {https://doi.org/10.1088/1367-2630/18/5/053008} {\bibfield  {journal}
  {\bibinfo  {journal} {New J. Phys.}\ }\textbf {\bibinfo {volume} {18}},\
  \bibinfo {pages} {1--13} (\bibinfo {year} {2016})}\BibitemShut {NoStop}%
\bibitem [{\citenamefont {Mikami}\ \emph {et~al.}(2016)\citenamefont {Mikami},
  \citenamefont {Kitamura}, \citenamefont {Yasuda}, \citenamefont {Tsuji},
  \citenamefont {Oka},\ and\ \citenamefont {Aoki}}]{Mikami2016}%
  \BibitemOpen
  \bibfield  {author} {\bibinfo {author} {\bibfnamefont {T.}~\bibnamefont
  {Mikami}}, \bibinfo {author} {\bibfnamefont {S.}~\bibnamefont {Kitamura}},
  \bibinfo {author} {\bibfnamefont {K.}~\bibnamefont {Yasuda}}, \bibinfo
  {author} {\bibfnamefont {N.}~\bibnamefont {Tsuji}}, \bibinfo {author}
  {\bibfnamefont {T.}~\bibnamefont {Oka}},\ and\ \bibinfo {author}
  {\bibfnamefont {H.}~\bibnamefont {Aoki}},\ }\bibfield  {title} {\bibinfo
  {title} {{Brillouin-Wigner theory for high-frequency expansion in
  periodically driven systems: Application to Floquet topological
  insulators}},\ }\href {https://doi.org/10.1103/PhysRevB.93.144307} {\bibfield
   {journal} {\bibinfo  {journal} {Phys. Rev. B}\ }\textbf {\bibinfo {volume}
  {93}},\ \bibinfo {pages} {144307} (\bibinfo {year} {2016})}\BibitemShut
  {NoStop}%
\bibitem [{\citenamefont {Blanes}\ \emph {et~al.}(2009)\citenamefont {Blanes},
  \citenamefont {Casas}, \citenamefont {Oteo},\ and\ \citenamefont
  {Ros}}]{Blanes2009}%
  \BibitemOpen
  \bibfield  {author} {\bibinfo {author} {\bibfnamefont {S.}~\bibnamefont
  {Blanes}}, \bibinfo {author} {\bibfnamefont {F.}~\bibnamefont {Casas}},
  \bibinfo {author} {\bibfnamefont {J.~A.}\ \bibnamefont {Oteo}},\ and\
  \bibinfo {author} {\bibfnamefont {J.}~\bibnamefont {Ros}},\ }\bibfield
  {title} {\bibinfo {title} {{The Magnus expansion and some of its
  applications}},\ }\href {https://doi.org/10.1016/j.physrep.2008.11.001}
  {\bibfield  {journal} {\bibinfo  {journal} {Phys. Rep.}\ }\textbf {\bibinfo
  {volume} {470}},\ \bibinfo {pages} {151--238} (\bibinfo {year}
  {2009})}\BibitemShut {NoStop}%
\bibitem [{\citenamefont {Rahav}\ \emph {et~al.}(2003)\citenamefont {Rahav},
  \citenamefont {Gilary},\ and\ \citenamefont {Fishman}}]{Rahav2003}%
  \BibitemOpen
  \bibfield  {author} {\bibinfo {author} {\bibfnamefont {S.}~\bibnamefont
  {Rahav}}, \bibinfo {author} {\bibfnamefont {I.}~\bibnamefont {Gilary}},\ and\
  \bibinfo {author} {\bibfnamefont {S.}~\bibnamefont {Fishman}},\ }\bibfield
  {title} {\bibinfo {title} {{Effective Hamiltonians for periodically driven
  systems}},\ }\href {https://doi.org/10.1103/PhysRevA.68.013820} {\bibfield
  {journal} {\bibinfo  {journal} {Phys. Rev. A}\ }\textbf {\bibinfo {volume}
  {68}},\ \bibinfo {pages} {013820} (\bibinfo {year} {2003})}\BibitemShut
  {NoStop}%
\bibitem [{\citenamefont {Goldman}\ and\ \citenamefont
  {Dalibard}(2014)}]{Goldman2014}%
  \BibitemOpen
  \bibfield  {author} {\bibinfo {author} {\bibfnamefont {N.}~\bibnamefont
  {Goldman}}\ and\ \bibinfo {author} {\bibfnamefont {J.}~\bibnamefont
  {Dalibard}},\ }\bibfield  {title} {\bibinfo {title} {{Periodically driven
  quantum systems: Effective Hamiltonians and engineered gauge fields}},\
  }\href {https://doi.org/10.1103/PhysRevX.4.031027} {\bibfield  {journal}
  {\bibinfo  {journal} {Phys. Rev. X}\ }\textbf {\bibinfo {volume} {4}},\
  \bibinfo {pages} {031027} (\bibinfo {year} {2014})}\BibitemShut {NoStop}%
\bibitem [{\citenamefont {Eckardt}\ and\ \citenamefont
  {Anisimovas}(2015)}]{Eckardt2015}%
  \BibitemOpen
  \bibfield  {author} {\bibinfo {author} {\bibfnamefont {A.}~\bibnamefont
  {Eckardt}}\ and\ \bibinfo {author} {\bibfnamefont {E.}~\bibnamefont
  {Anisimovas}},\ }\bibfield  {title} {\bibinfo {title} {{High-frequency
  approximation for periodically driven quantum systems from a Floquet-space
  perspective}},\ }\href {https://doi.org/10.1088/1367-2630/17/9/093039}
  {\bibfield  {journal} {\bibinfo  {journal} {New J. Phys.}\ }\textbf {\bibinfo
  {volume} {17}},\ \bibinfo {pages} {093039} (\bibinfo {year}
  {2015})}\BibitemShut {NoStop}%
\bibitem [{\citenamefont {Dunlap}\ and\ \citenamefont
  {Kenkre}(1986)}]{Dunlap1986}%
  \BibitemOpen
  \bibfield  {author} {\bibinfo {author} {\bibfnamefont {D.~H.}\ \bibnamefont
  {Dunlap}}\ and\ \bibinfo {author} {\bibfnamefont {V.~M.}\ \bibnamefont
  {Kenkre}},\ }\bibfield  {title} {\bibinfo {title} {{Dynamic localization of a
  charged particle moving under the influence of an electric field}},\ }\href
  {https://doi.org/10.1103/PhysRevB.34.3625} {\bibfield  {journal} {\bibinfo
  {journal} {Phys. Rev. B}\ }\textbf {\bibinfo {volume} {34}},\ \bibinfo
  {pages} {3625} (\bibinfo {year} {1986})}\BibitemShut {NoStop}%
\bibitem [{\citenamefont {Grossmann}\ \emph {et~al.}(1991)\citenamefont
  {Grossmann}, \citenamefont {Dittrich}, \citenamefont {Jung},\ and\
  \citenamefont {H{\"{a}}nggi}}]{Grossmann1991}%
  \BibitemOpen
  \bibfield  {author} {\bibinfo {author} {\bibfnamefont {F.}~\bibnamefont
  {Grossmann}}, \bibinfo {author} {\bibfnamefont {T.}~\bibnamefont {Dittrich}},
  \bibinfo {author} {\bibfnamefont {P.}~\bibnamefont {Jung}},\ and\ \bibinfo
  {author} {\bibfnamefont {P.}~\bibnamefont {H{\"{a}}nggi}},\ }\bibfield
  {title} {\bibinfo {title} {{Coherent destruction of tunneling}},\ }\href
  {https://doi.org/10.1103/PhysRevLett.67.516} {\bibfield  {journal} {\bibinfo
  {journal} {Phys. Rev. Lett.}\ }\textbf {\bibinfo {volume} {67}},\ \bibinfo
  {pages} {516} (\bibinfo {year} {1991})}\BibitemShut {NoStop}%
\bibitem [{\citenamefont {Lignier}\ \emph {et~al.}(2007)\citenamefont
  {Lignier}, \citenamefont {Sias}, \citenamefont {Ciampini}, \citenamefont
  {Singh}, \citenamefont {Zenesini}, \citenamefont {Morsch},\ and\
  \citenamefont {Arimondo}}]{Lignier2007}%
  \BibitemOpen
  \bibfield  {author} {\bibinfo {author} {\bibfnamefont {H.}~\bibnamefont
  {Lignier}}, \bibinfo {author} {\bibfnamefont {C.}~\bibnamefont {Sias}},
  \bibinfo {author} {\bibfnamefont {D.}~\bibnamefont {Ciampini}}, \bibinfo
  {author} {\bibfnamefont {Y.}~\bibnamefont {Singh}}, \bibinfo {author}
  {\bibfnamefont {A.}~\bibnamefont {Zenesini}}, \bibinfo {author}
  {\bibfnamefont {O.}~\bibnamefont {Morsch}},\ and\ \bibinfo {author}
  {\bibfnamefont {E.}~\bibnamefont {Arimondo}},\ }\bibfield  {title} {\bibinfo
  {title} {{Dynamical control of matter-wave tunneling in periodic
  potentials}},\ }\href {https://doi.org/10.1103/PhysRevLett.99.220403}
  {\bibfield  {journal} {\bibinfo  {journal} {Phys. Rev. Lett.}\ }\textbf
  {\bibinfo {volume} {99}},\ \bibinfo {pages} {220403} (\bibinfo {year}
  {2007})}\BibitemShut {NoStop}%
\bibitem [{\citenamefont {Eckardt}\ and\ \citenamefont
  {Holthaus}(2007)}]{Eckardt2007}%
  \BibitemOpen
  \bibfield  {author} {\bibinfo {author} {\bibfnamefont {A.}~\bibnamefont
  {Eckardt}}\ and\ \bibinfo {author} {\bibfnamefont {M.}~\bibnamefont
  {Holthaus}},\ }\bibfield  {title} {\bibinfo {title} {{AC-induced
  superfluidity}},\ }\href {https://doi.org/10.1209/0295-5075/80/50004}
  {\bibfield  {journal} {\bibinfo  {journal} {Europhys. Lett.}\ }\textbf
  {\bibinfo {volume} {80}},\ \bibinfo {pages} {50004} (\bibinfo {year}
  {2007})}\BibitemShut {NoStop}%
\bibitem [{\citenamefont {Goldman}\ \emph {et~al.}(2015)\citenamefont
  {Goldman}, \citenamefont {Dalibard}, \citenamefont {Aidelsburger},\ and\
  \citenamefont {Cooper}}]{Goldman2015}%
  \BibitemOpen
  \bibfield  {author} {\bibinfo {author} {\bibfnamefont {N.}~\bibnamefont
  {Goldman}}, \bibinfo {author} {\bibfnamefont {J.}~\bibnamefont {Dalibard}},
  \bibinfo {author} {\bibfnamefont {M.}~\bibnamefont {Aidelsburger}},\ and\
  \bibinfo {author} {\bibfnamefont {N.~R.}\ \bibnamefont {Cooper}},\ }\bibfield
   {title} {\bibinfo {title} {{Periodically driven quantum matter: The case of
  resonant modulations}},\ }\href {https://doi.org/10.1103/PhysRevA.91.033632}
  {\bibfield  {journal} {\bibinfo  {journal} {Phys. Rev. A}\ }\textbf {\bibinfo
  {volume} {91}},\ \bibinfo {pages} {033632} (\bibinfo {year}
  {2015})}\BibitemShut {NoStop}%
\bibitem [{\citenamefont {D'Alessio}\ and\ \citenamefont
  {Rigol}(2014)}]{DAlessio2014}%
  \BibitemOpen
  \bibfield  {author} {\bibinfo {author} {\bibfnamefont {L.}~\bibnamefont
  {D'Alessio}}\ and\ \bibinfo {author} {\bibfnamefont {M.}~\bibnamefont
  {Rigol}},\ }\bibfield  {title} {\bibinfo {title} {{Long-time behavior of
  isolated periodically driven interacting lattice systems}},\ }\href
  {https://doi.org/10.1103/PhysRevX.4.041048} {\bibfield  {journal} {\bibinfo
  {journal} {Phys. Rev. X}\ }\textbf {\bibinfo {volume} {4}},\ \bibinfo {pages}
  {041048} (\bibinfo {year} {2014})}\BibitemShut {NoStop}%
\bibitem [{\citenamefont {Lazarides}\ \emph {et~al.}(2014)\citenamefont
  {Lazarides}, \citenamefont {Das},\ and\ \citenamefont
  {Moessner}}]{Lazarides2014}%
  \BibitemOpen
  \bibfield  {author} {\bibinfo {author} {\bibfnamefont {A.}~\bibnamefont
  {Lazarides}}, \bibinfo {author} {\bibfnamefont {A.}~\bibnamefont {Das}},\
  and\ \bibinfo {author} {\bibfnamefont {R.}~\bibnamefont {Moessner}},\
  }\bibfield  {title} {\bibinfo {title} {{Equilibrium states of generic quantum
  systems subject to periodic driving}},\ }\href
  {https://doi.org/10.1103/PhysRevE.90.012110} {\bibfield  {journal} {\bibinfo
  {journal} {Phys. Rev. E}\ }\textbf {\bibinfo {volume} {90}},\ \bibinfo
  {pages} {012110} (\bibinfo {year} {2014})}\BibitemShut {NoStop}%
\bibitem [{\citenamefont {Kim}\ \emph {et~al.}(2014)\citenamefont {Kim},
  \citenamefont {Ikeda},\ and\ \citenamefont {Huse}}]{Kim2014}%
  \BibitemOpen
  \bibfield  {author} {\bibinfo {author} {\bibfnamefont {H.}~\bibnamefont
  {Kim}}, \bibinfo {author} {\bibfnamefont {T.~N.}\ \bibnamefont {Ikeda}},\
  and\ \bibinfo {author} {\bibfnamefont {D.~A.}\ \bibnamefont {Huse}},\
  }\bibfield  {title} {\bibinfo {title} {{Testing whether all eigenstates obey
  the eigenstate thermalization hypothesis}},\ }\href
  {https://doi.org/10.1103/PhysRevE.90.052105} {\bibfield  {journal} {\bibinfo
  {journal} {Phys. Rev. E}\ }\textbf {\bibinfo {volume} {90}},\ \bibinfo
  {pages} {052105} (\bibinfo {year} {2014})}\BibitemShut {NoStop}%
\bibitem [{\citenamefont {Mori}\ \emph {et~al.}(2018)\citenamefont {Mori},
  \citenamefont {Ikeda}, \citenamefont {Kaminishi},\ and\ \citenamefont
  {Ueda}}]{Mori2018_review}%
  \BibitemOpen
  \bibfield  {author} {\bibinfo {author} {\bibfnamefont {T.}~\bibnamefont
  {Mori}}, \bibinfo {author} {\bibfnamefont {T.~N.}\ \bibnamefont {Ikeda}},
  \bibinfo {author} {\bibfnamefont {E.}~\bibnamefont {Kaminishi}},\ and\
  \bibinfo {author} {\bibfnamefont {M.}~\bibnamefont {Ueda}},\ }\bibfield
  {title} {\bibinfo {title} {{Thermalization and prethermalization in isolated
  quantum systems: A theoretical overview}},\ }\href
  {https://doi.org/10.1088/1361-6455/aabcdf} {\bibfield  {journal} {\bibinfo
  {journal} {J. Phys. B At. Mol. Opt. Phys.}\ }\textbf {\bibinfo {volume}
  {51}},\ \bibinfo {pages} {112001} (\bibinfo {year} {2018})}\BibitemShut
  {NoStop}%
\bibitem [{\citenamefont {Das}(2010)}]{Das2010}%
  \BibitemOpen
  \bibfield  {author} {\bibinfo {author} {\bibfnamefont {A.}~\bibnamefont
  {Das}},\ }\bibfield  {title} {\bibinfo {title} {{Exotic freezing of response
  in a quantum many-body system}},\ }\href
  {https://doi.org/10.1103/PhysRevB.82.172402} {\bibfield  {journal} {\bibinfo
  {journal} {Phys. Rev. B}\ }\textbf {\bibinfo {volume} {82}},\ \bibinfo
  {pages} {172402} (\bibinfo {year} {2010})}\BibitemShut {NoStop}%
\bibitem [{\citenamefont {Haldar}\ \emph {et~al.}(2018)\citenamefont {Haldar},
  \citenamefont {Moessner},\ and\ \citenamefont {Das}}]{Haldar2018}%
  \BibitemOpen
  \bibfield  {author} {\bibinfo {author} {\bibfnamefont {A.}~\bibnamefont
  {Haldar}}, \bibinfo {author} {\bibfnamefont {R.}~\bibnamefont {Moessner}},\
  and\ \bibinfo {author} {\bibfnamefont {A.}~\bibnamefont {Das}},\ }\bibfield
  {title} {\bibinfo {title} {{Onset of Floquet thermalization}},\ }\href
  {https://doi.org/10.1103/PhysRevB.97.245122} {\bibfield  {journal} {\bibinfo
  {journal} {Phys. Rev. B}\ }\textbf {\bibinfo {volume} {97}},\ \bibinfo
  {pages} {245122} (\bibinfo {year} {2018})}\BibitemShut {NoStop}%
\bibitem [{\citenamefont {Haldar}\ \emph {et~al.}(2021)\citenamefont {Haldar},
  \citenamefont {Sen}, \citenamefont {Moessner},\ and\ \citenamefont
  {Das}}]{Haldar2021}%
  \BibitemOpen
  \bibfield  {author} {\bibinfo {author} {\bibfnamefont {A.}~\bibnamefont
  {Haldar}}, \bibinfo {author} {\bibfnamefont {D.}~\bibnamefont {Sen}},
  \bibinfo {author} {\bibfnamefont {R.}~\bibnamefont {Moessner}},\ and\
  \bibinfo {author} {\bibfnamefont {A.}~\bibnamefont {Das}},\ }\bibfield
  {title} {\bibinfo {title} {{Dynamical Freezing and Scar Points in Strongly
  Driven Floquet Matter: Resonance vs Emergent Conservation Laws}},\ }\href
  {https://doi.org/10.1103/PhysRevX.11.021008} {\bibfield  {journal} {\bibinfo
  {journal} {Phys. Rev. X}\ }\textbf {\bibinfo {volume} {11}},\ \bibinfo
  {pages} {021008} (\bibinfo {year} {2021})}\BibitemShut {NoStop}%
\bibitem [{\citenamefont {Mori}(2018)}]{Mori2018}%
  \BibitemOpen
  \bibfield  {author} {\bibinfo {author} {\bibfnamefont {T.}~\bibnamefont
  {Mori}},\ }\bibfield  {title} {\bibinfo {title} {{Floquet prethermalization
  in periodically driven classical spin systems}},\ }\href
  {https://doi.org/10.1103/PhysRevB.98.104303} {\bibfield  {journal} {\bibinfo
  {journal} {Phys. Rev. B}\ }\textbf {\bibinfo {volume} {98}},\ \bibinfo
  {pages} {104303} (\bibinfo {year} {2018})}\BibitemShut {NoStop}%
\bibitem [{\citenamefont {Rajak}\ \emph {et~al.}(2018)\citenamefont {Rajak},
  \citenamefont {Citro},\ and\ \citenamefont {{Dalla Torre}}}]{Rajak2018}%
  \BibitemOpen
  \bibfield  {author} {\bibinfo {author} {\bibfnamefont {A.}~\bibnamefont
  {Rajak}}, \bibinfo {author} {\bibfnamefont {R.}~\bibnamefont {Citro}},\ and\
  \bibinfo {author} {\bibfnamefont {E.~G.}\ \bibnamefont {{Dalla Torre}}},\
  }\bibfield  {title} {\bibinfo {title} {{Stability and pre-thermalization in
  chains of classical kicked rotors}},\ }\href
  {https://doi.org/10.1088/1751-8121/aae294} {\bibfield  {journal} {\bibinfo
  {journal} {J. Phys. A Math. Theor.}\ }\textbf {\bibinfo {volume} {51}},\
  \bibinfo {pages} {465001} (\bibinfo {year} {2018})}\BibitemShut {NoStop}%
\bibitem [{\citenamefont {Rajak}\ \emph {et~al.}(2019)\citenamefont {Rajak},
  \citenamefont {Dana},\ and\ \citenamefont {{Dalla Torre}}}]{Rajak2019}%
  \BibitemOpen
  \bibfield  {author} {\bibinfo {author} {\bibfnamefont {A.}~\bibnamefont
  {Rajak}}, \bibinfo {author} {\bibfnamefont {I.}~\bibnamefont {Dana}},\ and\
  \bibinfo {author} {\bibfnamefont {E.~G.}\ \bibnamefont {{Dalla Torre}}},\
  }\bibfield  {title} {\bibinfo {title} {{Characterizations of prethermal
  states in periodically driven many-body systems with unbounded chaotic
  diffusion}},\ }\href {https://doi.org/10.1103/PhysRevB.100.100302} {\bibfield
   {journal} {\bibinfo  {journal} {Phys. Rev. B}\ }\textbf {\bibinfo {volume}
  {100}},\ \bibinfo {pages} {100302(R)} (\bibinfo {year} {2019})}\BibitemShut
  {NoStop}%
\bibitem [{\citenamefont {Hodson}\ and\ \citenamefont
  {Jarzynski}(2021)}]{Hodson2021}%
  \BibitemOpen
  \bibfield  {author} {\bibinfo {author} {\bibfnamefont {W.}~\bibnamefont
  {Hodson}}\ and\ \bibinfo {author} {\bibfnamefont {C.}~\bibnamefont
  {Jarzynski}},\ }\bibfield  {title} {\bibinfo {title} {{Energy diffusion and
  absorption in chaotic systems with rapid periodic driving}},\ }\href
  {https://doi.org/10.1103/physrevresearch.3.013219} {\bibfield  {journal}
  {\bibinfo  {journal} {Phys. Rev. Res.}\ }\textbf {\bibinfo {volume} {3}},\
  \bibinfo {pages} {013219} (\bibinfo {year} {2021})}\BibitemShut {NoStop}%
\bibitem [{\citenamefont {Bukov}\ \emph
  {et~al.}(2015{\natexlab{b}})\citenamefont {Bukov}, \citenamefont
  {Gopalakrishnan}, \citenamefont {Knap},\ and\ \citenamefont
  {Demler}}]{Bukov2015}%
  \BibitemOpen
  \bibfield  {author} {\bibinfo {author} {\bibfnamefont {M.}~\bibnamefont
  {Bukov}}, \bibinfo {author} {\bibfnamefont {S.}~\bibnamefont
  {Gopalakrishnan}}, \bibinfo {author} {\bibfnamefont {M.}~\bibnamefont
  {Knap}},\ and\ \bibinfo {author} {\bibfnamefont {E.}~\bibnamefont {Demler}},\
  }\bibfield  {title} {\bibinfo {title} {{Prethermal floquet steady states and
  instabilities in the periodically driven, weakly interacting bose-hubbard
  model}},\ }\href {https://doi.org/10.1103/PhysRevLett.115.205301} {\bibfield
  {journal} {\bibinfo  {journal} {Phys. Rev. Lett.}\ }\textbf {\bibinfo
  {volume} {115}},\ \bibinfo {pages} {205301} (\bibinfo {year}
  {2015}{\natexlab{b}})}\BibitemShut {NoStop}%
\bibitem [{\citenamefont {{Dalla Torre}}\ and\ \citenamefont
  {Dentelski}(2021)}]{DallaTorre2021}%
  \BibitemOpen
  \bibfield  {author} {\bibinfo {author} {\bibfnamefont {E.~G.}\ \bibnamefont
  {{Dalla Torre}}}\ and\ \bibinfo {author} {\bibfnamefont {D.}~\bibnamefont
  {Dentelski}},\ }\bibfield  {title} {\bibinfo {title} {{Statistical Floquet
  prethermalization of the Bose-Hubbard model}},\ }\href
  {https://doi.org/10.21468/SCIPOSTPHYS.11.2.040} {\bibfield  {journal}
  {\bibinfo  {journal} {SciPost Phys.}\ }\textbf {\bibinfo {volume} {11}},\
  \bibinfo {pages} {040} (\bibinfo {year} {2021})}\BibitemShut {NoStop}%
\bibitem [{\citenamefont {Rubio-Abadal}\ \emph {et~al.}(2020)\citenamefont
  {Rubio-Abadal}, \citenamefont {Ippoliti}, \citenamefont {Hollerith},
  \citenamefont {Wei}, \citenamefont {Rui}, \citenamefont {Sondhi},
  \citenamefont {Khemani}, \citenamefont {Gross},\ and\ \citenamefont
  {Bloch}}]{Rubio-Abadal2020}%
  \BibitemOpen
  \bibfield  {author} {\bibinfo {author} {\bibfnamefont {A.}~\bibnamefont
  {Rubio-Abadal}}, \bibinfo {author} {\bibfnamefont {M.}~\bibnamefont
  {Ippoliti}}, \bibinfo {author} {\bibfnamefont {S.}~\bibnamefont {Hollerith}},
  \bibinfo {author} {\bibfnamefont {D.}~\bibnamefont {Wei}}, \bibinfo {author}
  {\bibfnamefont {J.}~\bibnamefont {Rui}}, \bibinfo {author} {\bibfnamefont
  {S.~L.}\ \bibnamefont {Sondhi}}, \bibinfo {author} {\bibfnamefont
  {V.}~\bibnamefont {Khemani}}, \bibinfo {author} {\bibfnamefont
  {C.}~\bibnamefont {Gross}},\ and\ \bibinfo {author} {\bibfnamefont
  {I.}~\bibnamefont {Bloch}},\ }\bibfield  {title} {\bibinfo {title} {{Floquet
  Prethermalization in a Bose-Hubbard System}},\ }\href
  {https://doi.org/10.1103/PhysRevX.10.021044} {\bibfield  {journal} {\bibinfo
  {journal} {Phys. Rev. X}\ }\textbf {\bibinfo {volume} {10}},\ \bibinfo
  {pages} {021044} (\bibinfo {year} {2020})}\BibitemShut {NoStop}%
\bibitem [{\citenamefont {Peng}\ \emph {et~al.}(2021)\citenamefont {Peng},
  \citenamefont {Yin}, \citenamefont {Huang}, \citenamefont {Ramanathan},\ and\
  \citenamefont {Cappellaro}}]{Peng2021}%
  \BibitemOpen
  \bibfield  {author} {\bibinfo {author} {\bibfnamefont {P.}~\bibnamefont
  {Peng}}, \bibinfo {author} {\bibfnamefont {C.}~\bibnamefont {Yin}}, \bibinfo
  {author} {\bibfnamefont {X.}~\bibnamefont {Huang}}, \bibinfo {author}
  {\bibfnamefont {C.}~\bibnamefont {Ramanathan}},\ and\ \bibinfo {author}
  {\bibfnamefont {P.}~\bibnamefont {Cappellaro}},\ }\bibfield  {title}
  {\bibinfo {title} {{Floquet prethermalization in dipolar spin chains}},\
  }\href {https://doi.org/10.1038/s41567-020-01120-z} {\bibfield  {journal}
  {\bibinfo  {journal} {Nat. Phys.}\ }\textbf {\bibinfo {volume} {17}},\
  \bibinfo {pages} {444} (\bibinfo {year} {2021})}\BibitemShut {NoStop}%
\bibitem [{\citenamefont {Mori}(2022)}]{Mori2022}%
  \BibitemOpen
  \bibfield  {author} {\bibinfo {author} {\bibfnamefont {T.}~\bibnamefont
  {Mori}},\ }\bibfield  {title} {\bibinfo {title} {{Heating Rates under Fast
  Periodic Driving beyond Linear Response}},\ }\href
  {https://doi.org/10.1103/physrevlett.128.050604} {\bibfield  {journal}
  {\bibinfo  {journal} {Phys. Rev. Lett.}\ }\textbf {\bibinfo {volume} {128}},\
  \bibinfo {pages} {050604} (\bibinfo {year} {2022})}\BibitemShut {NoStop}%
\bibitem [{\citenamefont {Breuer}\ and\ \citenamefont
  {Petruccione}(2002)}]{Breuer_text}%
  \BibitemOpen
  \bibfield  {author} {\bibinfo {author} {\bibfnamefont {H.~P.}\ \bibnamefont
  {Breuer}}\ and\ \bibinfo {author} {\bibfnamefont {F.}~\bibnamefont
  {Petruccione}},\ }\href@noop {} {\emph {\bibinfo {title} {{The theory of open
  quantum systems}}}}\ (\bibinfo  {publisher} {Oxford University Press, USA},\
  \bibinfo {year} {2002})\BibitemShut {NoStop}%
\bibitem [{\citenamefont {Redfield}(1957)}]{Redfield1957}%
  \BibitemOpen
  \bibfield  {author} {\bibinfo {author} {\bibfnamefont {A.~G.}\ \bibnamefont
  {Redfield}},\ }\bibfield  {title} {\bibinfo {title} {{On the theory of
  relaxation processes}},\ }\href {https://doi.org/10.1147/rd.11.0019}
  {\bibfield  {journal} {\bibinfo  {journal} {IBM J. Res. Dev.}\ }\textbf
  {\bibinfo {volume} {1}},\ \bibinfo {pages} {19} (\bibinfo {year}
  {1957})}\BibitemShut {NoStop}%
\bibitem [{\citenamefont {Lindblad}(1976)}]{Lindblad1976}%
  \BibitemOpen
  \bibfield  {author} {\bibinfo {author} {\bibfnamefont {G.}~\bibnamefont
  {Lindblad}},\ }\bibfield  {title} {\bibinfo {title} {{On the generators of
  quantum dynamical semigroups}},\ }\href {https://doi.org/10.1007/BF01608499}
  {\bibfield  {journal} {\bibinfo  {journal} {Commun. Math. Phys.}\ }\textbf
  {\bibinfo {volume} {48}},\ \bibinfo {pages} {119--130} (\bibinfo {year}
  {1976})}\BibitemShut {NoStop}%
\bibitem [{\citenamefont {Gorini}\ \emph {et~al.}(1976)\citenamefont {Gorini},
  \citenamefont {Kossakowski},\ and\ \citenamefont {Sudarshan}}]{Gorini1976}%
  \BibitemOpen
  \bibfield  {author} {\bibinfo {author} {\bibfnamefont {V.}~\bibnamefont
  {Gorini}}, \bibinfo {author} {\bibfnamefont {A.}~\bibnamefont
  {Kossakowski}},\ and\ \bibinfo {author} {\bibfnamefont {E.~C.~G.}\
  \bibnamefont {Sudarshan}},\ }\bibfield  {title} {\bibinfo {title}
  {{Completely positive dynamical semigroups of N-level systems}},\ }\href
  {https://doi.org/10.1063/1.522979} {\bibfield  {journal} {\bibinfo  {journal}
  {J. Math. Phys.}\ }\textbf {\bibinfo {volume} {17}},\ \bibinfo {pages}
  {821--825} (\bibinfo {year} {1976})}\BibitemShut {NoStop}%
\bibitem [{\citenamefont {van Kampen}(1992)}]{vanKampen_text}%
  \BibitemOpen
  \bibfield  {author} {\bibinfo {author} {\bibfnamefont {N.~G.}\ \bibnamefont
  {van Kampen}},\ }\href@noop {} {\emph {\bibinfo {title} {{Stochastic
  processes in physics and chemistry}}}}\ (\bibinfo  {publisher} {Elsevier},\
  \bibinfo {year} {1992})\BibitemShut {NoStop}%
\bibitem [{\citenamefont {Kohler}\ \emph {et~al.}(1997)\citenamefont {Kohler},
  \citenamefont {Dittrich},\ and\ \citenamefont {H{\"{a}}nggi}}]{Kohler1997}%
  \BibitemOpen
  \bibfield  {author} {\bibinfo {author} {\bibfnamefont {S.}~\bibnamefont
  {Kohler}}, \bibinfo {author} {\bibfnamefont {T.}~\bibnamefont {Dittrich}},\
  and\ \bibinfo {author} {\bibfnamefont {P.}~\bibnamefont {H{\"{a}}nggi}},\
  }\bibfield  {title} {\bibinfo {title} {{Floquet-Markovian description of the
  parametrically driven, dissipative harmonic quantum oscillator}},\ }\href
  {https://doi.org/10.1103/PhysRevE.55.300} {\bibfield  {journal} {\bibinfo
  {journal} {Phys. Rev. E}\ }\textbf {\bibinfo {volume} {55}},\ \bibinfo
  {pages} {300} (\bibinfo {year} {1997})}\BibitemShut {NoStop}%
\bibitem [{\citenamefont {Mori}(2014)}]{Mori2014}%
  \BibitemOpen
  \bibfield  {author} {\bibinfo {author} {\bibfnamefont {T.}~\bibnamefont
  {Mori}},\ }\bibfield  {title} {\bibinfo {title} {{Natural correlation between
  a system and a thermal reservoir}},\ }\href
  {https://doi.org/10.1103/PhysRevA.89.040101} {\bibfield  {journal} {\bibinfo
  {journal} {Phys. Rev. A}\ }\textbf {\bibinfo {volume} {89}},\ \bibinfo
  {pages} {040101} (\bibinfo {year} {2014})}\BibitemShut {NoStop}%
\bibitem [{\citenamefont {Mori}\ and\ \citenamefont
  {Miyashita}(2008)}]{Mori2008}%
  \BibitemOpen
  \bibfield  {author} {\bibinfo {author} {\bibfnamefont {T.}~\bibnamefont
  {Mori}}\ and\ \bibinfo {author} {\bibfnamefont {S.}~\bibnamefont
  {Miyashita}},\ }\bibfield  {title} {\bibinfo {title} {{Dynamics of the
  density matrix in contact with a thermal bath and the quantum master
  equation}},\ }\href {https://doi.org/10.1143/JPSJ.77.124005} {\bibfield
  {journal} {\bibinfo  {journal} {J. Phys. Soc. Japan}\ }\textbf {\bibinfo
  {volume} {77}},\ \bibinfo {pages} {1--9} (\bibinfo {year}
  {2008})}\BibitemShut {NoStop}%
\bibitem [{\citenamefont {Su{\'{a}}rez}\ \emph {et~al.}(1992)\citenamefont
  {Su{\'{a}}rez}, \citenamefont {Silbey},\ and\ \citenamefont
  {Oppenheim}}]{Suarez1992}%
  \BibitemOpen
  \bibfield  {author} {\bibinfo {author} {\bibfnamefont {A.}~\bibnamefont
  {Su{\'{a}}rez}}, \bibinfo {author} {\bibfnamefont {R.}~\bibnamefont
  {Silbey}},\ and\ \bibinfo {author} {\bibfnamefont {I.}~\bibnamefont
  {Oppenheim}},\ }\bibfield  {title} {\bibinfo {title} {{Memory effects in the
  relaxation of quantum open systems}},\ }\href
  {https://doi.org/10.1063/1.463831} {\bibfield  {journal} {\bibinfo  {journal}
  {J. Chem. Phys.}\ }\textbf {\bibinfo {volume} {97}},\ \bibinfo {pages}
  {5101--5107} (\bibinfo {year} {1992})}\BibitemShut {NoStop}%
\bibitem [{\citenamefont {Gaspard}\ and\ \citenamefont
  {Nagaoka}(1999)}]{Gaspard1999}%
  \BibitemOpen
  \bibfield  {author} {\bibinfo {author} {\bibfnamefont {P.}~\bibnamefont
  {Gaspard}}\ and\ \bibinfo {author} {\bibfnamefont {M.}~\bibnamefont
  {Nagaoka}},\ }\bibfield  {title} {\bibinfo {title} {{Slippage of initial
  conditions for the Redfield master equation}},\ }\href
  {https://doi.org/10.1063/1.479867} {\bibfield  {journal} {\bibinfo  {journal}
  {J. Chem. Phys.}\ }\textbf {\bibinfo {volume} {111}},\ \bibinfo {pages}
  {5668--5675} (\bibinfo {year} {1999})}\BibitemShut {NoStop}%
\bibitem [{\citenamefont {Kohler}\ \emph {et~al.}(1998)\citenamefont {Kohler},
  \citenamefont {Utermann}, \citenamefont {H{\"{a}}nggi},\ and\ \citenamefont
  {Dittrich}}]{Kohler1998}%
  \BibitemOpen
  \bibfield  {author} {\bibinfo {author} {\bibfnamefont {S.}~\bibnamefont
  {Kohler}}, \bibinfo {author} {\bibfnamefont {R.}~\bibnamefont {Utermann}},
  \bibinfo {author} {\bibfnamefont {P.}~\bibnamefont {H{\"{a}}nggi}},\ and\
  \bibinfo {author} {\bibfnamefont {T.}~\bibnamefont {Dittrich}},\ }\bibfield
  {title} {\bibinfo {title} {{Coherent and incoherent chaotic tunneling near
  singlet-doublet crossings}},\ }\href
  {https://doi.org/10.1103/PhysRevE.58.7219} {\bibfield  {journal} {\bibinfo
  {journal} {Phys. Rev. E}\ }\textbf {\bibinfo {volume} {58}},\ \bibinfo
  {pages} {7219} (\bibinfo {year} {1998})}\BibitemShut {NoStop}%
\bibitem [{\citenamefont {{Van Hove}}(1957)}]{vanHove1957}%
  \BibitemOpen
  \bibfield  {author} {\bibinfo {author} {\bibfnamefont {L.}~\bibnamefont {{Van
  Hove}}},\ }\bibfield  {title} {\bibinfo {title} {{The approach to equilibrium
  in quantum statistics. A perturbation treatment to general order}},\ }\href
  {https://doi.org/10.1016/S0031-8914(57)92891-4} {\bibfield  {journal}
  {\bibinfo  {journal} {Physica}\ }\textbf {\bibinfo {volume} {23}},\ \bibinfo
  {pages} {441--480} (\bibinfo {year} {1957})}\BibitemShut {NoStop}%
\bibitem [{\citenamefont {Spohn}(1980)}]{Spohn1980_review}%
  \BibitemOpen
  \bibfield  {author} {\bibinfo {author} {\bibfnamefont {H.}~\bibnamefont
  {Spohn}},\ }\bibfield  {title} {\bibinfo {title} {{Kinetic equations from
  Hamiltonian dynamics: Markovian limits}},\ }\href
  {https://doi.org/10.1103/RevModPhys.52.569} {\bibfield  {journal} {\bibinfo
  {journal} {Rev. Mod. Phys.}\ }\textbf {\bibinfo {volume} {52}},\ \bibinfo
  {pages} {569--615} (\bibinfo {year} {1980})}\BibitemShut {NoStop}%
\bibitem [{\citenamefont {Tindall}\ \emph {et~al.}(2019)\citenamefont
  {Tindall}, \citenamefont {Bu{\v{c}}a}, \citenamefont {Coulthard},\ and\
  \citenamefont {Jaksch}}]{Tindall2019}%
  \BibitemOpen
  \bibfield  {author} {\bibinfo {author} {\bibfnamefont {J.}~\bibnamefont
  {Tindall}}, \bibinfo {author} {\bibfnamefont {B.}~\bibnamefont {Bu{\v{c}}a}},
  \bibinfo {author} {\bibfnamefont {J.~R.}\ \bibnamefont {Coulthard}},\ and\
  \bibinfo {author} {\bibfnamefont {D.}~\bibnamefont {Jaksch}},\ }\bibfield
  {title} {\bibinfo {title} {{Heating-Induced Long-Range $\eta$ Pairing in the
  Hubbard Model}},\ }\href {https://doi.org/10.1103/PhysRevLett.123.030603}
  {\bibfield  {journal} {\bibinfo  {journal} {Phys. Rev. Lett.}\ }\textbf
  {\bibinfo {volume} {123}},\ \bibinfo {pages} {030603} (\bibinfo {year}
  {2019})}\BibitemShut {NoStop}%
\bibitem [{\citenamefont {Prosen}(2011)}]{Prosen2011}%
  \BibitemOpen
  \bibfield  {author} {\bibinfo {author} {\bibfnamefont {T.}~\bibnamefont
  {Prosen}},\ }\bibfield  {title} {\bibinfo {title} {{Open XXZ spin chain:
  Nonequilibrium steady state and a strict bound on ballistic transport}},\
  }\href {https://doi.org/10.1103/PhysRevLett.106.217206} {\bibfield  {journal}
  {\bibinfo  {journal} {Phys. Rev. Lett.}\ }\textbf {\bibinfo {volume} {106}},\
  \bibinfo {pages} {217206} (\bibinfo {year} {2011})}\BibitemShut {NoStop}%
\bibitem [{\citenamefont {{\v{Z}}nidari{\v{c}}}(2015)}]{Znidaric2015}%
  \BibitemOpen
  \bibfield  {author} {\bibinfo {author} {\bibfnamefont {M.}~\bibnamefont
  {{\v{Z}}nidari{\v{c}}}},\ }\bibfield  {title} {\bibinfo {title} {{Relaxation
  times of dissipative many-body quantum systems}},\ }\href
  {https://doi.org/10.1103/PhysRevE.92.042143} {\bibfield  {journal} {\bibinfo
  {journal} {Phys. Rev. E}\ }\textbf {\bibinfo {volume} {92}},\ \bibinfo
  {pages} {042143} (\bibinfo {year} {2015})}\BibitemShut {NoStop}%
\bibitem [{\citenamefont {Sponselee}\ \emph {et~al.}(2018)\citenamefont
  {Sponselee}, \citenamefont {Freystatzky}, \citenamefont {Abeln},
  \citenamefont {Diem}, \citenamefont {Hundt}, \citenamefont {Kochanke},
  \citenamefont {Ponath}, \citenamefont {Santra}, \citenamefont {Mathey},
  \citenamefont {Sengstock},\ and\ \citenamefont {Others}}]{Sponselee2018}%
  \BibitemOpen
  \bibfield  {author} {\bibinfo {author} {\bibfnamefont {K.}~\bibnamefont
  {Sponselee}}, \bibinfo {author} {\bibfnamefont {L.}~\bibnamefont
  {Freystatzky}}, \bibinfo {author} {\bibfnamefont {B.}~\bibnamefont {Abeln}},
  \bibinfo {author} {\bibfnamefont {M.}~\bibnamefont {Diem}}, \bibinfo {author}
  {\bibfnamefont {B.}~\bibnamefont {Hundt}}, \bibinfo {author} {\bibfnamefont
  {A.}~\bibnamefont {Kochanke}}, \bibinfo {author} {\bibfnamefont
  {T.}~\bibnamefont {Ponath}}, \bibinfo {author} {\bibfnamefont
  {B.}~\bibnamefont {Santra}}, \bibinfo {author} {\bibfnamefont
  {L.}~\bibnamefont {Mathey}}, \bibinfo {author} {\bibfnamefont
  {K.}~\bibnamefont {Sengstock}},\ and\ \bibinfo {author} {\bibnamefont
  {Others}},\ }\bibfield  {title} {\bibinfo {title} {{Dynamics of ultracold
  quantum gases in the dissipative Fermi--Hubbard model}},\ }\href
  {https://doi.org/10.1088/2058-9565/aadccd} {\bibfield  {journal} {\bibinfo
  {journal} {Quantum Sci. Technol.}\ }\textbf {\bibinfo {volume} {4}},\
  \bibinfo {pages} {14002} (\bibinfo {year} {2018})}\BibitemShut {NoStop}%
\bibitem [{\citenamefont {Shirai}\ \emph {et~al.}(2018)\citenamefont {Shirai},
  \citenamefont {Mori},\ and\ \citenamefont {Miyashita}}]{Shirai2018}%
  \BibitemOpen
  \bibfield  {author} {\bibinfo {author} {\bibfnamefont {T.}~\bibnamefont
  {Shirai}}, \bibinfo {author} {\bibfnamefont {T.}~\bibnamefont {Mori}},\ and\
  \bibinfo {author} {\bibfnamefont {S.}~\bibnamefont {Miyashita}},\ }\bibfield
  {title} {\bibinfo {title} {{Floquet^^e2^^80^^93Gibbs state in open quantum
  systems}},\ }\href {https://doi.org/10.1140/epjst/e2018-00087-1} {\bibfield
  {journal} {\bibinfo  {journal} {Eur. Phys. J. Spec. Top.}\ }\textbf {\bibinfo
  {volume} {227}},\ \bibinfo {pages} {323--333} (\bibinfo {year}
  {2018})}\BibitemShut {NoStop}%
\bibitem [{\citenamefont {Shirai}\ and\ \citenamefont
  {Mori}(2020)}]{Shirai2020}%
  \BibitemOpen
  \bibfield  {author} {\bibinfo {author} {\bibfnamefont {T.}~\bibnamefont
  {Shirai}}\ and\ \bibinfo {author} {\bibfnamefont {T.}~\bibnamefont {Mori}},\
  }\bibfield  {title} {\bibinfo {title} {{Thermalization in open many-body
  systems based on eigenstate thermalization hypothesis}},\ }\href
  {https://doi.org/10.1103/physreve.101.042116} {\bibfield  {journal} {\bibinfo
   {journal} {Phys. Rev. E}\ }\textbf {\bibinfo {volume} {101}},\ \bibinfo
  {pages} {042116} (\bibinfo {year} {2020})}\BibitemShut {NoStop}%
\bibitem [{\citenamefont {Iadecola}\ and\ \citenamefont
  {Chamon}(2015)}]{Iadecola2015_floquet}%
  \BibitemOpen
  \bibfield  {author} {\bibinfo {author} {\bibfnamefont {T.}~\bibnamefont
  {Iadecola}}\ and\ \bibinfo {author} {\bibfnamefont {C.}~\bibnamefont
  {Chamon}},\ }\bibfield  {title} {\bibinfo {title} {{Floquet systems coupled
  to particle reservoirs}},\ }\href
  {https://doi.org/10.1103/PhysRevB.91.184301} {\bibfield  {journal} {\bibinfo
  {journal} {Phys. Rev. B}\ }\textbf {\bibinfo {volume} {91}},\ \bibinfo
  {pages} {184301} (\bibinfo {year} {2015})}\BibitemShut {NoStop}%
\bibitem [{\citenamefont {Iwahori}\ and\ \citenamefont
  {Kawakami}(2016)}]{Iwahori2016}%
  \BibitemOpen
  \bibfield  {author} {\bibinfo {author} {\bibfnamefont {K.}~\bibnamefont
  {Iwahori}}\ and\ \bibinfo {author} {\bibfnamefont {N.}~\bibnamefont
  {Kawakami}},\ }\bibfield  {title} {\bibinfo {title} {{Long-time asymptotic
  state of periodically driven open quantum systems}},\ }\href
  {https://doi.org/10.1103/PhysRevB.94.184304} {\bibfield  {journal} {\bibinfo
  {journal} {Phys. Rev. B}\ }\textbf {\bibinfo {volume} {94}},\ \bibinfo
  {pages} {184304} (\bibinfo {year} {2016})}\BibitemShut {NoStop}%
\bibitem [{\citenamefont {Haddadfarshi}\ \emph {et~al.}(2015)\citenamefont
  {Haddadfarshi}, \citenamefont {Cui},\ and\ \citenamefont
  {Mintert}}]{Haddadfarshi2015}%
  \BibitemOpen
  \bibfield  {author} {\bibinfo {author} {\bibfnamefont {F.}~\bibnamefont
  {Haddadfarshi}}, \bibinfo {author} {\bibfnamefont {J.}~\bibnamefont {Cui}},\
  and\ \bibinfo {author} {\bibfnamefont {F.}~\bibnamefont {Mintert}},\
  }\bibfield  {title} {\bibinfo {title} {{Completely positive approximate
  solutions of driven open quantum systems}},\ }\href
  {https://doi.org/10.1103/PhysRevLett.114.130402} {\bibfield  {journal}
  {\bibinfo  {journal} {Phys. Rev. Lett.}\ }\textbf {\bibinfo {volume} {114}},\
  \bibinfo {pages} {130402} (\bibinfo {year} {2015})}\BibitemShut {NoStop}%
\bibitem [{\citenamefont {Dai}\ \emph {et~al.}(2016)\citenamefont {Dai},
  \citenamefont {Shi},\ and\ \citenamefont {Yi}}]{Dai2016}%
  \BibitemOpen
  \bibfield  {author} {\bibinfo {author} {\bibfnamefont {C.~M.}\ \bibnamefont
  {Dai}}, \bibinfo {author} {\bibfnamefont {Z.~C.}\ \bibnamefont {Shi}},\ and\
  \bibinfo {author} {\bibfnamefont {X.~X.}\ \bibnamefont {Yi}},\ }\bibfield
  {title} {\bibinfo {title} {{Floquet theorem with open systems and its
  applications}},\ }\href {https://doi.org/10.1103/PhysRevA.93.032121}
  {\bibfield  {journal} {\bibinfo  {journal} {Phys. Rev. A}\ }\textbf {\bibinfo
  {volume} {93}},\ \bibinfo {pages} {032121} (\bibinfo {year}
  {2016})}\BibitemShut {NoStop}%
\bibitem [{\citenamefont {Hartmann}\ \emph {et~al.}(2017)\citenamefont
  {Hartmann}, \citenamefont {Poletti}, \citenamefont {Ivanchenko},
  \citenamefont {Denisov},\ and\ \citenamefont {H{\"a}nggi}}]{Hartmann2017}%
  \BibitemOpen
  \bibfield  {author} {\bibinfo {author} {\bibfnamefont {M.}~\bibnamefont
  {Hartmann}}, \bibinfo {author} {\bibfnamefont {D.}~\bibnamefont {Poletti}},
  \bibinfo {author} {\bibfnamefont {M.}~\bibnamefont {Ivanchenko}}, \bibinfo
  {author} {\bibfnamefont {S.}~\bibnamefont {Denisov}},\ and\ \bibinfo {author}
  {\bibfnamefont {P.}~\bibnamefont {H{\"a}nggi}},\ }\bibfield  {title}
  {\bibinfo {title} {{Asymptotic Floquet states of open quantum systems: The
  role of interaction}},\ }\href {https://doi.org/10.1088/1367-2630/aa7ceb}
  {\bibfield  {journal} {\bibinfo  {journal} {New J. Phys.}\ }\textbf {\bibinfo
  {volume} {19}},\ \bibinfo {pages} {083011} (\bibinfo {year}
  {2017})}\BibitemShut {NoStop}%
\bibitem [{\citenamefont {Schnell}\ \emph {et~al.}(2020)\citenamefont
  {Schnell}, \citenamefont {Eckardt},\ and\ \citenamefont
  {Denisov}}]{Schnell2020}%
  \BibitemOpen
  \bibfield  {author} {\bibinfo {author} {\bibfnamefont {A.}~\bibnamefont
  {Schnell}}, \bibinfo {author} {\bibfnamefont {A.}~\bibnamefont {Eckardt}},\
  and\ \bibinfo {author} {\bibfnamefont {S.}~\bibnamefont {Denisov}},\
  }\bibfield  {title} {\bibinfo {title} {{Is there a Floquet Lindbladian?}},\
  }\href {https://doi.org/10.1103/PhysRevB.101.100301} {\bibfield  {journal}
  {\bibinfo  {journal} {Phys. Rev. B}\ }\textbf {\bibinfo {volume} {101}},\
  \bibinfo {pages} {100301} (\bibinfo {year} {2020})}\BibitemShut {NoStop}%
\bibitem [{\citenamefont {Mizuta}\ \emph {et~al.}(2021)\citenamefont {Mizuta},
  \citenamefont {Takasan},\ and\ \citenamefont {Kawakami}}]{Mizuta2021}%
  \BibitemOpen
  \bibfield  {author} {\bibinfo {author} {\bibfnamefont {K.}~\bibnamefont
  {Mizuta}}, \bibinfo {author} {\bibfnamefont {K.}~\bibnamefont {Takasan}},\
  and\ \bibinfo {author} {\bibfnamefont {N.}~\bibnamefont {Kawakami}},\
  }\bibfield  {title} {\bibinfo {title} {{Breakdown of markovianity by
  interactions in stroboscopic floquet-lindblad dynamics under high-frequency
  drive}},\ }\href {https://doi.org/10.1103/PhysRevA.103.L020202} {\bibfield
  {journal} {\bibinfo  {journal} {Phys. Rev. A}\ }\textbf {\bibinfo {volume}
  {103}},\ \bibinfo {pages} {L020202} (\bibinfo {year} {2021})}\BibitemShut
  {NoStop}%
\bibitem [{\citenamefont {Schnell}\ \emph {et~al.}(2021)\citenamefont
  {Schnell}, \citenamefont {Denisov},\ and\ \citenamefont
  {Eckardt}}]{Schnell2021}%
  \BibitemOpen
  \bibfield  {author} {\bibinfo {author} {\bibfnamefont {A.}~\bibnamefont
  {Schnell}}, \bibinfo {author} {\bibfnamefont {S.}~\bibnamefont {Denisov}},\
  and\ \bibinfo {author} {\bibfnamefont {A.}~\bibnamefont {Eckardt}},\
  }\bibfield  {title} {\bibinfo {title} {{High-frequency expansions for
  time-periodic Lindblad generators}},\ }\href
  {https://doi.org/10.1103/PhysRevB.104.165414} {\bibfield  {journal} {\bibinfo
   {journal} {Phys. Rev. B}\ }\textbf {\bibinfo {volume} {104}},\ \bibinfo
  {pages} {165414} (\bibinfo {year} {2021})}\BibitemShut {NoStop}%
\bibitem [{\citenamefont {Ikeda}\ \emph {et~al.}(2021)\citenamefont {Ikeda},
  \citenamefont {Chinzei},\ and\ \citenamefont
  {Sato}}]{Ikeda2021_nonequilibrium}%
  \BibitemOpen
  \bibfield  {author} {\bibinfo {author} {\bibfnamefont {T.}~\bibnamefont
  {Ikeda}}, \bibinfo {author} {\bibfnamefont {K.}~\bibnamefont {Chinzei}},\
  and\ \bibinfo {author} {\bibfnamefont {M.}~\bibnamefont {Sato}},\ }\bibfield
  {title} {\bibinfo {title} {{Nonequilibrium steady states in the
  Floquet-Lindblad systems: van Vleck's high-frequency expansion approach}},\
  }\href {https://doi.org/10.21468/scipostphyscore.4.4.033} {\bibfield
  {journal} {\bibinfo  {journal} {SciPost Phys. Core}\ }\textbf {\bibinfo
  {volume} {4}},\ \bibinfo {pages} {033} (\bibinfo {year} {2021})}\BibitemShut
  {NoStop}%
\bibitem [{\citenamefont {Wolf}\ \emph {et~al.}(2008)\citenamefont {Wolf},
  \citenamefont {Eisert}, \citenamefont {Cubitt},\ and\ \citenamefont
  {Cirac}}]{Wolf2008}%
  \BibitemOpen
  \bibfield  {author} {\bibinfo {author} {\bibfnamefont {M.~M.}\ \bibnamefont
  {Wolf}}, \bibinfo {author} {\bibfnamefont {J.}~\bibnamefont {Eisert}},
  \bibinfo {author} {\bibfnamefont {T.~S.}\ \bibnamefont {Cubitt}},\ and\
  \bibinfo {author} {\bibfnamefont {J.~I.}\ \bibnamefont {Cirac}},\ }\bibfield
  {title} {\bibinfo {title} {{Assessing non-markovian quantum dynamics}},\
  }\href {https://doi.org/10.1103/PhysRevLett.101.150402} {\bibfield  {journal}
  {\bibinfo  {journal} {Phys. Rev. Lett.}\ }\textbf {\bibinfo {volume} {101}},\
  \bibinfo {pages} {150402} (\bibinfo {year} {2008})}\BibitemShut {NoStop}%
\bibitem [{\citenamefont {Li}\ \emph {et~al.}(2018)\citenamefont {Li},
  \citenamefont {Chen},\ and\ \citenamefont {Fisher}}]{Li2018_quantum}%
  \BibitemOpen
  \bibfield  {author} {\bibinfo {author} {\bibfnamefont {Y.}~\bibnamefont
  {Li}}, \bibinfo {author} {\bibfnamefont {X.}~\bibnamefont {Chen}},\ and\
  \bibinfo {author} {\bibfnamefont {M.~P.}\ \bibnamefont {Fisher}},\ }\bibfield
   {title} {\bibinfo {title} {{Quantum Zeno effect and the many-body
  entanglement transition}},\ }\href
  {https://doi.org/10.1103/PhysRevB.98.205136} {\bibfield  {journal} {\bibinfo
  {journal} {Phys. Rev. B}\ }\textbf {\bibinfo {volume} {98}},\ \bibinfo
  {pages} {205136} (\bibinfo {year} {2018})}\BibitemShut {NoStop}%
\bibitem [{\citenamefont {Chan}\ \emph {et~al.}(2019)\citenamefont {Chan},
  \citenamefont {Nandkishore}, \citenamefont {Pretko},\ and\ \citenamefont
  {Smith}}]{Chan2019}%
  \BibitemOpen
  \bibfield  {author} {\bibinfo {author} {\bibfnamefont {A.}~\bibnamefont
  {Chan}}, \bibinfo {author} {\bibfnamefont {R.~M.}\ \bibnamefont
  {Nandkishore}}, \bibinfo {author} {\bibfnamefont {M.}~\bibnamefont
  {Pretko}},\ and\ \bibinfo {author} {\bibfnamefont {G.}~\bibnamefont
  {Smith}},\ }\bibfield  {title} {\bibinfo {title} {{Unitary-projective
  entanglement dynamics}},\ }\href {https://doi.org/10.1103/PhysRevB.99.224307}
  {\bibfield  {journal} {\bibinfo  {journal} {Phys. Rev. B}\ }\textbf {\bibinfo
  {volume} {99}},\ \bibinfo {pages} {224307} (\bibinfo {year}
  {2019})}\BibitemShut {NoStop}%
\bibitem [{\citenamefont {Skinner}\ \emph {et~al.}(2019)\citenamefont
  {Skinner}, \citenamefont {Ruhman},\ and\ \citenamefont
  {Nahum}}]{Skinner2019}%
  \BibitemOpen
  \bibfield  {author} {\bibinfo {author} {\bibfnamefont {B.}~\bibnamefont
  {Skinner}}, \bibinfo {author} {\bibfnamefont {J.}~\bibnamefont {Ruhman}},\
  and\ \bibinfo {author} {\bibfnamefont {A.}~\bibnamefont {Nahum}},\ }\bibfield
   {title} {\bibinfo {title} {{Measurement-Induced Phase Transitions in the
  Dynamics of Entanglement}},\ }\href
  {https://doi.org/10.1103/PhysRevX.9.031009} {\bibfield  {journal} {\bibinfo
  {journal} {Phys. Rev. X}\ }\textbf {\bibinfo {volume} {9}},\ \bibinfo {pages}
  {031009} (\bibinfo {year} {2019})}\BibitemShut {NoStop}%
\bibitem [{\citenamefont {Li}\ \emph {et~al.}(2019)\citenamefont {Li},
  \citenamefont {Chen},\ and\ \citenamefont {Fisher}}]{Li2019_measurement}%
  \BibitemOpen
  \bibfield  {author} {\bibinfo {author} {\bibfnamefont {Y.}~\bibnamefont
  {Li}}, \bibinfo {author} {\bibfnamefont {X.}~\bibnamefont {Chen}},\ and\
  \bibinfo {author} {\bibfnamefont {M.~P.}\ \bibnamefont {Fisher}},\ }\bibfield
   {title} {\bibinfo {title} {{Measurement-driven entanglement transition in
  hybrid quantum circuits}},\ }\href
  {https://doi.org/10.1103/PhysRevB.100.134306} {\bibfield  {journal} {\bibinfo
   {journal} {Phys. Rev. B}\ }\textbf {\bibinfo {volume} {100}},\ \bibinfo
  {pages} {134306} (\bibinfo {year} {2019})}\BibitemShut {NoStop}%
\bibitem [{\citenamefont {Cao}\ \emph {et~al.}(2019)\citenamefont {Cao},
  \citenamefont {Tilloy},\ and\ \citenamefont
  {de~Luca}}]{Cao2019_entanglement}%
  \BibitemOpen
  \bibfield  {author} {\bibinfo {author} {\bibfnamefont {X.}~\bibnamefont
  {Cao}}, \bibinfo {author} {\bibfnamefont {A.}~\bibnamefont {Tilloy}},\ and\
  \bibinfo {author} {\bibfnamefont {A.}~\bibnamefont {de~Luca}},\ }\bibfield
  {title} {\bibinfo {title} {{Entanglement in a free fermion chain under
  continuous monitoring}},\ }\href
  {https://doi.org/10.21468/SciPostPhys.7.2.024} {\bibfield  {journal}
  {\bibinfo  {journal} {SciPost Phys.}\ }\textbf {\bibinfo {volume} {7}},\
  \bibinfo {pages} {024} (\bibinfo {year} {2019})}\BibitemShut {NoStop}%
\bibitem [{\citenamefont {Bao}\ \emph {et~al.}(2020)\citenamefont {Bao},
  \citenamefont {Choi},\ and\ \citenamefont {Altman}}]{Bao2020}%
  \BibitemOpen
  \bibfield  {author} {\bibinfo {author} {\bibfnamefont {Y.}~\bibnamefont
  {Bao}}, \bibinfo {author} {\bibfnamefont {S.}~\bibnamefont {Choi}},\ and\
  \bibinfo {author} {\bibfnamefont {E.}~\bibnamefont {Altman}},\ }\bibfield
  {title} {\bibinfo {title} {{Theory of the phase transition in random unitary
  circuits with measurements}},\ }\href
  {https://doi.org/10.1103/PhysRevB.101.104301} {\bibfield  {journal} {\bibinfo
   {journal} {Phys. Rev. B}\ }\textbf {\bibinfo {volume} {101}},\ \bibinfo
  {pages} {104301} (\bibinfo {year} {2020})}\BibitemShut {NoStop}%
\bibitem [{\citenamefont {Gullans}\ and\ \citenamefont
  {Huse}(2020)}]{Gullans2020}%
  \BibitemOpen
  \bibfield  {author} {\bibinfo {author} {\bibfnamefont {M.~J.}\ \bibnamefont
  {Gullans}}\ and\ \bibinfo {author} {\bibfnamefont {D.~A.}\ \bibnamefont
  {Huse}},\ }\bibfield  {title} {\bibinfo {title} {{Dynamical Purification
  Phase Transition Induced by Quantum Measurements}},\ }\href
  {https://doi.org/10.1103/PhysRevX.10.041020} {\bibfield  {journal} {\bibinfo
  {journal} {Phys. Rev. X}\ }\textbf {\bibinfo {volume} {10}},\ \bibinfo
  {pages} {041020} (\bibinfo {year} {2020})}\BibitemShut {NoStop}%
\bibitem [{\citenamefont {Fuji}\ and\ \citenamefont {Ashida}(2020)}]{Fuji2020}%
  \BibitemOpen
  \bibfield  {author} {\bibinfo {author} {\bibfnamefont {Y.}~\bibnamefont
  {Fuji}}\ and\ \bibinfo {author} {\bibfnamefont {Y.}~\bibnamefont {Ashida}},\
  }\bibfield  {title} {\bibinfo {title} {{Measurement-induced quantum
  criticality under continuous monitoring}},\ }\href
  {https://doi.org/10.1103/PhysRevB.102.054302} {\bibfield  {journal} {\bibinfo
   {journal} {Phys. Rev. B}\ }\textbf {\bibinfo {volume} {102}},\ \bibinfo
  {pages} {054302} (\bibinfo {year} {2020})}\BibitemShut {NoStop}%
\bibitem [{\citenamefont {Ippoliti}\ \emph {et~al.}(2021)\citenamefont
  {Ippoliti}, \citenamefont {Gullans}, \citenamefont {Gopalakrishnan},
  \citenamefont {Huse},\ and\ \citenamefont {Khemani}}]{Ippoliti2021}%
  \BibitemOpen
  \bibfield  {author} {\bibinfo {author} {\bibfnamefont {M.}~\bibnamefont
  {Ippoliti}}, \bibinfo {author} {\bibfnamefont {M.~J.}\ \bibnamefont
  {Gullans}}, \bibinfo {author} {\bibfnamefont {S.}~\bibnamefont
  {Gopalakrishnan}}, \bibinfo {author} {\bibfnamefont {D.~A.}\ \bibnamefont
  {Huse}},\ and\ \bibinfo {author} {\bibfnamefont {V.}~\bibnamefont
  {Khemani}},\ }\bibfield  {title} {\bibinfo {title} {{Entanglement Phase
  Transitions in Measurement-Only Dynamics}},\ }\href
  {https://doi.org/10.1103/PhysRevX.11.011030} {\bibfield  {journal} {\bibinfo
  {journal} {Phys. Rev. X}\ }\textbf {\bibinfo {volume} {11}},\ \bibinfo
  {pages} {011030} (\bibinfo {year} {2021})}\BibitemShut {NoStop}%
\bibitem [{\citenamefont {Alberton}\ \emph {et~al.}(2021)\citenamefont
  {Alberton}, \citenamefont {Buchhold},\ and\ \citenamefont
  {Diehl}}]{Alberton2021}%
  \BibitemOpen
  \bibfield  {author} {\bibinfo {author} {\bibfnamefont {O.}~\bibnamefont
  {Alberton}}, \bibinfo {author} {\bibfnamefont {M.}~\bibnamefont {Buchhold}},\
  and\ \bibinfo {author} {\bibfnamefont {S.}~\bibnamefont {Diehl}},\ }\bibfield
   {title} {\bibinfo {title} {{Entanglement Transition in a Monitored
  Free-Fermion Chain: From Extended Criticality to Area Law}},\ }\href
  {https://doi.org/10.1103/PhysRevLett.126.170602} {\bibfield  {journal}
  {\bibinfo  {journal} {Phys. Rev. Lett.}\ }\textbf {\bibinfo {volume} {126}},\
  \bibinfo {pages} {170602} (\bibinfo {year} {2021})}\BibitemShut {NoStop}%
\end{thebibliography}%

\end{document}